\documentclass[showpacs,aps,superscriptaddress,letterpaper,nofootinbib]{revtex4}

\usepackage{graphicx}
\usepackage{epsfig}
\usepackage{float}

\newcommand{\be}{\begin{equation}}
\newcommand{\ee}{\end{equation}}
\newcommand{\ba}{\begin{eqnarray}}
\newcommand{\ea}{\end{eqnarray}}

\def\vec#1{{\mbox{\boldmath$#1$}}}

\newcommand{\ep}{\epsilon}

\newcommand{\Et}{\ensuremath{E_\mathrm{T}}}                                     
\newcommand{\met}{\ensuremath{\Et^{\mathrm{miss}}}}                             
\newcommand{\ifb}{\ensuremath{\mathrm{fb^{-1}}}}                                

\def\sss{\scriptscriptstyle}

\begin{document}

\begin{flushright}
\vbox{
\begin{tabular}{l}
{ ANL-HEP-PR-12-62, FERMILAB-PUB-12-475-PPD, arXiv:1208.4018 [hep-ph]}
\end{tabular}
}
\end{flushright}

\vspace{0.6cm}

\title{On the spin and parity of a single-produced resonance at the LHC}
\author{Sara Bolognesi \thanks{e-mail:  sbologne@pha.jhu.edu}}
\affiliation{Department of Physics and  Astronomy, Johns Hopkins University, Baltimore, MD 21218, USA}
\author{Yanyan Gao \thanks{e-mail:  ygao@fnal.gov}}
\affiliation{Fermi National Accelerator Laboratory (FNAL), Batavia, IL 60510, USA}
\author{Andrei V. Gritsan \thanks{e-mail:  gritsan@pha.jhu.edu}}
\affiliation{Department of Physics and  Astronomy, Johns Hopkins University, Baltimore, MD 21218, USA}
\author{Kirill Melnikov \thanks{e-mail:  melnikov@pha.jhu.edu}}
\affiliation{Department of Physics and  Astronomy, Johns Hopkins University, Baltimore, MD 21218, USA}
\author{Markus Schulze \thanks{e-mail:  markus.schulze@anl.gov}}
\affiliation{Argonne National Laboratory (ANL), Lemont, IL 60439, USA}
\author{Nhan V. Tran  \thanks{e-mail:  ntran@fnal.gov}}
\affiliation{Fermi National Accelerator Laboratory (FNAL), Batavia, IL 60510, USA}
\author{Andrew Whitbeck  \thanks{e-mail:  whitbeck@pha.jhu.edu}}
\affiliation{Department of Physics and  Astronomy, Johns Hopkins University, Baltimore, MD 21218, USA}

\date{August 20, 2012}

\begin{abstract}
\vspace{2mm}
The experimental determination of the properties of the newly discovered boson
at the Large Hadron Collider is currently the most crucial task in high energy physics.
We show how information about the spin, parity, and, more generally, the tensor structure 
of the boson couplings can be obtained by studying angular and mass distributions of events 
in which the resonance decays to pairs of gauge bosons,  $ZZ, WW$, and $\gamma \gamma$.
A complete Monte Carlo simulation of the process $pp \to X \to VV \to 4f$ is performed and 
verified by comparing it to an analytic calculation of the decay amplitudes $X \to VV \to 4f$.
Our studies account  for all spin correlations and include general couplings 
of a spin $J=0,1,2$  resonance to Standard Model  particles. 
We also discuss how to use angular and mass distributions of the resonance 
decay products for optimal background rejection. 
It is shown that by the end of the 8 TeV run of the LHC, it might be possible to separate
extreme hypotheses of the spin and parity of the new boson with a confidence level of 
99\% or better for a wide range of models. 
We briefly discuss the feasibility of testing scenarios where the resonances is not a parity eigenstate. 
\end{abstract}


\maketitle

\thispagestyle{empty}


\section{Introduction}
\label{sect1}
The discovery of the new boson~\cite{discovery-atlas, discovery-cms} at the LHC,  
which is further corroborated by the strong evidence from the Tevatron~\cite{evidence},
is the culmination of the hunt for the elusive Higgs boson. 
Three primary decay channels\footnote{Throughout this paper, 
we will use a uniform notation for both on-shell and off-shell massive gauge bosons.}
 $X\to ZZ$, $WW$, and $\gamma \gamma$ 
were observed experimentally by the CMS and ATLAS collaborations.
However, not much is currently 
known about detailed properties of the new boson beyond its mass, $m_X \sim 125~{\rm GeV}$, 
although some information can be reasoned from data.  
We know that the width of the new particle is consistent with being smaller than the experimental 
resolution of about a ${\rm GeV}$. 
We also know that, as a consequence of the Landau-Yang theorem \cite{landau, yang},
the new boson cannot have spin one because it decays to two on-shell photons.
Finally, we know that the relative decay branching fractions 
and production cross-sections of the new particle 
are generally consistent with the Standard Model (SM) Higgs boson 
hypothesis \cite{fits},  although current accuracy of experimental measurements does not allow for 
an unambiguous conclusion.

Since the new boson interacts with massive gauge bosons, we expect it to 
play some role 
in electroweak symmetry breaking. However, this needs to be verified by direct measurements of  its properties. 
In particular, it is  important to experimentally study  
the tensor structure of 
couplings of the new boson to SM fields and its $SU(2) \times U(1)$ quantum 
numbers (if any), avoiding  theoretical prejudice.  
For example, 
we may wonder if the relatively strong interaction of the new particle with electroweak gauge 
bosons {\it already observed} implies that this new boson is not a pseudoscalar. 
One may argue that this is the case because a pseudoscalar must interact with gauge bosons by 
means of higher-dimensional operator whose significant contributions to $X \to VV$ would 
imply low scale physics beyond the SM which should have already been observed 
experimentally. Since no beyond the SM physics has been observed at the LHC, the scale of new physics
cannot be low and it is tempting to 
conclude that the pseudoscalar nature of the new boson is excluded\footnote{
Even stronger arguments about $XVV$ coupling by means of higher-dimensional operators are possible 
if one assumes that  $X$ is a singlet under electroweak $SU(2) \times U(1)$ \cite{zeppenfeld}.
}. 
While such arguments are appealing and may in fact be valid, 
it is important to test them experimentally, {\it especially} when 
such tests are within reach. In fact, as we show in this paper, it is entirely possible 
to achieve that with  data from the 8 TeV run of the LHC using di-boson final states. 
Hence, it is realistic to expect that a clear profile of the new boson can be established 
by purely experimental means in a short period of time.  

The determination of the quantum numbers of a Higgs-like particle was discussed in great detail 
in the literature, see  
Refs.~\cite{DellAquila1986,Nelson1988,Soni1993,Barger1994,Allanach2002,Choi2003,Buszello2004,
Godbole2007,Keung2008,Antipin2008,Hagiwara2009,Cao2010,Gao:2010qx,DeRujula2010,Englert2010,Gainer2011,Ellis2012}.
The strategy that we use in this paper is similar to what has already been 
discussed in Ref.~\cite{Gao:2010qx}. 
In that reference we demonstrated that $X$
decaying to two vector bosons provides an excellent channel 
to study the tensor structure of its couplings and outlined the general way to do so.
Since Ref.~\cite{Gao:2010qx} was written {\it before} the new particle 
was discovered, its mass and production rates were unknown. As a result, 
many 
examples studied in Ref.~\cite{Gao:2010qx} are, by now, of an 
academic interest.   
The discovery of the new boson allows us to extend the discussion  presented 
in  Ref.~\cite{Gao:2010qx}
and arrive at realistic predictions about the prospects for measuring 
its spin and couplings at the LHC.

We extend the analysis reported in Ref.~\cite{Gao:2010qx} in several important ways.  First, 
since the mass of the new resonance is $m_{\sss X}\simeq 125~{\rm  GeV}$, at least one of the 
bosons in the $X \to ZZ/WW$ decay is off-shell. 
Calculations reported in Ref.~\cite{Gao:2010qx} employed general structures 
of scattering amplitudes and general angular distributions  
but did not fully include the off-shell kinematics of the vector bosons;
we improve on this in the current paper. 
We also extend those earlier results by including $WW$ 
and $\gamma \gamma$ final states in the Monte Carlo simulation. 
As we stressed earlier~\cite{Gao:2010qx}, the optimal analysis for the new boson discovery 
and its property measurements requires utilization of the full kinematic information about the 
process. Analysis based on matrix elements or multivariate per-event likelihoods,
such as MELA (Matrix Element Likelihood Analysis), 
adopted by CMS~\cite{discovery-cms, cmshzz2l2q},
allows for optimal  background suppression. 
The same techniques  also guarantee 
the best performance when applied to measurements of the new boson's properties.   

In this paper, we consider the gluon fusion, $gg$, and quark-antiquark annihilation, $q\bar{q}$,
production mechanisms. 
The primary production mode of the SM Higgs boson is expected to be gluon fusion. 
The inclusion of the $q\bar{q}$ production process 
completes all the possible 
initial state polarization scenarios for spin-one and spin-two resonance 
hypotheses  thus allowing for
the most general treatment of kinematics, and inclusion of  all relevant 
 spin correlations. 
We also note that weak Vector Boson Fusion (VBF) is expected to account for $7\%$ 
to the SM Higgs boson production rate.  
Since jet tagging identification would reduce the experimentally observable rate
even further, the contribution from the VBF topology is at the level of a few percent. As a result,
we leave dedicated analysis of the VBF topology as well as the analysis of other final states in the
decay of the new boson to future work.

The paper is organized as follows. 
In Section~\ref{sec:kinematics} we review kinematics in resonance production and decay,
expanding on our earlier work in Ref.~\cite{Gao:2010qx} and focusing on the case relevant
to the observed boson mass $m_X < 2 m_{Z(W)}$. 
In Section~\ref{sec:mc} we discuss the  Monte-Carlo event 
generator for simulating  production and decay of  a new boson 
with different hypotheses for spin and tensor structure of interactions, 
expanding~Ref.~\cite{Gao:2010qx} to include 
new final states covered in this paper.
In Section~\ref{sec:analysis} we discuss the analysis methods.
We summarize the results and conclude in Section~\ref{sec:summary}.
Detailed formulas for angular distributions 
and some numerical results  are given in the Appendix.


\section{Kinematics in Resonance Production and Decay}
\label{sec:kinematics}

Before going into the discussion of how properties of 
the boson $X$ can be studied, it is interesting to point out that the 
determination of the spin-parity of a resonance through 
its decays to two gauge bosons, that subsequently decay to four leptons, 
was first attempted more than a half-century ago, with the study of neutral pion decays
$\pi^0\to\gamma\gamma$ and $\pi^0\to \gamma^*\gamma^*\to e^+e^-e^+e^-$.
Photon polarization in $\pi^0\to\gamma\gamma$ 
can be used to determine $\pi^0$  parity~\cite{yang},
but it is more practical to use the orientation of the planes 
of the Dalitz pairs in the decay
$\pi^0 \to e^+e^-e^+e^-$~\cite{dalitz}.  Further developments of these 
techniques were discussed in Refs.~\cite{Jacob1959, Cabibbo1965}
and additional  refinements were suggested in the context of 
Higgs physics in~Refs.~\cite{DellAquila1986,Nelson1988,Soni1993,
Barger1994,Allanach2002,Choi2003,Buszello2004,
Godbole2007,Keung2008,Antipin2008,Hagiwara2009,Cao2010,Gao:2010qx,DeRujula2010,Englert2010,
Gainer2011,Ellis2012}
and in the context of $B$-physics in Refs.~\cite{Dunietz1991,Kramer1992,pdg,bpolarization}. 
By analogy, the decay $X\to ZZ\to 4\ell$ is an excellent channel to measure the spin, parity, and tensor structure 
of couplings of the new boson since the full decay kinematics are experimentally accessible.
In the channels $X\to WW\to 2\ell2\nu$ and $X\to\gamma\gamma$  less kinematic information is available,
but they can complement the measurements of the resonance properties.  
Other final states of $ZZ$ and $WW$ could be considered, but they typically suffer from higher backgrounds.

We begin by discussing kinematics of the process. Consider a sequence of processes
\begin{equation} 
\label{eqk1}
gg / q \bar q \to X (q) \to V_1(q_1)  V_2(q_2),
\;\;\;
V_1 \to  f(q_{11}) \bar{f}(q_{12}),\;\;\; V_2 \to f(q_{21}) \bar{f}(q_{22}),
\ee
that correspond to the production of a resonance $X$, followed by its decay to two vector bosons, 
followed by their decays to four fermions.  The four-momenta of all particles are shown in parentheses. 
Momentum conservation implies  $q_i = q_{i1}+q_{i2}$, $q = q_1 + q_2$.  We denote the 
invariant mass of the $i$-th 
gauge-boson by  $m_i^2 = q_i^2$ and stress that it can differ from its 
mass $m_{\sss V}^2$.  We assume that the particle $X$ is produced on the mass shell, so that 
$q^2 = (q_1 + q_2)^2 = m_{\sss X}^2$.  In what follows, we will refer to the heavier (lighter)  
of the two gauge bosons as $V_1$ ($V_2$), $m_1>m_2$.

\begin{figure}[t]
\centerline{
\setlength{\epsfxsize}{0.5\linewidth}\leavevmode\epsfbox{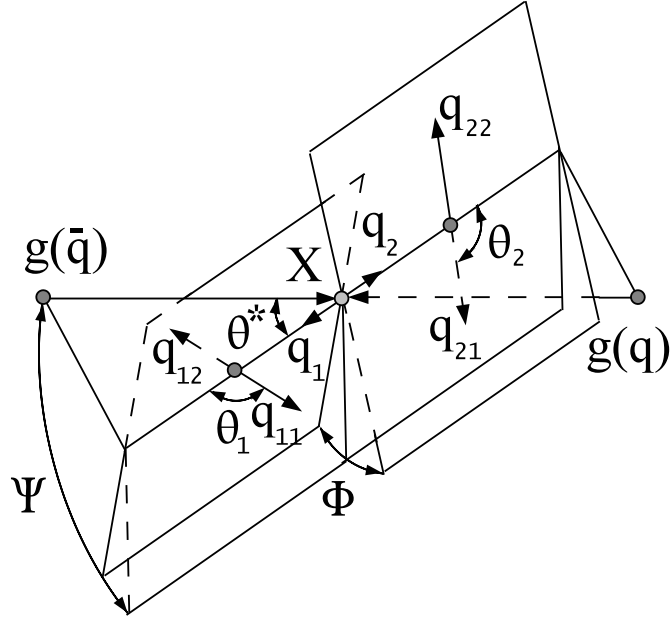}
}
\caption{
Illustration of a $X$ particle production and decay in $pp$ collision
$gg$ or $q\bar{q}\to X\to V_1(q_1) V_2(q_2)$, $V_1 \to  f(q_{11}) \bar{f}(q_{12})$, $V_2 \to f(q_{21}) \bar{f}(q_{22})$.
The three-momenta of the fermions ($f$) and antifermions ($\bar{f}$), $\vec q_{11}$,   $\vec q_{12}$,   
$\vec q_{21}$,  and $\vec q_{22}$, 
are shown in their parent $V_i$ rest-frames, and the three-momenta of the $V_i$ bosons, $\vec q_{i}$, 
are shown in the $X$ rest-frame. For sign convention of the angles between planes see text.}
\label{fig:decay}
\end{figure}

As was already described in Ref.~\cite{Gao:2010qx}, three invariant masses 
$m_{\sss V_1\! V_2}$, $m_1$, and $m_2$, and six angles  
fully characterize the kinematics of the process in Eq.~(\ref{eqk1}) 
in the rest frame of the resonance $X$. Five of these angles are 
illustrated in Fig.~\ref{fig:decay}, 
while  the sixth angle defines the global rotation of an event 
in the plane transverse to the collision axis and, for this reason, 
it is not shown.
We define these angles explicitly through the momenta of the leptons that are directly measurable experimentally.
\begin{itemize} 
\item 
The angles $\theta^*  \in [0,\pi]$ and $\Phi^*  \in [-\pi,\pi]$ are defined through the unit vector 
of $V_1$ direction, $\vec{\hat q}_1 = ( \sin \theta^* \cos \Phi^*, \sin \theta^* \sin \Phi^*, \cos \theta^*)$,
in the rest frame of $X$. In this reference frame, the collision axis is aligned with the $z$-axis, 
$\vec{\hat n}_z = (0,0,1)$, taken as the direction of a colliding quark or one of the colliding gluons.
Note, however, that the angle $\Phi^*$ offset is arbitrary and it is not used in the final analysis. 
Also, when sequential decay of the vector bosons is not available, 
which is  the case for $X\to\gamma\gamma$, only the angle $\theta^*$ is accessible experimentally. 
\item  
The angles 
$\Phi  \in [-\pi,\pi]$ and $\Phi_1 \in [-\pi,\pi]$ 
are the two azimuthal angles between the three planes constructed from 
the $X$ decay products and the two $V_i$-boson decay products in the $X$ rest frame.
The angle $\Psi\in [-\pi,\pi]$ can be used in place of $\Phi_1$,
it is defined as $\Psi=\Phi_1+\Phi/2$ and can be interpreted as the angle between the parton-scattering plane 
and the average between the two decay planes shown in Fig.~\ref{fig:decay}.
These angles are explicitly defined as
\begin{eqnarray}
&& \Phi = \frac{{ \vec{q}_{1} \cdot  (\vec{\hat n}_{1} \times \vec{\hat n}_{2}) }}{|{ \vec{q}_{1} \cdot  (\vec{\hat n}_{1} \times \vec{\hat n}_{2}) }|}
 \times \cos^{-1} \left( - \vec{\hat{n}}_1 \cdot \vec{\hat{n}}_2  \right)
\,, \nonumber \\
&& \Phi_1 = \frac{{ \vec{q}_{1} \cdot  (\vec{\hat n}_{1} \times \vec{\hat n}_{\rm sc}) }}{|{ \vec{q}_{1} \cdot  (\vec{\hat n}_{1} \times \vec{\hat n}_{\rm sc}) }|}
 \times \cos^{-1} \left( \vec{\hat{n}}_1 \cdot \vec{\hat{n}}_{\rm sc}  \right)
\,,
 \label{eq:angles-1}
 \end{eqnarray}
where the normal vectors to the three planes are defined as
\begin{eqnarray}
  \vec{\hat{n}}_1 = \frac{ \vec{q}_{11} \times \vec{q}_{12} } { |  \vec{q}_{11} \times \vec{q}_{12} | }
\,,
\quad \mathrm{~} \quad
  \vec{\hat{n}}_2 = \frac{ \vec{q}_{21}\times \vec{q}_{22} } { |  \vec{q}_{21} \times \vec{q}_{22} | }
\,,
\quad \mathrm{and} \quad
  \vec{\hat{n}}_{\mathrm{sc}} = \frac{ \vec{\hat{n}}_z \times \vec{q}_{1}}{ | \vec{\hat{n}}_z \times \vec{q}_{1} | }
  \,.
 \label{eq:angles-2}
 \end{eqnarray}
In the above equations, $\vec q_{i1(2)}$ is the three-momentum of a fermion (antifermion) 
in the decay of the $V_i$, and $\vec q_{1}=\vec q_{11}+\vec q_{12}$ is the $V_1$ three-momentum,
where all three-momenta are defined in the $X$ rest frame.
\item Finally, the angles $\theta_1$ and $\theta_2 \in [0,\pi]$ are defined as 
\be
\theta_{1} =   \cos^{-1}\left(-\frac {\vec q_2 \cdot \vec q_{11}}{|\vec q_2| |\vec q_{11}|} \right)
\,,\quad
 \theta_{2} =  \cos^{-1}\left(-\frac {\vec q_1 \cdot \vec q_{21}}{|\vec q_1| |\vec q_{21}|} \right)
 \,,
 \label{eq:angles-3}
\ee
where all three-momenta are taken in the rest frame of $V_i$ for the angle $\theta_i$. 
\end{itemize}

The invariant masses of the two-fermion final states, the six angles defined above,
and four-momentum of the initial partonic state exhaust the twelve degrees of freedom 
available to the four particles in the final 
state\footnote{Throughout the paper, we take fermions in the final state to be massless.}. 
The initial state four-momentum defines
the $X$ invariant mass $m_{\sss V_1\! V_2}$ and
the motion of the $X$ system in  
the longitudinal (rapidity $Y$) and  transverse (${\vec p}_T$) directions. 
Both $Y$ and ${\vec p}_T$ distributions depend on the production mechanism
and therefore could help to further differentiate production models either for signal or background.
However, these observables have little discrimination power 
between different signal hypotheses 
once production and decay channels are fixed 
and  they introduce additional systematic 
uncertainties due to QCD effects. It is important to point out that 
the transverse momentum of the $X$ particle introduces smearing in the determination of the 
production angles $\theta^*$  and $\Psi$. The Collins-Soper frame~\cite{Collins1977} 
is designed to minimize the impact of the $X$ transverse momentum on the angular 
measurements. However, the 
effect is expected to be small compared to statistical uncertainties for 
the  luminosity expected in the 8 TeV run of the LHC and, for this reason, 
we do not study it in this paper. 

The full differential mass 
and angular distribution can be expressed using Eq.~(\ref{eq:mixedJtotal}),
where we can factorize the phase-space and propagator terms
%
\begin{eqnarray}
&&\frac{d \Gamma_{J}(m_1,m_2,\cos\theta^\ast,\Psi,\cos\theta_1,\cos\theta_2,\Phi)}
{ \; dm_1 dm_2 d\cos\theta^\ast d\Psi d\cos\theta_1d\cos\theta_2 d\Phi } \propto
     \frac{d \Gamma_{J}(m_1,m_2,\cos\theta^\ast,\Psi,\cos\theta_1,\cos\theta_2,\Phi)}
     { \; d\cos\theta^\ast d\Psi d\cos\theta_1d\cos\theta_2 d\Phi } \times P(m_1,m_2),
\label{eq:differential-1}
\end{eqnarray}
%
which are defined in Ref.~\cite{Choi2003} as
%
\begin{eqnarray}
 && P(m_1,m_2) =
       \Bigl[1-\frac{(m_1+m_2)^2}{m_{\sss X}^2}\Bigr]^\frac{1}{2} 
       \times\Bigl[1-\frac{(m_1-m_2)^2}{m_{\sss X}^2}\Bigr]^\frac{1}{2} 
     \times\frac{m_1^3}{(m_1^2-m_{\sss V}^2)^2+m_{\sss V}^2\Gamma_{\sss V}^2}
     \times\frac{m_2^3}{(m_2^2-m_{\sss V}^2)^2+m_{\sss V}^2\Gamma_{\sss V}^2}.
\label{eq:differential-2}
\end{eqnarray}
%
After integration over the five angles, the differential mass distribution takes the form
%
\begin{eqnarray}
&&\frac{d \Gamma_{J}}{ \; dm_1 dm_2} \propto
    \sum_{\alpha,\beta=-,0,+} \left|A_{\alpha\beta}(m_1,m_2)\right|^2 \times P(m_1,m_2).
\label{eq:differential-3}
\end{eqnarray}
%

Below we discuss how to calculate $A_{\alpha\beta}(m_1,m_2)$ for each spin and coupling hypothesis 
after a brief comment on the notation that we use throughout the paper. 
The polarization vectors of spin-one bosons are denoted by $\epsilon_{i}$; 
we assume them to be transverse, $q_i \epsilon_i = 0$.  
Fermion wave functions are conventional Dirac spinors.
The spin-two $X$  wave function is given by a symmetric
traceless tensor $t_{\mu \nu} $, transverse to its 
momentum $t_{\mu \nu} q^\nu = 0$; its explicit form 
can be found in  Ref.~\cite{Gao:2010qx}. We will often use the notation  
$f^{(i),{\mu \nu}} = \epsilon_i^{\mu}q_i^{\nu} - \epsilon_{i}^\nu q_i^{\mu} $ 
to denote   the field strength tensor of a gauge boson 
with momentum $q_i$ and polarization vector $\epsilon_i$. 
Assuming that momenta of the two bosons, $V_{1,2}$, are along the $z$-axis
$q_{1,2} = (E_{1,2},0, 0, \pm |\vec q|)$, the polarization vectors read 
\begin{eqnarray}
e_{1,2}^{\mu}(0) = \frac{1}{m_{1,2}} \left( \pm |\vec q|, 0, 0, E_{1,2} \right ),
\;\;\;\;\;
e_1^{\mu}(\pm) =  e_2^{\mu}(\mp) = \frac{1}{\sqrt{2}}(0, \mp 1, -i, 0).
\label{eq:polarization} 
\end{eqnarray}
The  conjugate field strength tensor is defined as 
${\tilde f}^{(i)}_{\mu \nu}  = 
1/2\; \epsilon_{\mu \nu \alpha \beta} f^{(i),\alpha \beta} 
= \epsilon_{\mu \nu \alpha \beta} \epsilon_i^{\alpha} q_i^{\beta}$. 
We use ${\tilde q} = q_1 - q_2$ to denote difference of momenta of 
the two gauge bosons. 

\subsection{Spin zero}

Suppose that the new boson is a  spin-zero particle. The general 
scattering amplitude that describes the interaction of this boson with gauge bosons reads 
\begin{eqnarray}
A(X \to V_1 V_2) && = \frac{1}{v} \left ( 
  g^{(0)}_{\sss 1} m_{\sss V}^2 \epsilon_1^* \epsilon_2^* 
+ g^{(0)}_{\sss 2} f_{\mu \nu}^{*(1)}f^{*(2),\mu \nu}
+ g^{(0)}_{\sss 3} f^{*(1),\mu \nu} f^{*(2)}_{\mu \alpha}\frac{ q_{2\nu} q_1^{\alpha}}{\Lambda^2} 
+ g^{(0)}_{\sss 4}  f^{*(1)}_{\mu \nu} {\tilde f}^{*(2),\mu  \nu}
\right )\,,
\label{eq:fullampl-spin0} 
\end{eqnarray}
where $\Lambda$ denotes the scale where new physics could appear.
We insert an explicit factor $m_{\sss V}^2$ in the amplitude 
to allow for a smooth massless limit consistent with generic requirements 
of gauge invariance which is relevant in case $V = \gamma$ or $g$.  

It is instructive to discuss 
the connection between the amplitude in Eq.~(\ref{eq:fullampl-spin0}) 
and the concept of the effective Lagrangian which is often used to discuss properties of the new boson.
While the two approaches are related, 
the amplitude $A(X \to V_1 V_2)$ provides a more general description of the properties of the new 
boson than any effective Lagrangian because the couplings $g_i^{(0)}$ are momentum-dependent form-factors 
that, for example, can have both real and imaginary parts.  We do not expect  this issue to be important for 
the new boson with a mass of $125~{\rm GeV}$, discovered at the LHC, 
but it may be essential for heavier resonances that may be discovered later, 
so we prefer to stick to this description.  
On the other hand, it is also true that effective Lagrangians lead to  streamlined 
prediction for scattering amplitudes, since they 
provide  an opportunity to order contributions of operators 
of different mass dimensions by their relevance,
 thereby reducing the number of terms that contribute to 
 scattering amplitudes.  
Of course, given the scattering amplitude and assuming that form-factors are momentum-independent 
constants, the corresponding Lagrangian can always be constructed.  For example, in case of 
Eq.~(\ref{eq:fullampl-spin0}), the following correspondence is valid 
\begin{eqnarray} 
&& \frac{g^{(0)}_{\sss 1} m_{\sss V}^2}{v}  \epsilon_1^* \epsilon_2^* 
\Leftrightarrow {\cal L} \sim g^{(0)}_{\sss 1} X Z_\mu Z^{\mu},
\;\;\;\;
\frac{g^{(0)}_{\sss 2}}{v} f_{\mu \nu}^{*(1)}f^{*(2),\mu \nu}  
\Leftrightarrow {\cal L} \sim  \frac{g^{(0)}_{\sss 2}}{v} 
X Z_{\mu \nu} Z^{\mu \nu},
\nonumber \\  
&& g^{(0)}_{\sss 3} f^{*(1),\mu \nu} 
f^{*(2)}_{\mu \alpha}\frac{ q_{\nu} q^{\alpha}}{\Lambda^2}
 \Leftrightarrow {\cal L} \sim g^{(0)}_{\sss 3} 
Z^{\mu\nu} Z_{\mu\alpha} \left [ \partial_\nu \partial_\alpha  X \right ] ,
\;\;\;\;\;
g^{(0)}_{\sss 4}  f^{*(1)}_{\mu \nu} {\tilde f}^{*(2),\mu  \nu}
 \Leftrightarrow {\cal L} \sim 
g^{(0)}_{\sss 4} X  Z^{\mu\nu} {\tilde Z}_{\mu \nu},
\end{eqnarray}
where $v$ is the vacuum expectation value of the $X$ field.
Therefore,  terms with $g_{1}^{(0)}$ in $A(X \to V_1 V_2 )$ are associated with  dimension-three operators in the Lagrangian, 
terms with $g_{2}^{(0)}$ and $g_4^{(0)}$ with dimension-five, 
and terms with $g_3^{(0)}$ with dimension seven.
As mentioned above, 
power-counting arguments suggest that lower-dimensional operators give larger 
contributions to the amplitude.  

We can re-write Eq.~(\ref{eq:fullampl-spin0}) as 
\begin{equation}
A(X \to V_1V_2)  = 
 v^{-1} \epsilon_{1}^{*\mu} \epsilon_{2}^{*\nu}
\left (
  a_{1} g_{\mu \nu} m_{\sss X}^2 
+ a_{2} \,q_\mu q_\nu 
+ a_{3} \epsilon_{\mu\nu\alpha\beta}\,q_1^\alpha q_2^\beta
\right)
\,.
\label{eq:ampl-spin0} 
\end{equation}
The coefficients $a_{1,2,3}$ are related to $g^{(0)}_{\sss 1,2,3,4}$ by
\begin{equation}
a_{1} =  g^{(0)}_{\sss 1}\frac{m_{\sss V}^2}{m_{\sss X}^2} 
+ \frac{s}{m_{\sss X}^2} \left(2 g^{(0)}_{\sss 2} + g^{(0)}_{\sss 3} \frac{s}{\Lambda^2}\right)\,,
\;\;\;\;
a_{2} = -\left(2 g^{(0)}_{\sss 2} + g^{(0)}_{\sss 3} \frac{s}{\Lambda^2}\right)\,,\;\;\;\; 
a_{3} = -2 g^{(0)}_{\sss 4},
\label{eq13}
\end{equation}
where $s$ is defined as 
\begin{eqnarray}
&& s = q_1 q_2 = \frac{m_{\sss X}^2 - m_{1}^2 - m_2^2}{2}.
\label{eq:ampl-spin0-s}
\end{eqnarray}

For a spin-zero resonance with couplings shown in Eq.~(\ref{eq:ampl-spin0}),
the three contributing helicity amplitudes are
\begin{eqnarray}
{A_{00}}  &=&  -\frac{m_{\sss X}^2}{v} \left ( { a_1} \sqrt{1+x}+ { a_2}\frac{m_{1}m_{2}}{m_{\sss X}^2}{x}\right ) 
  \,, \nonumber \\ 
 {A_{++}}  &=& \frac{m_{\sss X}^2}{v} \left( { a_1} {+} i{ a_3} \frac{m_{1}m_{2}}{m_{\sss X}^2}\sqrt{x} \right)
 \,, \nonumber \\ 
 {A_{--}}  &=& \frac{m_{\sss X}^2}{v} \left( { a_1} {-} i{ a_3} \frac{m_{1}m_{2}}{m_{\sss X}^2}\sqrt{x} \right)
\, ,
\label{eq:relate-spin0}
\end{eqnarray}
where $x$ is defined as 
\begin{eqnarray}
&& {x}=\left( \frac{m_{\sss X}^2-m_{1}^2-m_{2}^2}{2m_{1}m_{2}} \right)^2-1
\label{eq:ampl-spin0-x}.
\end{eqnarray}

For a SM Higgs boson decaying to two massive vector bosons, $ZZ$ or $WW$,
the values of the couplings are $g^{(0)}_{1}=1$, and $g^{(0)}_{2}=g^{(0)}_{3}=g^{(0)}_{4}=0$. 
A small value  of  
$g^{(0)}_2 \sim  {\cal O}(\alpha_{\rm  EW}) \sim 10^{-2}$
is generated in the SM by electroweak 
 radiative corrections. 
The  $C\!P$-violating constant   $g^{(0)}_{4}$ is tiny in the SM since 
it appears only  at the three-loop level.
For the SM Higgs boson decays $\gamma\gamma$, $Z\gamma$, or $gg$,
only loop-induced couplings are possible so that 
$g^{(0)}_{2}\ne 0$ while the other couplings  are zero.
However, allowing for 
beyond the SM scenarios, values of the $g^{(0)}_{i}$ need 
to be determined experimentally. 
For example, for a pseudoscalar Higgs boson one would expect
$g^{(0)}_{4}\ne 0$ while the other $g^{(0)}_{i}= 0$. It 
is also interesting to consider the model $g^{(0)}_{2}\ne 0$ 
as an alternative to the SM scalar hypothesis, or a mixture of any of the
above contributions.

\subsection{Spin one}

For a spin-one resonance the amplitude depends on two independent terms
%
\begin{eqnarray}
A(X \to V_1 V_2)  = 
 b_1 \left[  (\epsilon_1^* q) (\epsilon_2^* \epsilon_{\sss X}) 
  +  (\epsilon_2^* q) (\epsilon_1^* \epsilon_{\sss X})  \right]
+b_{2} \epsilon_{\alpha\mu\nu\beta} \epsilon_{\sss X}^{\alpha} 
\epsilon_1^{*,\mu} \epsilon_2^{*,\nu} {\tilde q}^\beta
\,,
\label{eq:ampl-spin1} 
\end{eqnarray}
%
where $\epsilon_{\sss X}$ is the polarization vector of particle $X$. The decay into two massless identical vector
bosons is not allowed. 
The helicity amplitudes in the spin-one case corresponding to Eq.~(\ref{eq:ampl-spin1}) are the following
%
\begin{eqnarray} 
\label{eq:relate-spin1}
A_{00} &=&  b_1 \frac{(m_1^2-m_2^2)}{m_{\sss X}} \sqrt{x}
\,, \nonumber \\ 
A_{++} &=&  i\,b_2 \frac{(m_1^2-m_2^2)}{m_{\sss X}} 
\,, \nonumber \\ 
A_{--} &=&  -i\,b_2 \frac{(m_1^2-m_2^2)}{m_{\sss X}} 
\,, \nonumber \\ 
A_{+0} &=&  b_1 m_1 \sqrt{x}
+ i\, b_2 \frac{m_2}{m_{\sss X}^2} \left[
\frac{1}{2}\left( {m_{\sss X}^2} - {m_1^2}+ {m_2^2}\right) \left( \frac{m_1^2}{m_2^2}-1\right)
+ 2 m_1^2 x
\right]
\,, \nonumber \\ 
A_{0+} &=&  - b_1 m_2 \sqrt{x}
- i\, b_2 \frac{m_1}{m_{\sss X}^2} \left[
\frac{1}{2}\left( {m_{\sss X}^2} + {m_1^2} - {m_2^2}\right) \left( \frac{m_2^2}{m_1^2}-1\right)
+ 2 m_2^2 x
\right]
\,, \nonumber \\ 
A_{-0} &=&  b_1 m_1 \sqrt{x}
- i\, b_2 \frac{m_2}{m_{\sss X}^2} \left[
\frac{1}{2}\left( {m_{\sss X}^2} - {m_1^2}+ {m_2^2}\right) \left( \frac{m_1^2}{m_2^2}-1\right)
+ 2 m_1^2 x
\right]
\,, \nonumber \\ 
A_{0-} &=&  - b_1 m_2 \sqrt{x}
+ i\, b_2 \frac{m_1}{m_{\sss X}^2} \left[
\frac{1}{2}\left({m_{\sss X}^2} + {m_1^2} - {m_2^2}\right) \left( \frac{m_2^2}{m_1^2}-1\right)
+ 2 m_2^2 x
\right]
\,.
\end{eqnarray}
%

The model $b_1=g^{(1)}_{\sss 1}\ne0$ corresponds to a vector particle and $b_2=g^{(1)}_{\sss 2}\ne0$ 
to pseudovector particle, assuming parity-conserving interactions. 
Even though the spin-one hypothesis is  rejected
by the observation of $X \to \gamma \gamma$ decay, 
it is still interesting to consider the spin-one models in the decay to
massive vector bosons. 
Indeed, there could be two nearby resonances at 125~GeV, 
one decaying to massive gauge bosons and the other to $\gamma\gamma$,
and there have been models suggested~\cite{Gunion:2012gc} which 
predict the presence of two resonances.

\subsection{Spin two}

For a decay of a spin-two resonance  to two vector bosons, including $ZZ$, $WW$, and $\gamma\gamma$,
the scattering amplitude has the following general form
%
\begin{eqnarray}
\label{eq:fullampl-spin2} 
&&  A(X \to V_1 V_2)  = \Lambda^{-1} \left [ 
2 g^{(2)}_{\sss 1} t_{\mu \nu} f^{*(1)\mu \alpha} f^{*(2)\nu \alpha} 
+ 2 g^{(2)}_{\sss 2} t_{\mu \nu} \frac{q_\alpha q_\beta }{\Lambda^2} 
f^{*(1)\mu \alpha}  f^{*(2)\nu\beta}
+ g^{(2)}_{\sss 3} 
 \frac{{\tilde q}^\beta {\tilde q}^{\alpha}}{\Lambda^2}
t_{\beta \nu} 
\left( f^{*(1)\mu \nu} f^{*(2)}_{\mu \alpha} + f^{*(2)\mu \nu} f^{*(1)}_{\mu \alpha} \right)
\right.  \nonumber \\
&& \left. + g^{(2)}_{\sss 4}\frac{{\tilde q}^{\nu} {\tilde q}^\mu}{{\Lambda^2} } 
t_{\mu \nu} f^{*(1)\alpha \beta} f^{*(2)}_{\alpha \beta}
 + m_{\sss V}^2  \left ( 
2 g^{(2)}_{\sss 5}  t_{\mu\nu} \ep_1^{*\mu} \ep_2^{*\nu} 
+2 g^{(2)}_{\sss 6} \frac{{\tilde q}^\mu q_\alpha}{\Lambda^2}  t_{\mu \nu}
\left ( \ep_1^{*\nu} \ep_2^{*\alpha} - 
\ep_1^{*\alpha} \ep_2^{*\nu} \right ) 
+g^{(2)}_{\sss 7}
 \frac{{\tilde q}^\mu {\tilde q}^\nu}{\Lambda^2}  
t_{\mu \nu} \ep^*_1 \ep^*_2
\right) 
\right.  \nonumber \\
&& \left. 
+g^{(2)}_{\sss 8} \frac{{\tilde q}_{\mu} {\tilde q}_{\nu}}{\Lambda^2} 
 t_{\mu \nu} f^{*(1)\alpha \beta} {\tilde f}^{*(2)}_{\alpha \beta}
+ m_{\sss V}^2  \left ( 
g^{(2)}_{\sss 9} \frac{t_{\mu \alpha} {\tilde q}^\alpha}{\Lambda^2} \epsilon_{\mu \nu \rho \sigma} \epsilon_1^{*\nu} \epsilon_2^{*\rho} q^{\sigma} 
+\frac{g^{(2)}_{\sss 10} t_{\mu \alpha} {\tilde q}^\alpha}{\Lambda^4} \epsilon_{\mu \nu \rho \sigma} q^\rho {\tilde q}^{\sigma} 
\left ( \epsilon_1^{*\nu}(q\epsilon_2^*)+ \epsilon_2^{*\nu}(q\epsilon_1^*) \right )
\right)
\right ] \, ,
\end{eqnarray}
%
where $t_{\mu \nu}$ is the $X$ wave function given by a symmetric traceless tensor~\cite{Gao:2010qx}. This amplitude can be re-written as 
%
\begin{eqnarray}
&&  A(X \to V_1 V_2)  =   \Lambda^{-1} e_1^{*\mu} \, e_2^{*\nu} 
\left[  \, c_{ 1} \, (q_1 q_2) t_{\mu\nu}
+  c_{ 2} \, g_{\mu\nu} t_{\alpha\beta} {\tilde q}^\alpha {\tilde q}^\beta
 +  c_{ 3} \, \frac{q_{2\mu} q_{1\nu}}{m_{\sss X}^2} 
t_{\alpha\beta} {\tilde q}^\alpha {\tilde q}^\beta
+  2 c_{ 41} \, q_{1\nu} q_{2}^{\alpha} t_{\mu\alpha} 
+  2 c_{ 42} \, q_{2\mu} q_{1}^{\alpha} t_{\nu\alpha} 
\right. \nonumber \\
&& \left. 
+ c_{ 5} t_{\alpha \beta} \frac{{\tilde q}^\alpha {\tilde q}^\beta}{m_{\sss X}^2}
 \epsilon_{\mu \nu \rho \sigma} q_1^{\rho} q_2^\sigma
+  c_{ 6} t^{\alpha \beta} {\tilde q}_\beta
 \epsilon_{\mu \nu \alpha  \rho } q^\rho
+\frac{c_{ 7} t^{\alpha \beta} {\tilde q}_\beta}{m_{\sss X}^2}
\left ( 
\epsilon_{\alpha \mu \rho \sigma} q^\rho {\tilde q}^{\sigma}q_\nu
+ \epsilon_{\alpha \nu \rho \sigma} q^\rho {\tilde q}^{\sigma}q_\mu
\right )
\right ].
\label{eq:ampl-spin2} 
\end{eqnarray}
%
In case of massless bosons, like $\gamma\gamma$ or $gg$, the terms with 
$m_{\sss V}$ in Eq.~(\ref{eq:fullampl-spin2}) vanish.
The coefficients $c_{ 1-7}$ can be expressed through $g^{(2)}_{\sss 1,..,10}$  
\begin{eqnarray}
&&c_{ 1} = 2  g^{(2)}_{\sss 1} +  2 g^{(2)}_{\sss 2} \frac{s}{\Lambda^2}\left (1+ \frac{m_{1}^2}{s} \right ) \left (1+ \frac{m_{2}^2}{s} \right )
+2 g^{(2)}_{\sss 5} \frac{m_{\sss V}^2}{s}\,,\;\;\; 
\nonumber \\
&&
c_{ 2} = -\frac{g^{(2)}_{\sss 1}}{2} +g^{(2)}_{\sss 3} \frac{s}{\Lambda^2}\left ( 1 - \frac{m_{1}^2+m_{2}^2}{2s} \right ) + 2 g^{(2)}_{\sss 4} \frac{s}{\Lambda^2}
 + g^{(2)}_{\sss 7}\frac{m_{\sss V}^2}{\Lambda^2}\,, \nonumber \\
&&c_{ 3} = - \left ( \frac{g^{(2)}_{\sss 2}}{2} + g^{(2)}_{\sss 3} + 2 g^{(2)}_{\sss 4} \right ) \frac{m_{\sss X}^2}{\Lambda^2}\,,\;\;\;\;\;\;\; 
\nonumber \\
&&
c_{ 41} = -g^{(2)}_{\sss 1} - g^{(2)}_{\sss 2} \frac{s+m_1^2}{\Lambda^2} - g^{(2)}_{\sss 3}\frac{m_{2}^2}{\Lambda^2}-2g^{(2)}_{\sss 6}\frac{m_{\sss V}^2}{\Lambda^2}\,,  \nonumber \\
&&
c_{ 42} = -g^{(2)}_{\sss 1} - g^{(2)}_{\sss 2} \frac{s+m_2^2}{\Lambda^2} - g^{(2)}_{\sss 3}\frac{m_{1}^2}{\Lambda^2}-2g^{(2)}_{\sss 6}\frac{m_{\sss V}^2}{\Lambda^2}\,,  \nonumber \\
&&
c_{ 5} = 2 g^{(2)}_{\sss 8} \frac{m_{\sss X}^2}{\Lambda^2}\,,\;\;\;\;\;\; 
c_{ 6} = g^{(2)}_{\sss 9}\frac{m_{\sss V}^2}{\Lambda^2}\,,\;\;\;\;\;\;\;\;
c_{ 7} = g^{(2)}_{\sss 10}\frac{m_{\sss X}^2m_{\sss V}^2}{\Lambda^4}.
\label{eq:ampl-spin2-c} 
\end{eqnarray}
We note that when constructing parametrizations of parity-odd amplitudes in Eq.~(\ref{eq:fullampl-spin2}), 
we should carefully exploit Schouten identities to remove mutually-dependent Lorentz structures. 
Such dependences lead to an interesting result -- it turns out that a potentially contributing term   
$t_{\mu \nu} f^{\mu}_{ \alpha} {\tilde f}^{\nu \alpha}$ 
 {\it vanishes} for traceless symmetric 
tensors, $t^{\mu \nu}$.\footnote{We are grateful to  S.~Palmer and  
 M.~Baumgart for useful discussions of this point.}
This cancellation implies that  contributions due to 
$t^{\mu \alpha} {\tilde q}_\alpha \epsilon_{\mu \nu \rho \sigma} \epsilon_1^{*\nu} \epsilon_2^{*\rho} q^{\sigma} $
and 
$t^{\mu \alpha} \epsilon_{1,\alpha}\epsilon_{\mu \nu \rho \sigma} {\tilde q}^{\nu}  \epsilon_2^{*\rho} q^{\sigma}$
are related.  
Therefore, if we do not assume that the amplitude  depends on 
$f_{\mu \nu}$ only, we could have  had two Lorentz structures contributing  to the amplitude.
However, because the Schouten identity connects these structures, 
we choose to keep only one of them in Eq.~(\ref{eq:fullampl-spin2}).

We are now in a position to write down 
the helicity amplitudes for  the spin-two case, using the parametrization 
shown  in Eq.~(\ref{eq:ampl-spin2}). 
For simplicity, we omit $\Lambda$ in the following equations; 
the dependence on $\Lambda$ can be restored on dimensional grounds. 
The amplitudes read 
%
\begin{eqnarray} 
\label{eq:relate-spin2}
&& A_{00} = \frac{m_{\sss X}^4}{ m_1m_2\sqrt{6} } \frac{c_1}{8}
+ \frac{m_1m_2}{ \sqrt{6} } \left [    c_1  \frac{1}{2}\left( 1 + x  \right) - c_2 2x + c_{41} 2x+ c_{42} 2x  \right ] 
- \frac{(m_1^4+m_2^4)}{    m_1m_2\sqrt{6} }\frac{c_1}{4}
+ \frac{m_1m_2(m_1^2-m_2^2)}{ m_{\sss X}^2\sqrt{6} }(c_{41}-c_{42})2x
\nonumber \\ 
&& +\frac{(m_1^8+m_2^8)}{ m_{\sss X}^4 m_1m_2 \sqrt{6} }\frac{c_1}{8}
+ \frac{m_1^3m_2^3}{    m_{\sss X}^4\sqrt{6} } \left [  c_1 \left( \frac{3}{4} + x \right) - c_2 \left( 4x + 8x^2 \right) - c_3 8x^2 \right ] 
+ \frac{m_1m_2(m_1^4+m_2^4)}{    m_{\sss X}^4\sqrt{6} } \left [ - c_1 \frac{1}{2}\left( 1 + x  \right) + c_2 2x \right ]
\,, \nonumber \\ 
&& A_{++} = \frac{m_{\sss X}^2}{\sqrt{6}}\frac{c_1}{4} 
-\frac{(m_1^4+m_2^4)}{m_{\sss X}^2\sqrt{6}}\frac{c_1}{4} 
+\frac{m_1^2 m_2^2}{m_{\sss X}^2\sqrt{6}} \left [ c_1 \left( \frac{1}{2} + x  \right) + c_2 8x \right ]
-i\, \frac{m_1 m_2}{\sqrt{6}} c_6 4\sqrt{x} 
+i\, \frac{m_1^3 m_2^3}{m_{\sss X}^4\sqrt{6}} c_5 8x\sqrt{x}
\,, \nonumber \\ 
&& A_{--} = \frac{m_{\sss X}^2}{\sqrt{6}}\frac{c_1}{4} 
-\frac{(m_1^4+m_2^4)}{m_{\sss X}^2\sqrt{6}}\frac{c_1}{4} 
+\frac{m_1^2 m_2^2}{m_{\sss X}^2\sqrt{6}} \left [ c_1 \left( \frac{1}{2} + x  \right) + c_2 8x \right ]
+i\, \frac{m_1 m_2}{\sqrt{6}} c_6 4\sqrt{x} 
-i\, \frac{m_1^3 m_2^3}{m_{\sss X}^4\sqrt{6}} c_5 8x\sqrt{x}
\,, \nonumber \\ 
&& A_{+0} = \frac{m_{\sss X}^3}{m_2\sqrt{2}}\frac{c_1}{8} 
+\frac{m_{\sss X} m_2}{\sqrt{2}} \left( 1-\frac{m_1^2}{m_2^2} \right) \frac{c_1}{8}  
+\frac{ m_1^2m_2}{m_{\sss X}\sqrt{2}} \left[ c_1\left(\frac{1}{4}+\frac{1}{2}x - \frac{m_2^2}{8m_1^2} - \frac{m_1^2}{8m_2^2} \right) +c_{41} 2x \right] 
\nonumber \\ 
&& +\frac{ m_1^2m_2^3}{m_{\sss X}^3\sqrt{2}} 
c_1 \left[ \frac{1}{8} \left( \frac{m_1^4}{m_2^4}  - \frac{m_2^2}{m_1^2} \right) + \left( 1- \frac{m_1^2}{m_2^2} \right) \left( \frac{3}{8}+\frac{1}{2}x \right) \right] 
-i \, \frac{m_{\sss X} m_1}{\sqrt{2}} c_6 \sqrt{x}
+i \, \frac{m_1^3}{m_{\sss X} \sqrt{2}} c_6 \left( 1-\frac{m_2^2}{m_1^2} \right) \sqrt{x}
-i \, \frac{m_1^3m_2^2}{m_{\sss X}^3 \sqrt{2}} c_7 4x\sqrt{x}
\,, \nonumber \\ 
&& A_{0+} = \frac{m_{\sss X}^3}{m_1\sqrt{2}}\frac{c_1}{8} 
+\frac{m_{\sss X} m_1}{\sqrt{2}} \left( 1-\frac{m_2^2}{m_1^2} \right) \frac{c_1}{8} 
+\frac{ m_1m_2^2}{m_{\sss X}\sqrt{2}} \left[ c_1\left(\frac{1}{4}+\frac{1}{2}x - \frac{m_2^2}{8m_1^2} - \frac{m_1^2}{8m_2^2} \right) +c_{42} 2x \right] 
\nonumber \\ 
&& +\frac{ m_1^3m_2^2}{m_{\sss X}^3\sqrt{2}}  
c_1 \left[ \frac{1}{8} \left( \frac{m_2^4}{m_1^4}  - \frac{m_1^2}{m_2^2} \right) + \left( 1- \frac{m_2^2}{m_1^2} \right) \left( \frac{3}{8}+\frac{1}{2}x \right) \right] 
 -i \, \frac{m_{\sss X} m_2}{\sqrt{2}} c_6 \sqrt{x}
+i \, \frac{m_2^3}{m_{\sss X} \sqrt{2}} c_6 \left( 1-\frac{m_1^2}{m_2^2} \right) \sqrt{x}
-i \, \frac{m_1^2m_2^3}{m_{\sss X}^3 \sqrt{2}} c_7 4x\sqrt{x}
\,, \nonumber \\ 
&& A_{-0} = \frac{m_{\sss X}^3}{m_2\sqrt{2}}\frac{c_1}{8} 
+\frac{m_{\sss X} m_2}{\sqrt{2}} \left( 1-\frac{m_1^2}{m_2^2} \right) \frac{c_1}{8}  
+\frac{ m_1^2m_2}{m_{\sss X}\sqrt{2}} \left[ c_1\left(\frac{1}{4}+\frac{1}{2}x - \frac{m_2^2}{8m_1^2} - \frac{m_1^2}{8m_2^2} \right) +c_{41} 2x \right] 
\nonumber \\ 
&& +\frac{ m_1^2m_2^3}{m_{\sss X}^3\sqrt{2}} 
c_1 \left[ \frac{1}{8} \left( \frac{m_1^4}{m_2^4}  - \frac{m_2^2}{m_1^2} \right) + \left( 1- \frac{m_1^2}{m_2^2} \right) \left( \frac{3}{8}+\frac{1}{2}x \right) \right] 
 +i \, \frac{m_{\sss X} m_1}{\sqrt{2}} c_6 \sqrt{x}
-i \, \frac{m_1^3}{m_{\sss X} \sqrt{2}} c_6 \left( 1-\frac{m_2^2}{m_1^2} \right) \sqrt{x}
+i \, \frac{m_1^3m_2^2}{m_{\sss X}^3 \sqrt{2}} c_7 4x\sqrt{x}
\,, \nonumber \\ 
&& A_{0-}  = \frac{m_{\sss X}^3}{m_1\sqrt{2}}\frac{c_1}{8} 
+\frac{m_{\sss X} m_1}{\sqrt{2}} \left( 1-\frac{m_2^2}{m_1^2} \right) \frac{c_1}{8}  
+\frac{ m_1m_2^2}{m_{\sss X}\sqrt{2}} \left[ c_1\left(\frac{1}{4}+\frac{1}{2}x - \frac{m_2^2}{8m_1^2} - \frac{m_1^2}{8m_2^2} \right) +c_{42} 2x \right] 
\nonumber \\ 
&& +\frac{ m_1^3m_2^2}{m_{\sss X}^3\sqrt{2}}  
c_1 \left[ \frac{1}{8} \left( \frac{m_2^4}{m_1^4}  - \frac{m_1^2}{m_2^2} \right) + \left( 1- \frac{m_2^2}{m_1^2} \right) \left( \frac{3}{8}+\frac{1}{2}x \right) \right] 
 +i \, \frac{m_{\sss X} m_2}{\sqrt{2}} c_6 \sqrt{x}
-i \, \frac{m_2^3}{m_{\sss X} \sqrt{2}} c_6 \left( 1-\frac{m_1^2}{m_2^2} \right) \sqrt{x}
+i \, \frac{m_1^2m_2^3}{m_{\sss X}^3 \sqrt{2}} c_7 4x\sqrt{x}
\,, \nonumber \\ 
&& A_{+-} =  A_{-+} =  m_{\sss X}^2\frac{c_1}{4}
+ \frac{m_1^2 m_2^2}{m_{\sss X}^2} c_1 x
- \frac{\left(m_1^2-m_2^2\right)^2}{m_{\sss X}^2}\frac{c_1}{4}
\,.
\end{eqnarray} 
%

The minimal coupling scenario corresponds to the
case $g^{(2)}_1=g^{(2)}_5\ne0$. However, when higher-dimension operators are considered,
a broader range of options 
becomes available, analogous to $g^{(0)}_2$ and $g^{(0)}_4$ 
in the spin-zero case. This variety of couplings corresponds to the complete set of vector boson 
$V_1$ and $V_2$ polarization states for the given $m_1$ and $m_2$.
Among non-minimal couplings, 
the $g^{(2)}_4$ term provides an interesting
Lorentz structure with the field strength tensors of the two gauge bosons appearing similarly
to the $g^{(0)}_2$ term in the spin-zero case.

We note that, in principle, all couplings that we employ in the paper should be considered functions of 
kinematic invariants, e.g. $g^{(J)}_i(m_{\sss X}^2,q_1^2, q_2^2)$.  
Since the generic  functional form of these couplings is unknown, 
accounting for dependences of $g^{(J)}_i$ on $q_{1,2}^2$
 introduces additional complications that 
are  beyond the scope of this paper. Instead, we prefer to  start the 
spin-parity determination program  by 
treating   couplings 
as constants to understand the ``big picture'' from data.  Once this is  
accomplished,   many further refinements and, in particular, kinematic dependences 
of the coupling constants,  can be examined.  We also note that we use the same 
parametrization for the amplitudes that describe decays  $X \to ZZ$ and $X \to W^+W^-$.
In principle, since $W$'s are not identical particles, the number of independent form factors 
required to describe the most general 
$X \to W^+W^-$ amplitude should be larger. We neglect this effect 
for the reasons explained above. 
Similarly, we point out that for spin-one and spin-two particles, the most general parametrization 
of the amplitude involves terms that depend on the {\it difference} of  invariant masses of 
two vector  bosons so that such terms vanish on the mass shell due to Bose symmetry. 
We do not include such terms in the present calculation and only employ those 
Lorentz structures in all amplitudes that have non-vanishing on-shell limit 
$q_1^2  = q_2^2 = m_{\sss V}^2$.  
While this approximation is not parametric, we believe that the  current 
parametrization already provides sufficient variety of Lorentz structures 
of couplings for hypothesis testing. More sophisticated parameterizations are 
only warranted if credible evidence shows that non-minimal couplings 
that we already introduced are insufficient to describe properties of the new particle. 


\section{Monte Carlo Simulation}
\label{sec:mc}

\begin{table}[b]
\caption{
List of scenarios chosen for the analysis of the production and decay of an exotic $X$ particle
with quantum numbers $J^P$. 
The subscripts $m$ (minimal couplings)  and $h$ (couplings with higher-dimension operators) 
distinguish different scenarios, as discussed in the last column. 
The spin-zero and spin-one $X$ production parameters do not affect the angular and mass distributions,
and therefore are not specified.
}
\begin{tabular}{cccc}
\hline\hline
\vspace{0.1cm}
scenario & $X$ production  & $X\to VV$ decay   & comments \\
\hline
$0_m^+$ &  $gg\to X$ & $g_1^{(0)}\ne0$ in Eq.~(\ref{eq:fullampl-spin0})  & SM Higgs boson scalar \\
$0_h^+$  &  $gg\to X$ & $g_2^{(0)}\ne0$ in Eq.~(\ref{eq:fullampl-spin0}) & scalar with higher-dimension operators\\
$0^-$  &  $gg\to X$ & $g_4^{(0)}\ne0$ in Eq.~(\ref{eq:fullampl-spin0})  & pseudo-scalar  \\
$1^+$  &  $q\bar{q}\to X$  & $b_2\ne0$ in Eq.~(\ref{eq:ampl-spin1})      & exotic pseudo-vector \\
$1^-$   &  $q\bar{q}\to X$  & $b_1\ne0$ in Eq.~(\ref{eq:ampl-spin1})      & exotic vector \\
$2_m^+$ 
      & ~~$g^{(2)}_{\sss 1}\ne0$ in Eq.~(\ref{eq:fullampl-spin2})~~  
      & ~~$g^{(2)}_{\sss 1}=g^{(2)}_{\sss 5}\ne0$ in Eq.~(\ref{eq:fullampl-spin2})~~ 
      & ~~graviton-like tensor with minimal couplings~~ \\
$2_h^+$ 
      & $g^{(2)}_{\sss 4}\ne0$ in Eq.~(\ref{eq:fullampl-spin2}) 
      & $g^{(2)}_{\sss 4}\ne0$  in Eq.~(\ref{eq:fullampl-spin2}) 
      & tensor with higher-dimension operators \\
\vspace{0.1cm}
$2_h^-$ 
      &  $g^{(2)}_{\sss 8}\ne0$ in Eq.~(\ref{eq:fullampl-spin2})  
      & $g^{(2)}_{\sss 8}\ne0$ in Eq.~(\ref{eq:fullampl-spin2})
      & ``pseudo-tensor''\\
\hline\hline
\end{tabular}
\label{table-scenarios}
\end{table}

We have extended the simulation program~\cite{Gao:2010qx, support} to allow for various di-boson 
final states and to include the option of resonances decaying to off-shell gauge bosons. 
This program simulates the production and decay to two vector bosons of the spin-zero, spin-one, 
and spin-two resonances in hadron-hadron collisions, including all spin correlations.
The processes $gg / q\bar q \to X\to ZZ$ and  $WW\to 4f$,
as well as  $gg / q\bar q\to X\to \gamma\gamma$, are implemented.
It includes the general couplings of the $X$ particle 
to gluons and quarks in production and to vector bosons in decay.
The program can be interfaced to parton shower simulation (e.g. PYTHIA~\cite{pythia}) 
as well as full detector simulation through the Les Houches Event file format.  

As we discussed in Sec.~\ref{sec:kinematics}, in principle there is  
a large number of coupling constants to be determined.  To illustrate 
the main idea of spin-parity determination, 
we  pick several scenarios listed in Table~\ref{table-scenarios}.
Among them we include the  SM Higgs boson spin-zero hypothesis ($0^+_m$) and the  
graviton-like minimal coupling 
hypothesis  for spin-two ($2^+_m$). Other, more exotic, hypotheses are
also considered. 
We note that for the spin-two scenarios, we assume that gluon fusion dominates the production 
mechanism, which is the case for the minimal coupling Kaluza-Klein 
graviton ($2^+_m$)~\cite{Antipin2008},
and this assumption may have an impact on the final 
results for the achievable significance of spin hypotheses separation. 
On the other hand, 
for  the spin-zero scenarios, the production mechanism
does not affect the angular and mass distributions.
The chosen scenarios listed in Table~\ref{table-scenarios} are similar to those 
considered in our earlier paper~\cite{Gao:2010qx}.

Distributions of some of the representative observables are 
shown in  Fig.~\ref{fig:simulated} for $m_{\sss X}=125$~GeV. 
A complete set of distributions in the $ZZ$ and $WW$ final states 
is shown in Appendix~\ref{sec:appendix-b}
in Figs.~\ref{fig:simulated-zz-mass}, \ref{fig:simulated-zz-angles},~\ref{fig:simulated-ww}.
Throughout the paper we consider $\sqrt{s}= 8~{\rm TeV}$ proton-proton collisions 
and use the CTEQ6L1 parton distribution functions \cite{cteq}.

In the following we describe a simplified treatment of the detector effects 
which is not meant to reproduce exactly any of the LHC experiments, 
but still allows us to reliably understand feasibility of spin-parity 
studies  at the LHC. We introduce smearing of the track  
momentum transverse to the collision axis, $p_T$, and photon cluster energy.
However, the exact resolution parameterization is not crucial  
as long as the overall signal-to-background separation is reproduced well.
We mimic detector acceptance effects by cutting on geometric 
and kinematic parameters, such as $p_T$ and pseudorapidity, $\eta=-\ln\tan(\theta/2)$. 
Both leptons and photons are required to be in the effective acceptance range $|\eta|<2.5$.

The main backgrounds in the $X\to ZZ$, $WW$, and $\gamma\gamma$ analyses are the continuum 
di-boson production, including $Z\gamma^*$ for $ZZ$~\cite{discovery-atlas, discovery-cms}.
These are modeled with 
POWHEG~\cite{powheg} ($ZZ$) and MadGraph~\cite{madgraph} ($WW, \gamma\gamma$).
Additional contributions of backgrounds with fake vector boson reconstruction 
requires special treatment. However, their contributions are smaller and observable
distributions are similar to the $VV$ background, so their contributions can be effectively accounted for by
rescaling the di-boson background rate to match total background rates observed 
by the LHC experiments. 

\begin{figure}[t]
\centerline{
\epsfig{figure=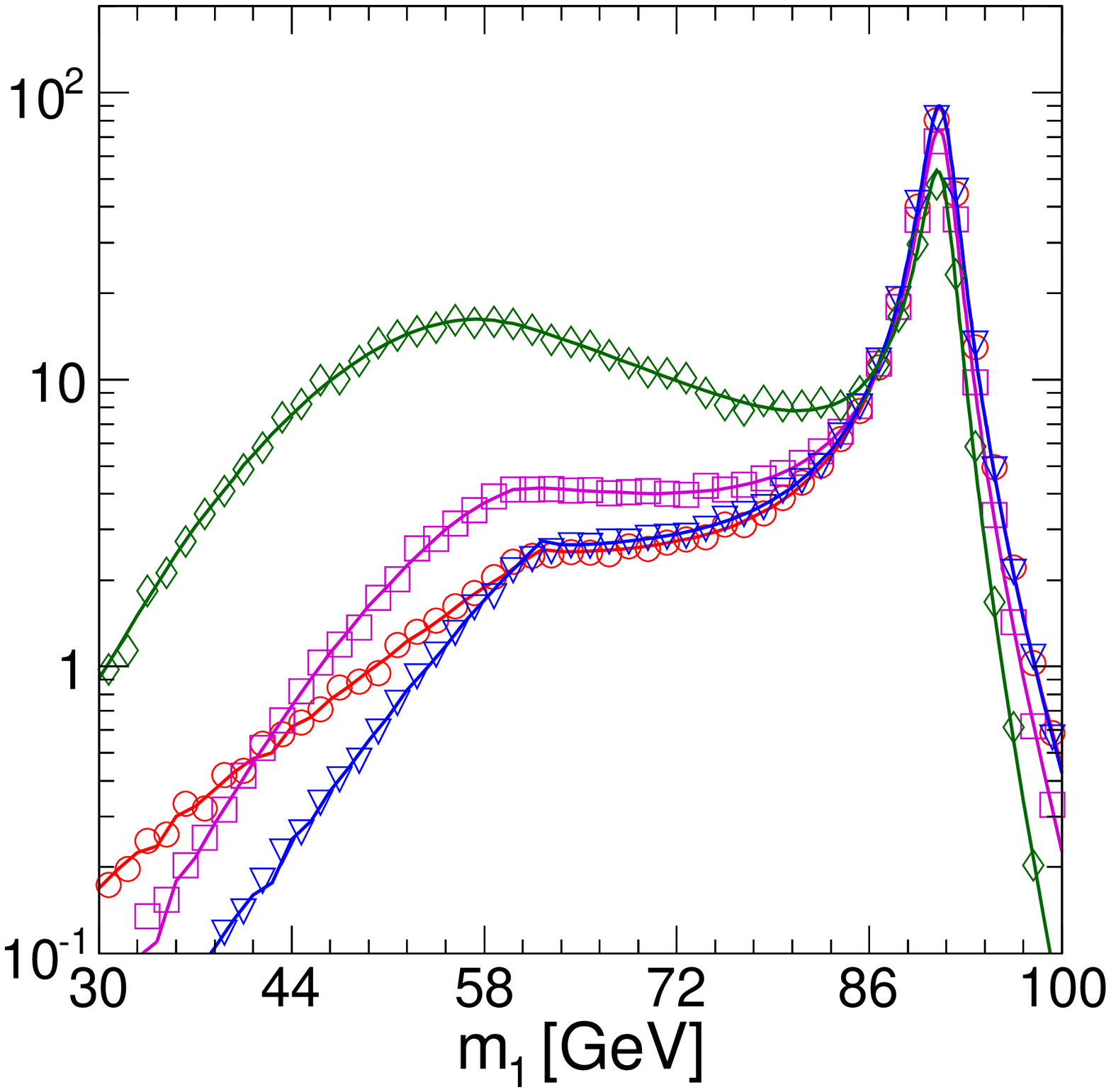,width=0.33\linewidth}
\epsfig{figure=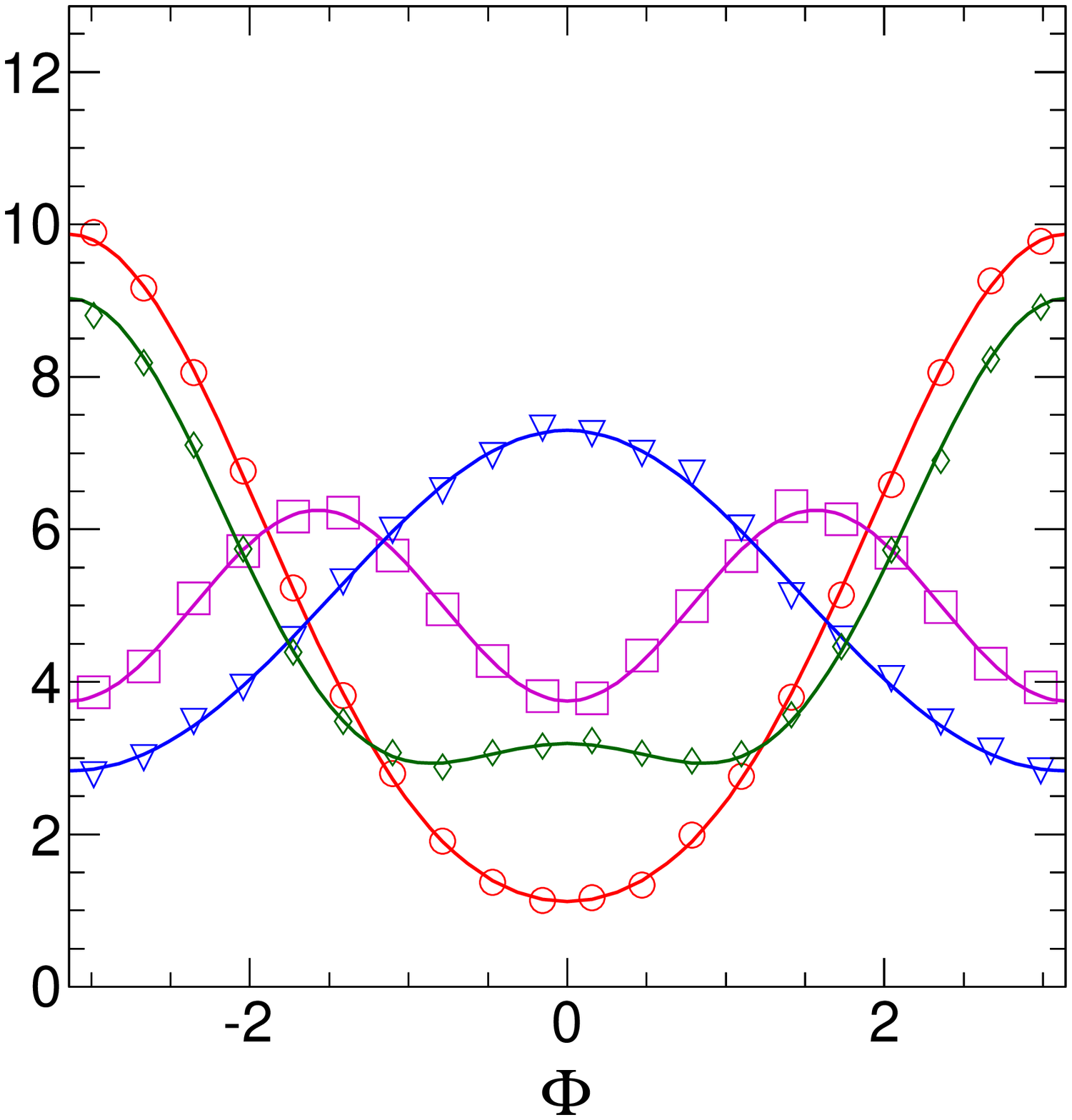,width=0.33\linewidth}
\epsfig{figure=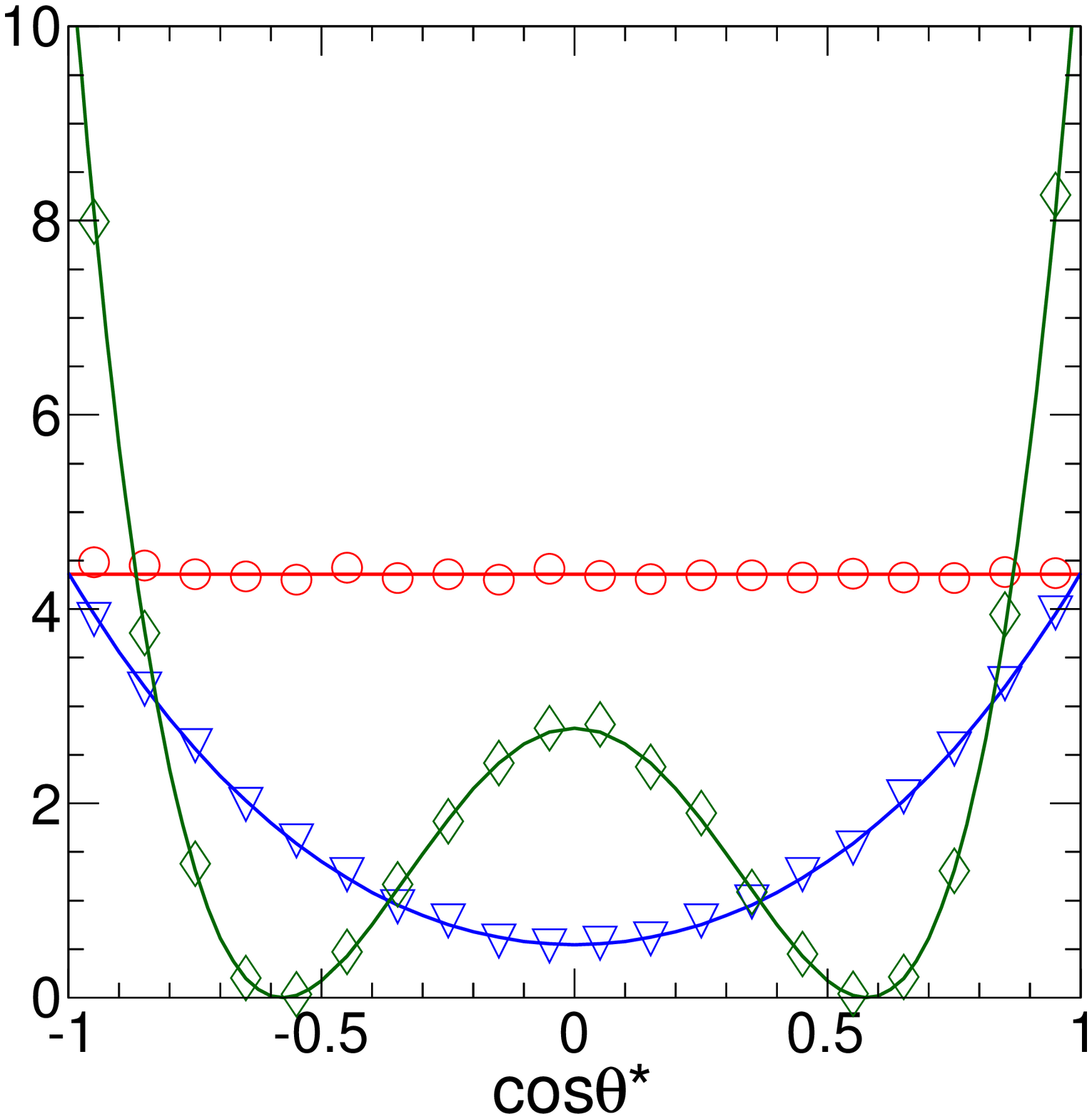,width=0.33\linewidth}
}
\caption{
Distributions of some of the representative observables:
$m_1$ in the  $X\to ZZ$ analysis (left),
$\Phi$ in the  $X\to WW$ analysis (middle), and
$\cos\theta^*$ in the $X\to\gamma\gamma $ analysis.
Four signal hypotheses are shown:
SM Higgs boson (red circles), $0^-$ (magenta squares), $2^+_m$ (blue triangles), 
$2^+_h$ (green diamonds), as defined in Table~\ref{table-scenarios}.
Points show simulated events and lines show projections of analytical distributions.
Here and throughout the paper, where only shapes of the distributions
are illustrated and unless otherwise noted, units on the $y$ axis are arbitrary. 
}
\label{fig:simulated}
\end{figure}


\section{\boldmath Analysis Methods}
\label{sec:analysis}

In this Section, we illustrate the application of the matrix element 
analysis formalism to  distinguishing  different spin-parity hypotheses 
for the observed boson near 125 GeV.
We illustrate this with 
the seven  scenarios defined in Table~\ref{table-scenarios}
and comment on future direction of the measurements.

In  Ref.~\cite{Gao:2010qx} 
we pointed out that the ultimate goal of the analysis should be 
the experimental determination of  all helicity amplitudes that involve 
$X$ and two gauge  bosons. 
The techniques discussed here and in Ref.~\cite{Gao:2010qx} are ideally 
suited for 
such measurements since  parameters in the angular and mass distributions
become fit parameters in analysis of data. However, such multi-parameter 
fits require large samples of the signal events which are not yet available. 
Therefore, in our opinion, the 
first step in understanding the spin-parity of the  
resonance should be distinguishing  between different hypotheses. 
For such a goal, a simplified, but still optimal,  analysis approach 
can be developed that employs just two observables. 
A simple extension of 
this analysis,
which naturally arises 
if we assume, for example, that the observed resonance is a mixed 
spin-parity state, 
 is to fit for ratios of couplings.   Ultimately, this approach will  lead 
to a complete multi-dimensional fit of all coupling parameters
using a complete set of kinematic observables.

Going back to the two-dimensional fit, we note that 
one of the two observables is related to the resonance mass as it 
typically has the largest discriminating power against the background.
This observable depends on the final state; for example, it is the four-lepton 
invariant mass $m_{4\ell}$ in the $X\to ZZ\to 4\ell$ analysis, 
the transverse mass $m_{T}$~\cite{discovery-atlas, discovery-cms} 
in the $X\to WW\to 2\ell 2\nu$ analysis, 
or the two-photon invariant mass $m_{\gamma\gamma}$ 
in the $X\to\gamma\gamma$ analysis.

The second observable combines other  kinematic 
information that is available, and it is designed 
to distinguish  between  
different signal spin-parity hypotheses 
in the optimal way. 
In the $X\to ZZ\to 4\ell$ analysis we build the kinematic discriminant,  defined in the MELA 
approach adopted by the CMS experiment~\cite{discovery-cms, cmshzz2l2q},
which combines the five angular and two mass observables in the optimal way.
In the $X\to WW\to 2\ell 2\nu$ analysis, the complete matrix element information
cannot be exploited because of the neutrinos in the final state. 
Therefore, we adopt a simplified approach by picking one observable that 
is most sensitive to the spin-parity of $X$. We found this observable to be 
the di-lepton invariant mass $m_{\ell\ell}$  while  the opening angle between
the two leptons in the transverse plain provides less sensitivity. 
Finally, in the $X\to\gamma\gamma$ analysis, the only available observable is $\cos\theta^*$
since there is no further sequential decay chain involved.

We note that it is not our goal in this paper to demonstrate how the analysis should be optimized for the 
signal-background separation. Doing so requires simulation of detector performance and 
of all background processes~\cite{discovery-atlas, discovery-cms}. 
Instead, we assume an excess of signal events over background in each of the
three channels, $X\to ZZ$, $WW$, and $\gamma\gamma$,
and calculate the achievable level of separation power between different 
signal spin-parity hypotheses. While precise prediction
of spin-parity separation significances
also requires detailed simulation, as long as the phase-space of the discriminating observables is well-modeled, 
such predictions  are less sensitive to details of the analysis once a given signal significance is observed.

We present results for the expected separation significance between the SM Higgs boson scenario 
($0_m^+$) and various $J^P$ and coupling hypotheses defined in Table~\ref{table-scenarios} 
for a fixed hypothesis of a signal excess, which we take to be $5\sigma$ for the SM Higgs-like resonance.
The performance quoted in Table~\ref{table-separation} follows from the studies presented 
in the following subsections and can be interpreted in terms of integrated luminosity 
and $pp$ collision energy at the LHC  for each 
of the three channels $X\to ZZ$, $WW$, and $\gamma\gamma$.
We observe that a simple rule of scaling with luminosity $L$, ${\rm significance} \sim \sqrt{L} $, 
is a very good approximation in these studies as long as the uncertainties 
are dominated by statistical errors. 

\begin{table}[t]
\caption{
Expected separation significance (Gaussian $\sigma$)
between the SM Higgs boson scenario ($0_m^+$) and 
various $J^P_x$ hypotheses defined in Table~\ref{table-scenarios}.
Expectations are given for the scenario of a 5.0\,$\sigma$ signal-to-background separation
observed in the search for the SM Higgs boson in each channel, 
and therefore interpretation in terms of integrated luminosity 
and $pp$ collision energy at the LHC may differ significantly 
between the three channels $X\to ZZ$, $WW$, and $\gamma\gamma$.
}
\begin{tabular}{lccc}
\hline\hline
\vspace{0.1cm}
scenario & ~~$X\to ZZ$~~ &  ~~$X\to WW$~~ & ~~$X\to\gamma\gamma$~~  \\
\hline
\vspace{0.1cm}
$0_m^+$ vs background & 5.0 & 5.0 & 5.0 \\
\hline
$0_m^+$ vs $0_h^+$      & 1.7 &  1.1 &  0.0  \\
$0_m^+$ vs $0^-$            &  2.9 &  1.2 &  0.0   \\
$0_m^+$ vs $1^+$           &  1.9 &  2.0 &  -- \\
$0_m^+$ vs $1^-$            &  2.6 &  3.2 & --  \\
$0_m^+$ vs $2_m^+$     &  1.5  &  2.8 &  2.4  \\
$0_m^+$ vs $2_h^+$      &  $\sim$5 & 1.1 &   3.1   \\
\vspace{0.1cm}
$0_m^+$ vs $2_h^-$       &  $\sim$5 &  2.5 &  3.1   \\
\hline\hline
\end{tabular}
\label{table-separation}
\end{table}

We use an extended maximum-likelihood fit~\cite{Gao:2010qx}
to extract simultaneously the signal and background yields. The likelihood is defined as
%
\begin{equation}
{{\cal L}_k} =  \exp\left( - n_{\rm sig}-n_{\rm bkg}  \right) 
\prod_i\left( n_{\rm sig} \times{\cal P}^{k}_{\rm sig}(\vec{x}_{i};~\vec{\alpha};~\vec{\beta})  
+n_{\rm bkg} \times{\cal P}_{\rm bkg}(\vec{x}_{i};~\vec{\beta})  
\right)\,,
\label{eq:likelihood}
\end{equation}
where $n_{\rm sig}$ is the number of signal events, $n_{{\rm bkg}}$ is the number of background events and 
${\cal P}(\vec{x}_{i};\vec{\alpha};\vec{\beta})$ is the probability density function for background or signal
for different spin hypotheses, $k$.
Each event candidate, $i$, is characterized by a set of two observables $\vec{x}_{i}=(m, D)$. The signal coupling 
parameters are collectively denoted by $\vec{\alpha}$, and the remaining parameters by $\vec{\beta}$. 
The correlated $(m, D)$ distribution is parameterized with a binned histogram (template) using simulation. 
The likelihood ${\cal L}_k$ in Eq.~(\ref{eq:likelihood}) is evaluated independently for each spin hypothesis $k$. 
Two sets of pseudo-experiments are generated, each with same average number 
of signal events  of a particular type embedded
into the expected background.  

Examples of 
distributions 
of   $-{2\ln({\cal L}_1/{\cal L}_2)}$ 
are shown for a large number
of generated experiments in Fig.~\ref{fig:separation}, 
where one of the signal types  is chosen to be the SM Higgs boson.
The probability for an alternative signal 
to produce a value of $-{2\ln({\cal L}_1/{\cal L}_2)}$
below the median value for the SM Higgs boson hypothesis
is taken as the one-sided Gaussian probability 
and interpreted as the number of Gaussian 
standard deviations, ${\cal S}$. The value of ${\cal S}$ 
corresponds to an effective separation
between the two distributions in the symmetric case, or equivalently to the expected separation 
between the two hypotheses. 
However, a certain amount of asymmetry between the distributions is possible,
as we note in some cases below, and the expected significance of separating type 2 signal from type 1 
may differ from separation of type 1 from type 2. An approximate average of the two values could be 
obtained from the point beyond which the right-side tail of the left histogram and the left-side tail 
of the right histogram have equal areas (corresponding to ${\cal S}/2$). 
We choose to quote the first of the three values as more relevant for separation of alternative
hypotheses from the SM Higgs boson. A similar technique can be employed for the significance calculation 
of the signal excess over background.
Below we discuss details of the analysis methods that are particular to each channel.

\begin{figure}[t]
\centerline{
\epsfig{figure=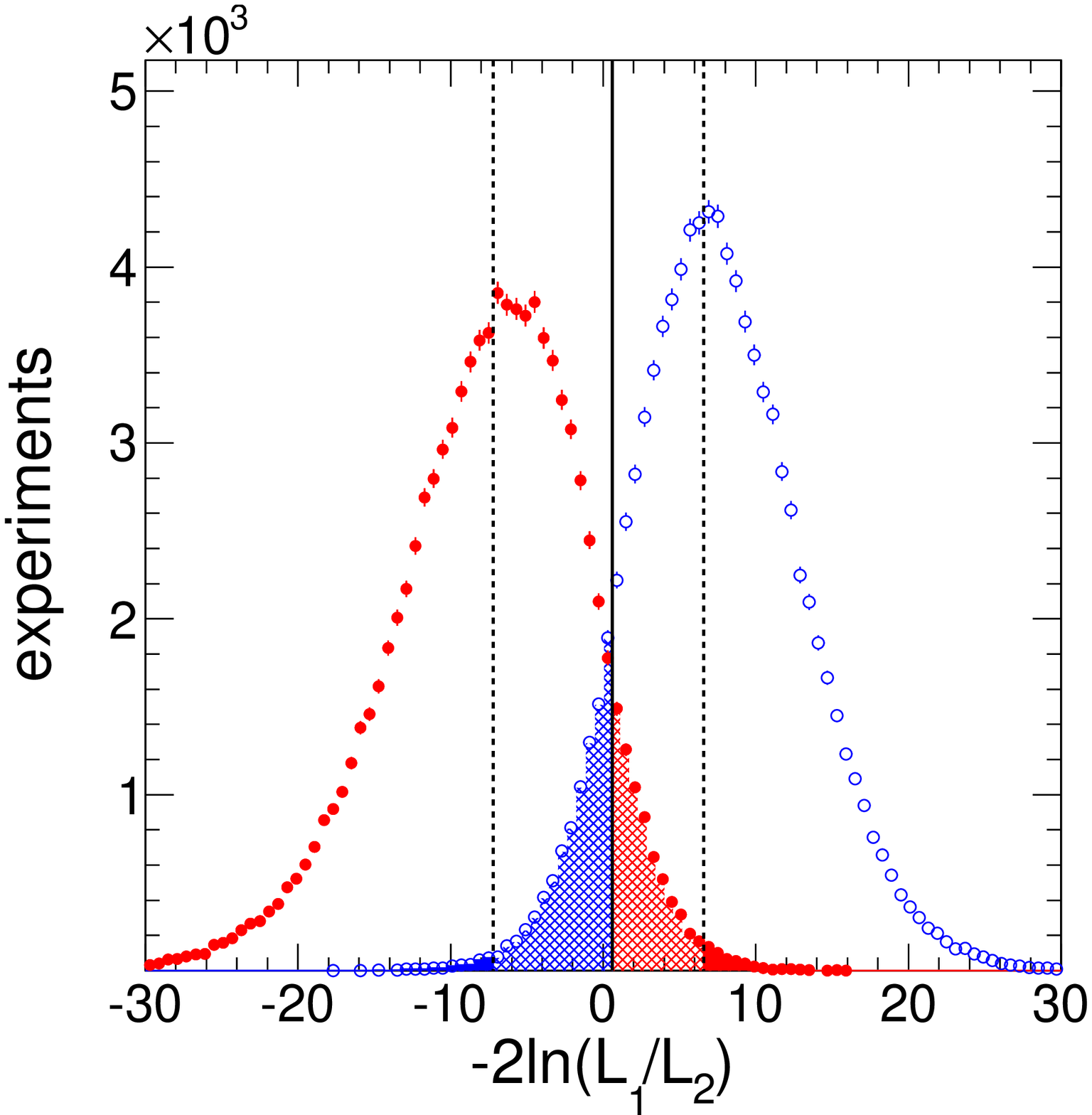,width=0.33\linewidth}
\epsfig{figure=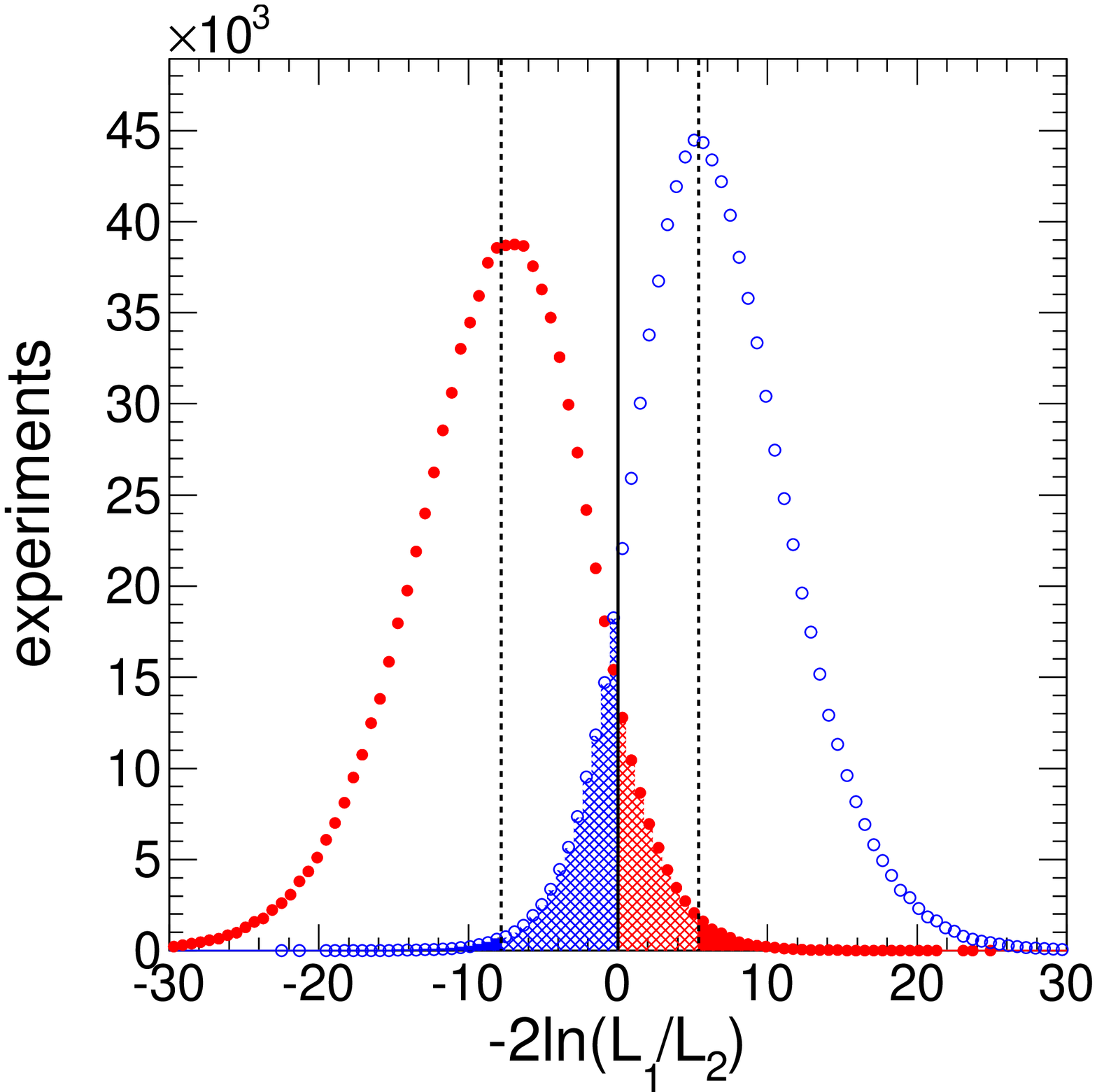,width=0.33\linewidth}
\epsfig{figure=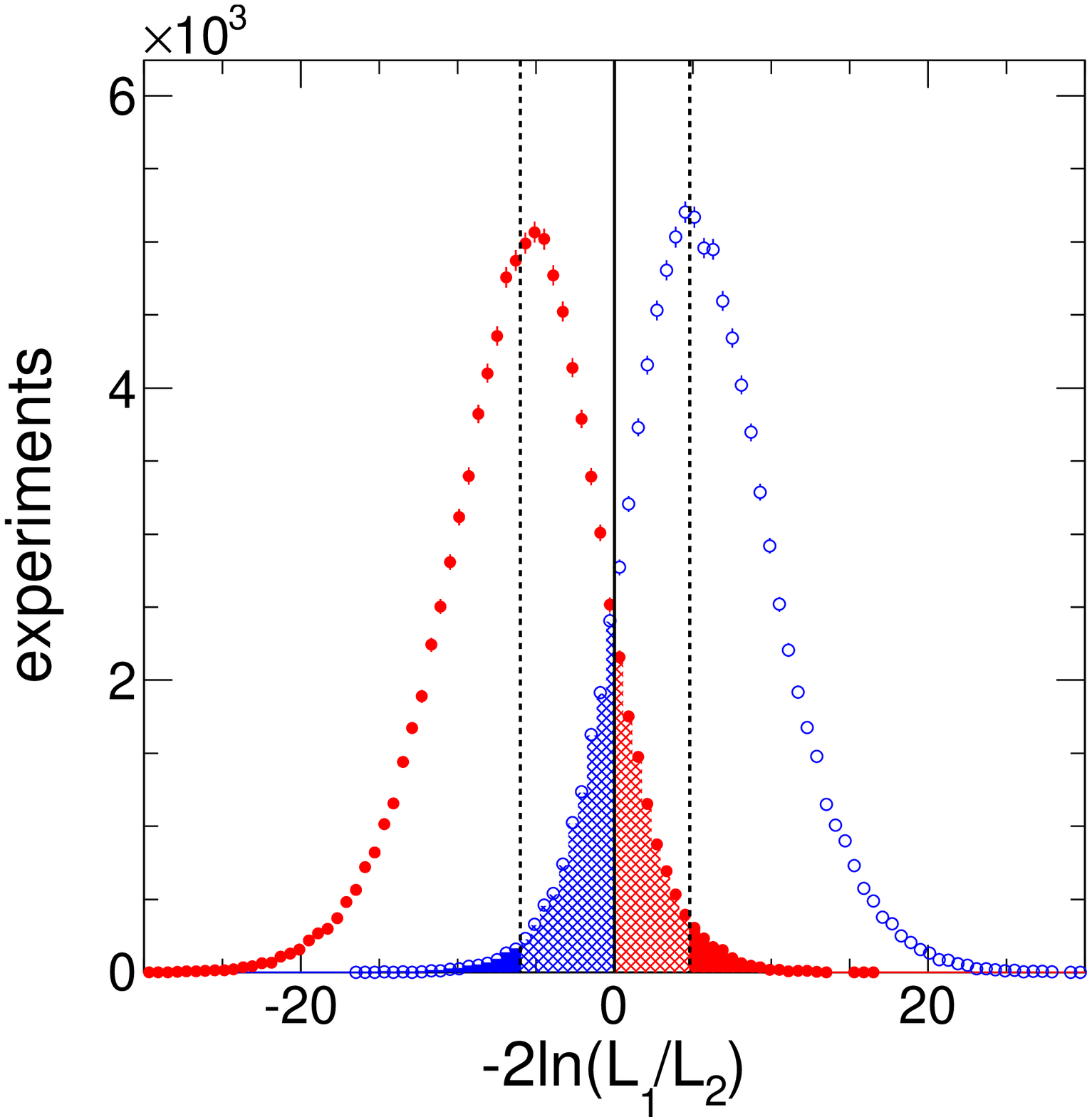,width=0.33\linewidth}
}
\caption{
Distributions of $-{2\ln({\cal L}_1/{\cal L}_2)}$ with the likelihood 
${\cal L}_k$ evaluated for two models and shown for a large number of generated experiments
in the analysis of $X\to ZZ$ (left), $WW$ (middle), and $\gamma\gamma$ (right).
The models shown are the SM Higgs boson $0^+$ (red solid points) 
and the pseudoscalar  $0^-$ for $ZZ$ or the graviton-like $2^+_m$ 
for $WW$ and $\gamma\gamma$ (blue open points).
The scenarios correspond to those shown in Table~\ref{table-separation}.
}
\label{fig:separation}
\end{figure}


\subsection{\boldmath $X\to ZZ$ }
\label{sec:analysis-zz}

In the $X\to ZZ\to 4\ell$ channel, the dominant background is the continuous production of 
$Z\gamma^*/ZZ$\,\footnote{We will collectively refer to these two processes as $ZZ$ in what follows.}.  
The $ZZ$ production cross section is  comparable to 
that of the SM Higgs boson in the four-lepton invariant
mass window comparable to detector resolution. Loose selection requirements are applied to 
simulated signal and $q\bar{q}\to ZZ$ background events to model detector effects of CMS and ATLAS.  
For lepton track  transverse momentum, we apply Gaussian random smearing with an rms 
$\Delta p_T=0.014\times p_T$ (GeV) for 90\% of the core of the distribution
and a wider smearing for the 10\% tail. 
Leptons are required to have pseudo-rapidity in the range $|\eta|<2.4$ and 
$p_T$  greater than $7$~GeV. In addition, leptons with the highest and next-to-highest 
transverse momentum are required to also have $p_T>20$ and $10$~GeV, respectively. 
To reject the non-$ZZ$ background, the invariant masses of the
di-lepton pairs are required to satisfy $50<m_1<120$~GeV and 
$12<m_2<120$~GeV, where $m_1>m_2$.
The overall $ZZ$ rate is then scaled to be consistent with the total background  
observed in LHC experiments, including Drell-Yan and top events with jets faking leptons.  
We do not attempt to model this instrumental background shape and implicitly assume 
that shapes are well-modeled by $q\bar{q}\to ZZ$ events.
Only events with $110<m_{4\ell}<160$ GeV are considered in the final analysis.  
The number of signal events after all selections is $0.8$ events/fb$^{-1}$,
while the number of background events is $1.9$ events/fb$^{-1}$.
Using only the $m_{4\ell}$ shape of signal and background, we find an expected significance of 3.3\,$\sigma$ 
with 10~fb$^{-1}$ of data, 
comparable to that observed at the  LHC\footnote{We disregard the difference between
the 7 and 8 TeV collision energies of LHC for simplicity.}.

As was pointed out earlier in Ref.~\cite{Gao:2010qx, Gainer2011}, using full kinematic 
information in the $X\to ZZ$ channel improves signal-to-background separation 
by about 20\% compared to a one-dimensional analysis of the invariant mass $m_{4\ell}$. 
This has been exploited by the CMS experiment in the discovery of the new boson~\cite{discovery-cms}. 
One can perform either a multidimensional fit or create a kinematic discriminant 
(MELA)~\cite{discovery-cms, cmshzz2l2q} which is constructed from the ratio of probabilities 
for signal and background hypotheses
\begin{eqnarray}
\label{eq:mela}
{D_{\rm bkg}}=\left[1+\frac{{\cal P}_{\rm bkg} (m_{4\ell}; m_1, m_2, \vec\Omega) }
{{\cal P}_{\rm sig} (m_{4\ell}; m_1, m_2, \vec\Omega) } \right]^{-1}
\,.
\end{eqnarray}
Here ${\cal P}_{\rm sig}$ and ${\cal P}_{\rm bkg}$ are the probabilities, 
as a function of masses $m_i$ and angular observables $\vec\Omega$
for a given value of invariant mass $m_{4\ell}$, as defined in
Eq.~(\ref{eq:differential-1}), for the SM Higgs boson signal and $ZZ$ background, respectively.
Although analytic computation of the 
matrix element for continuum $ZZ$ production~\cite{Gainer2011} exists,
it  neglects the $Z\gamma^*$  process and therefore it can not yet 
be  applied  to the region below $m_{4\ell}\sim 180$~GeV.
Instead, we use a large sample of 
POWHEG simulated events to fill a multi-dimensional histogram 
(template), where the most
important correlations between up to 
three observables are taken into account.  

The above approach to background rejection is illustrated in Fig.~\ref{fig:zzobs}, where
we plot $m_{4\ell}$ and $D_{\rm bkg}$, which are mostly uncorrelated
in the small $m_{4\ell}$ region considered. 
As shown in  Fig.~\ref{fig:zzobs}, for a wide range of 
different signal spin-parity hypotheses, 
the $D_{\rm bkg}$ distributions do not differ considerably. 
However, all signal distributions of $D_{\rm bkg}$ differ considerably from background. 
We confirm that significance of the signal observation in the two-dimensional analysis of 
$(m_{4\ell}, D_{\rm bkg})$ increases by more than 20\% compared to a one-dimensional 
analysis of the invariant mass $m_{4\ell}$. 
To simplify the fitting model, in the rest of the paper we will not use the additional background suppression power
of the $D_{\rm bkg}$ observable but we note that the effective 
significance can be increased by either including $D_{\rm bkg}$ in the multivariate fit, 
or, equivalently, using information contained in $D_{\rm bkg}$ in the fit.

\begin{figure}[b]
\centerline{
\epsfig{figure=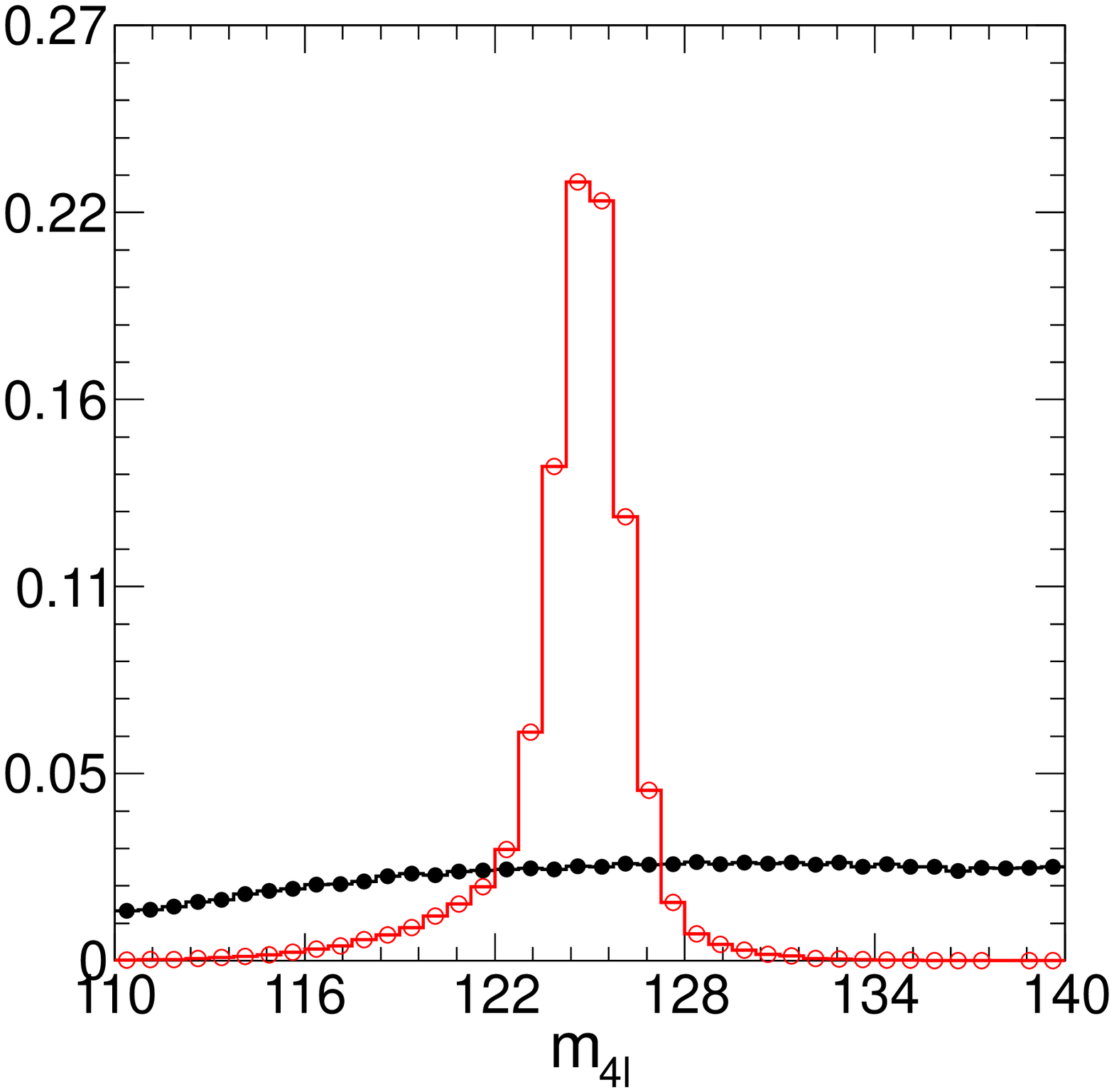,width=0.33\linewidth}
\epsfig{figure=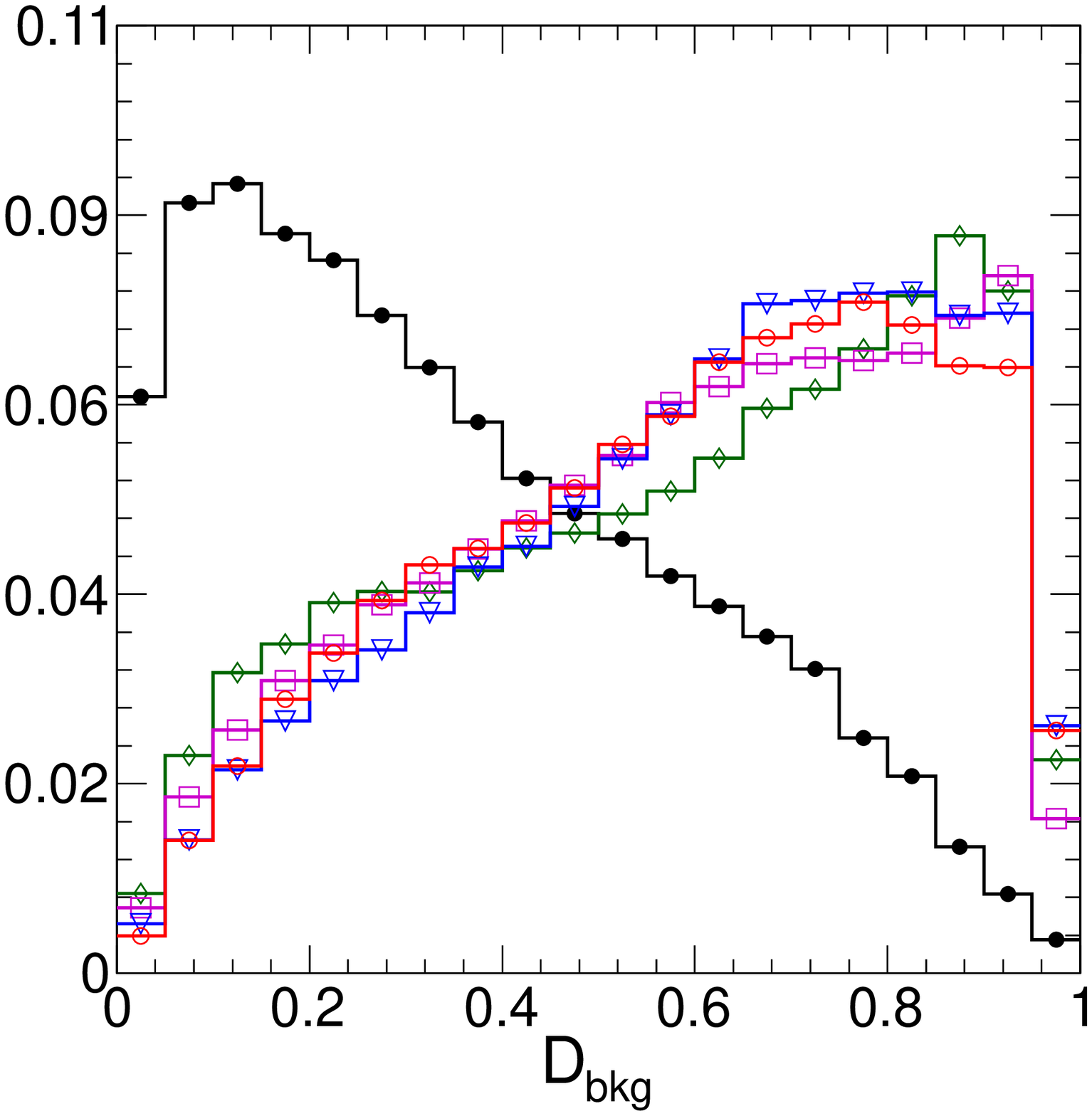,width=0.33\linewidth}
\epsfig{figure=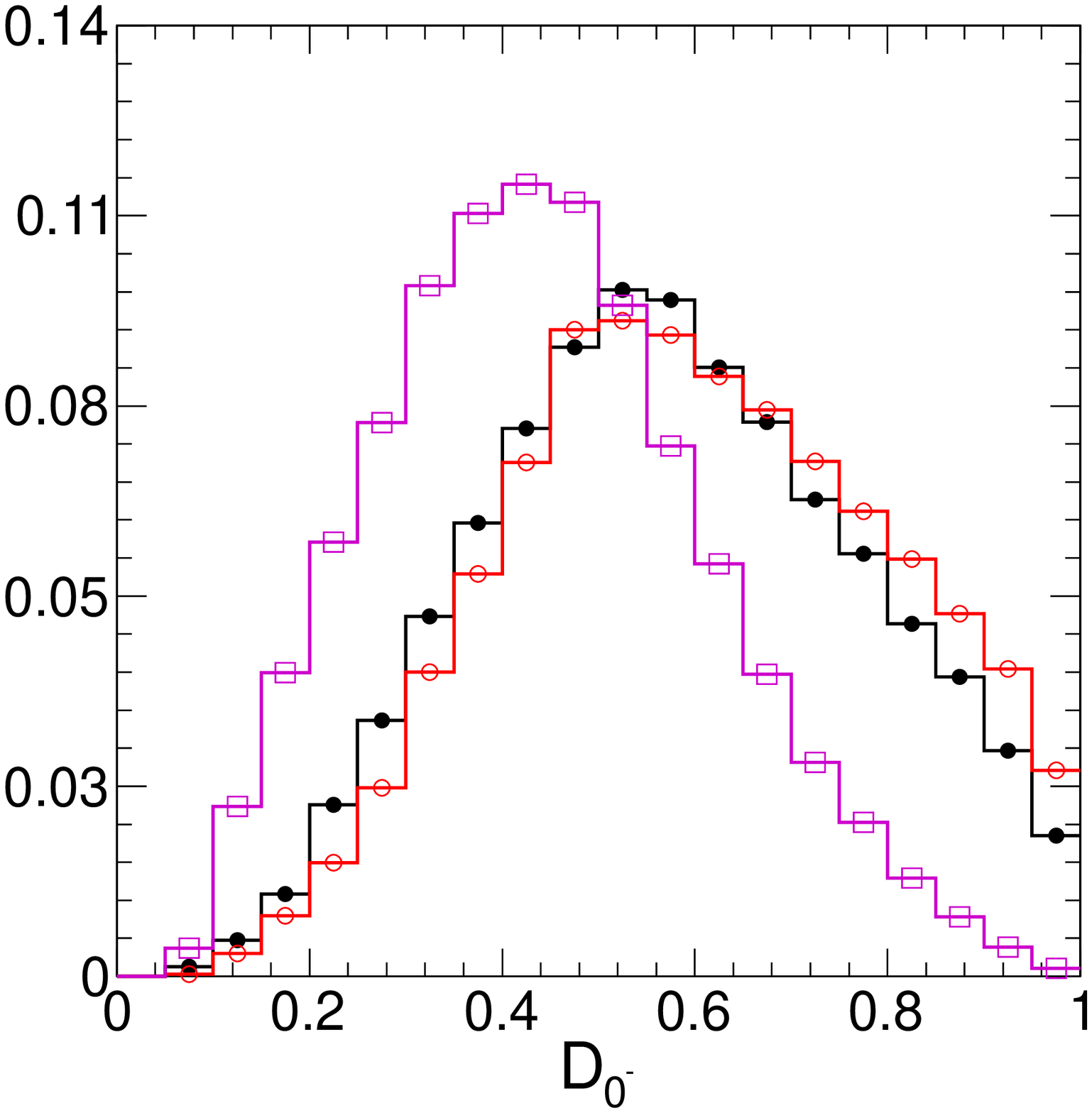,width=0.33\linewidth}
}
\caption{
Distributions of $m_{4\ell}$  (left), $D_{\rm bkg}$  (middle), and $D_{0^-}$ (right) in the $X\to ZZ$ analysis 
for the non-resonant $ZZ$ background (black solid circles), and four signal hypotheses: 
SM Higgs boson (red open circles), $0^-$ (magenta squares), $2^+_m$ (blue triangles), 
and $2^+_h$ (green diamonds).
Not all signal hypotheses are shown on all plots.
The mass range $120<m_{4\ell}<130$ GeV is shown in the $D$ distributions.
}
\label{fig:zzobs}
\centerline{
\epsfig{figure=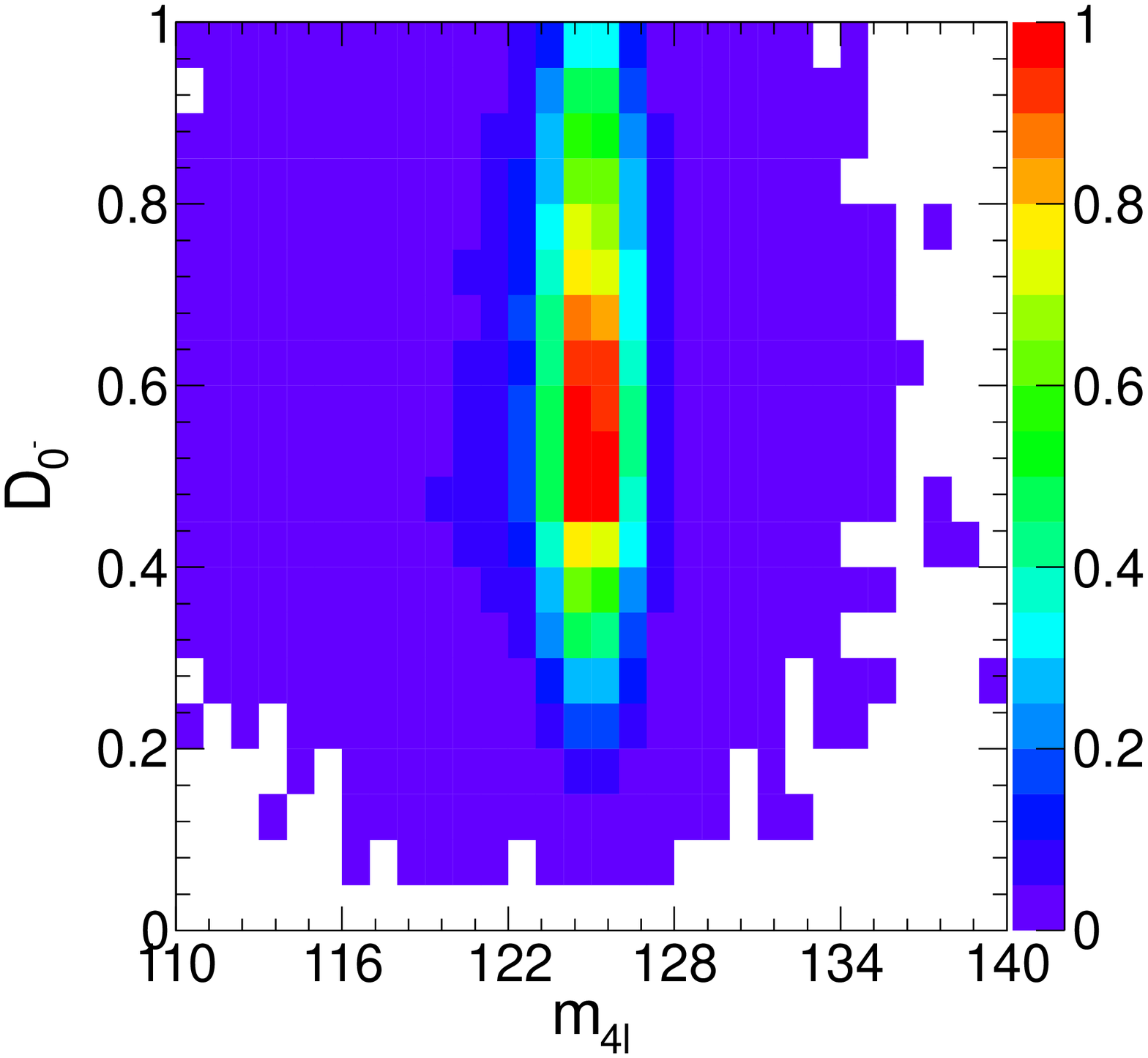,width=0.33\linewidth}
\epsfig{figure=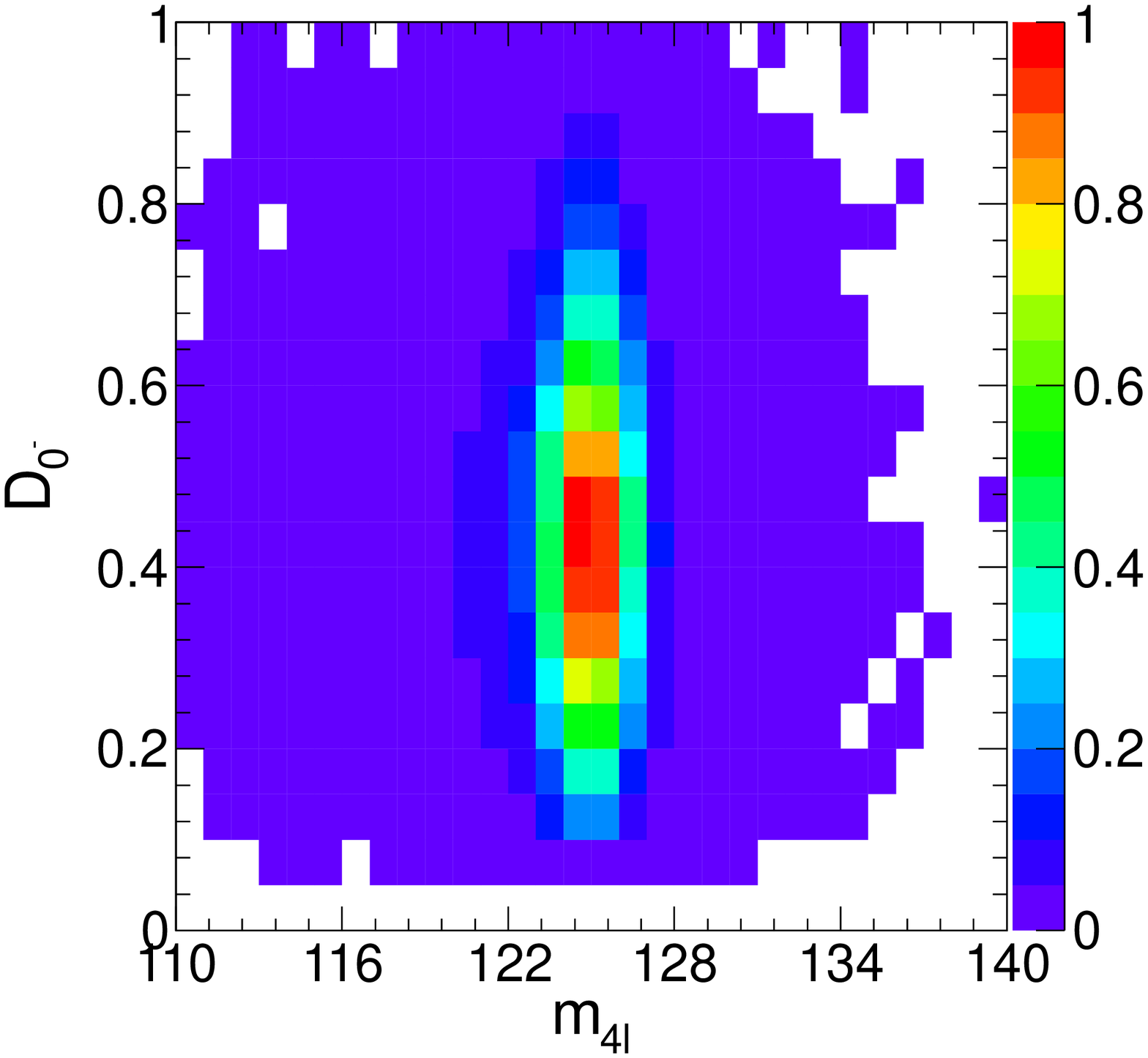,width=0.33\linewidth}
\epsfig{figure=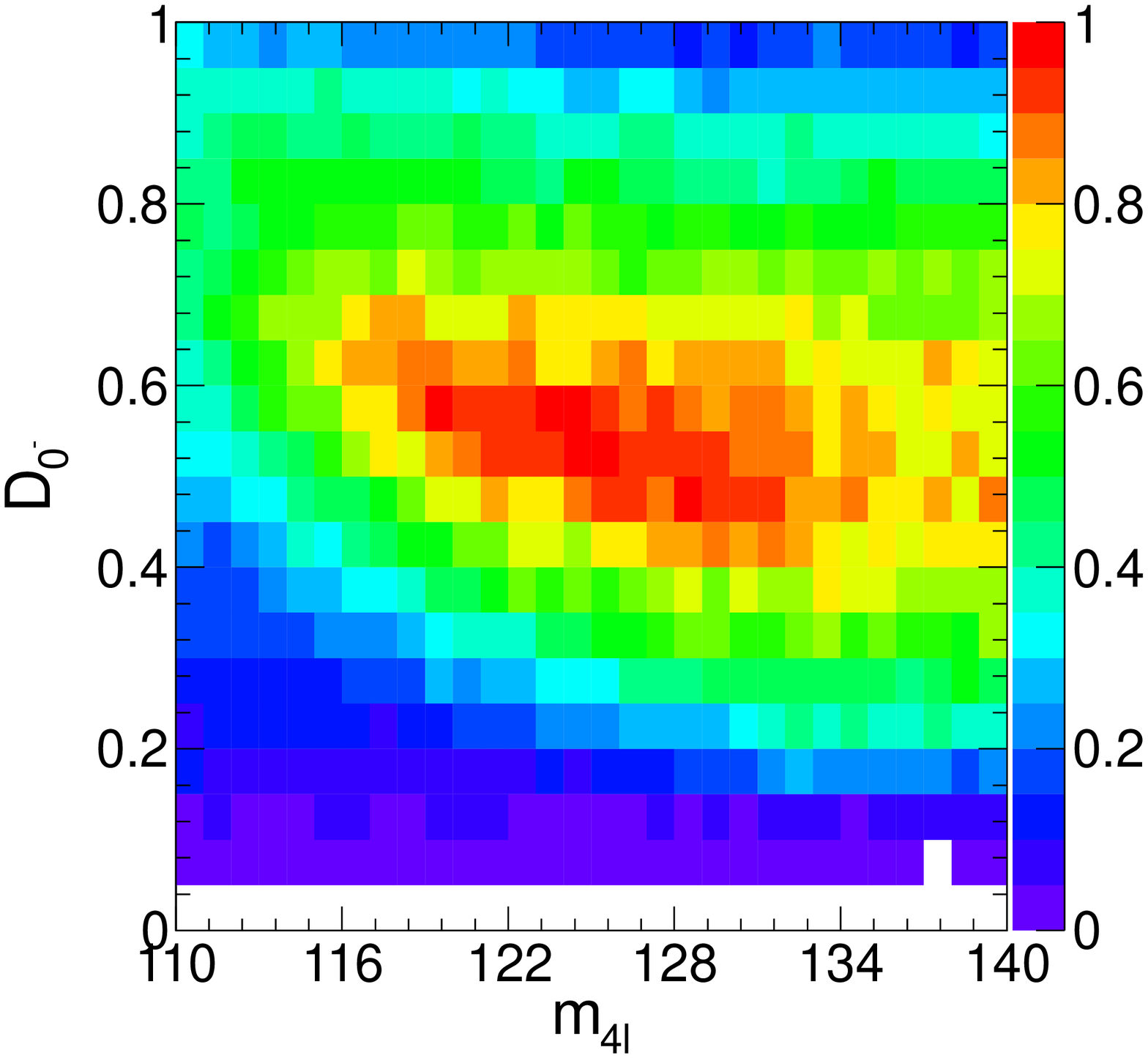,width=0.33\linewidth}
}
\caption{
Template distributions 
of $D_{0^-}$ vs  $m_{4\ell}$ in the $X\to ZZ$ analysis for the 
SM Higgs boson (left), pseudoscalar resonance (middle), and
non-resonant $ZZ$ background (right).
}
\label{fig:zzobs_2D}
\end{figure}

Separation between different spin-parity scenarios of the observed resonance can also be obtained using 
kinematic information. We can re-write Eq.~(\ref{eq:mela}) as 
\begin{eqnarray}
\label{eq:melaSig}
{D_{J^P_x}}=\left[1+\frac{{\cal P}_{2} (m_{4\ell}; m_1, m_2, \vec\Omega) }
{{\cal P}_{1} (m_{4\ell}; m_1, m_2, \vec\Omega) } \right]^{-1}
\,.
\end{eqnarray}
where  ${\cal P}_{1}$ and ${\cal P}_{2}$ are the probabilities as defined in Eq.~(\ref{eq:differential-1}), 
for two different hypotheses of spin-parity and tensor structure 
of interactions of the signal resonance.
Equation~(\ref{eq:melaSig}) is indeed the optimal way to combine all relevant
kinematic information into a single observable $D_{J^P_x}$ for
separating the SM Higgs boson scenario from other ${J^P_x}$ hypotheses without loss of information.

As an example, in the right plot of Fig.~\ref{fig:zzobs} we show a kinematic 
discriminant $D_{0^-}$  optimized for separating $0_{m}^+$ and $0^-$ signal hypotheses.  
$D_{0^-}$ is calculated in Eq.~(\ref{eq:melaSig}) with ${\cal P}_{2}$ taken as the 
probability density for  $0^-$. Since $m_{4\ell}$ is still the most 
powerful observable to discriminate any type of signal from the  background, we perform a two-dimensional  
fit of $(m_{4\ell}, D_{0^-})$. The probability densities for signal and background are parameterized
as two-dimensional template histograms using simulation, as shown in Fig.~\ref{fig:zzobs_2D}.

In the scenario described above, with an expected SM Higgs boson signal
significance of 3.3\,$\sigma$ with 10~fb$^{-1}$ of data,
we estimate an average separation of 1.9\,$\sigma$ between the $0_{m}^+$ 
and $0^-$ signal hypotheses. Equivalently, assuming that the integrated 
luminosity is high enough to ensure 
$5\sigma$
signal-to-background separation, 
the average expected separation of $0_{m}^+$ and $0^-$ is 
2.9\,$\sigma$, see Fig.~\ref{fig:separation}.
This and other results for several other signal hypotheses are shown in Table~\ref{table-separation}.
We find the $0_{m}^+$ and $0^-$ separation results consistent with those predicted by CMS~\cite{incandela2012},
taking into account the assumed signal significance (expected vs  observed).

The separation power depends on information contained 
in kinematic distributions; we show illustrative  examples in 
Figs.~\ref{fig:simulated}, \ref{fig:simulated-zz-mass}, 
and \ref{fig:simulated-zz-angles}.
For example, separation of the SM Higgs boson hypothesis from 
$0^-$ is better than from $2^+_m$
since a number of mass and angular distributions are more distinct. 
Also, we note that both $2^+_h$ and $2^-_h$ are even more different
from $0^+_m$ than any other hypothesis considered. 
One of the kinematic distributions that shows important differences 
between the SM Higgs boson and the pseudoscalar, 
as well as between the SM Higgs boson and $2^+_h$ or $2^-_h$,
is the low-mass tail of $m_1$ distribution, see the left plot in Fig.~\ref{fig:simulated}.  
As shown in Ref.~\cite{incandela2012}, there is a rather large fraction of the $X\to ZZ$ 
events in CMS with both $Z$'s off-shell. If this feature persists in data, it may reveal contributions 
of more exotic couplings shown in Eqs.~(\ref{eq:fullampl-spin0}) and~(\ref{eq:fullampl-spin2}).
We note that a particular feature which may enhance the $m_1$ tail considerably,
as shown for the $2^+_h$ hypothesis in Fig.~\ref{fig:simulated}, 
is the presence of a large power of the parameter $x$ in Eq.~(\ref{eq:relate-spin2}).  
In turn, the appearance of this parameter is related to the terms 
$c_2$ and $c_3$ in $A_{00}$ which are sensitive to $g^{(2)}_{\sss 4}$ coupling. 

We comment about an interesting feature in Table~\ref{table-separation}. 
On general grounds one can expect that significance of hypotheses separation
between two types of signal  to be smaller than the observation significance of the signal. 
However, in a situation when the kinematic discriminant itself provides substantial background 
rejection power, significance of the observation of the alternative signal may become
higher  than that for the SM Higgs-like resonance. This phenomenon occurs  in the study of 
$2_h^+$ and $2_h^-$ hypotheses, where 
the $m_1$ mass distribution becomes a particularly  
powerful discriminating observable. 
As a result, for the corresponding signal types,
$D_{2^+_h}$ and $D_{2^-_h}$ may become even stronger background rejection observables 
than $m_{4\ell}$. We do not see this in Fig.~\ref{fig:zzobs} because the SM Higgs boson hypothesis is used 
for the computation of $D_{\rm bkg}$, but an alternative signal hypothesis could have 
been considered as well. The $m_1$-distribution then also leads to a very strong signal hypothesis 
separation, approaching the values of 5\,$\sigma$ for SM Higgs boson vs  $2_h^+$ and $2_h^-$ 
in Table~\ref{table-separation}.

We note that analysis of the spin-parity hypotheses 
should not be limited to just discrete hypothesis testing. 
In Ref.~\cite{Gao:2010qx} we showed how a continuous spectrum of parameters
can be obtained from a multidimensional fit. 
As an intermediate step, one could consider determination
of the fraction of a certain component in a mixed state. 
For the spin-zero particle, this can be modeled by  
non-vanishing  $g^{(0)}_{\sss 1}$
and $g^{(0)}_{\sss 4}$ couplings  in Eq.~(\ref{eq:fullampl-spin0}). 
We note that, in this scenario, there is an interference term in the amplitude 
which is correctly described by our simulation. 
Results presented in Table~\ref{table-separation}  can be used 
to illustrate the typical precision on the fraction by dividing
the full range between the two extreme hypotheses by the number of standard 
deviations between them. For example, this implies  that by the end of the 
8 TeV run, the LHC experiments may be able to constrain an approximately 50\% 
admixture of the $C\!P$-violating amplitude at 95\% confidence level.


\subsection{\boldmath $X\to WW$ }
\label{sec:analysis-ww}

Compared to the $ZZ$ final state, the $X\to WW\to 2\ell 2\nu$ channel is expected to 
have a larger rate due to a larger branching fraction of $WW\to 2\ell 2\nu$, 
provided that decay rates $X \to ZZ$ and $X \to WW$ are comparable.  
However the analysis suffers from large backgrounds and the fact that neutrino momenta  
cannot be reconstructed.  In place of the four-lepton invariant mass,  the transverse mass defined as
$m_T = ({2p_T^{\ell\ell}\met(1-\cos\Delta\phi_{\ell\ell-\met})})^{1/2}$,
where $\Delta\phi_{\ell\ell-\met}$ is the angle between the direction of 
the di-lepton pair and the missing energy $\met$ vector in the transverse plane, 
is exploited to disentangle signal from background~\cite{discovery-atlas, discovery-cms}. 
In our simplified study $\met$ is calculated from the $2\nu$ momentum.
The one-jet $e\mu$ and zero-jet same-flavor categories only contribute to the signal sensitivity 
at the 10\% level because of larger backgrounds from top-quark decays and Drell-Yan production, 
respectively~\cite{discovery-atlas, discovery-cms}. Therefore,
we only select events with different lepton flavors ($e\mu$) and 
little jet activity to enhance  signal-to-background ratio, so all events  
with jets with transverse energy greater than 30 GeV are rejected.
In this category, the main background comes from the non-resonant $WW$ production~\cite{discovery-atlas, discovery-cms}. 
To further reject the reducible backgrounds such as Drell-Yan and $W+\text{jets}/\gamma$ 
processes, we require 
$p_T > 20(10)$~GeV for the leading (sub-leading) lepton,
$p_T^{\ell\ell} > 30$~GeV for the di-lepton system,
$\met > 20$~GeV,
$60 < m_T < 130 $~GeV,
and $10 < m_{\ell\ell} < 90 $~GeV.

\begin{figure}[t]
\centerline{
\epsfig{figure=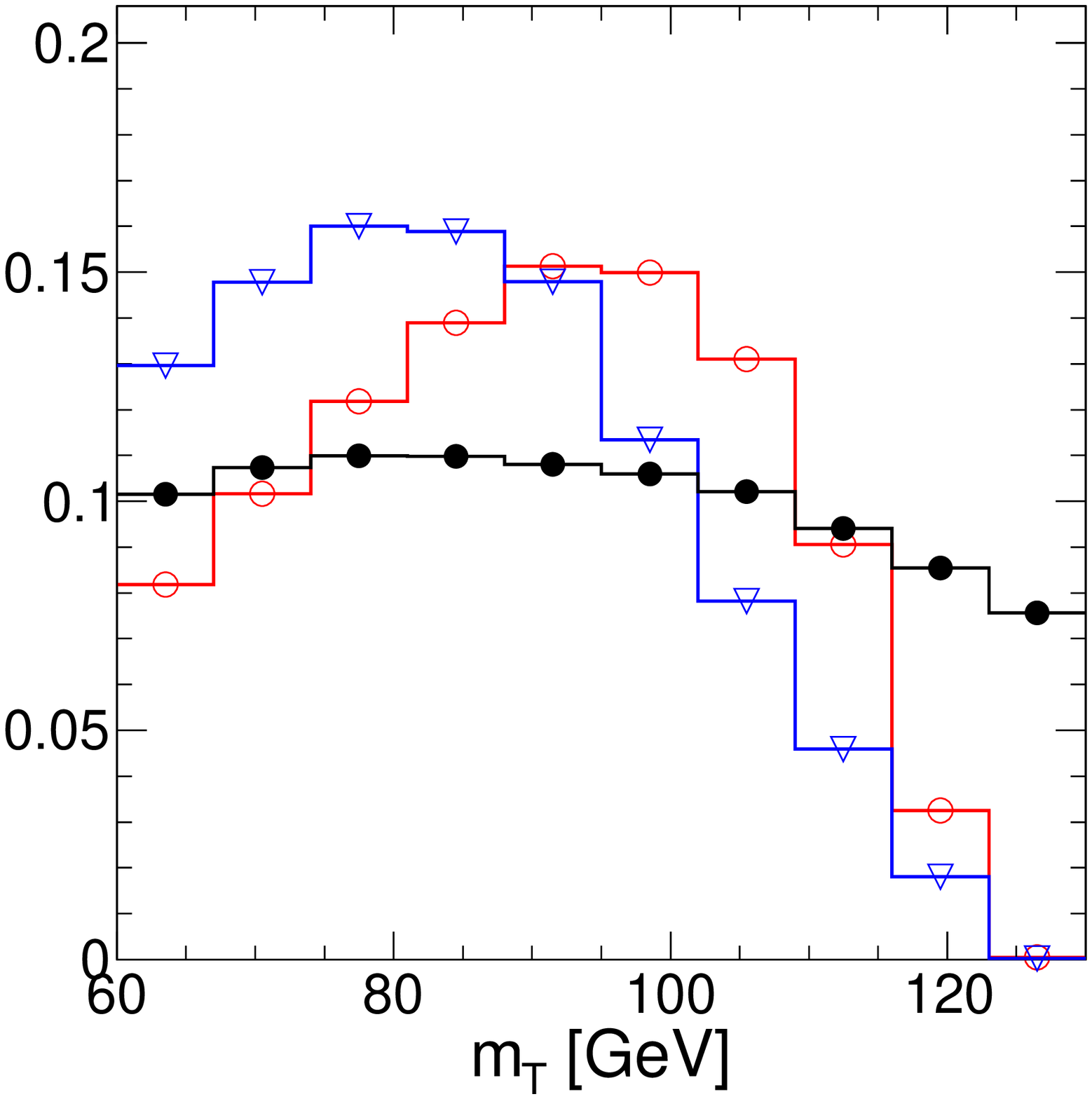,width=0.33\linewidth}
\epsfig{figure=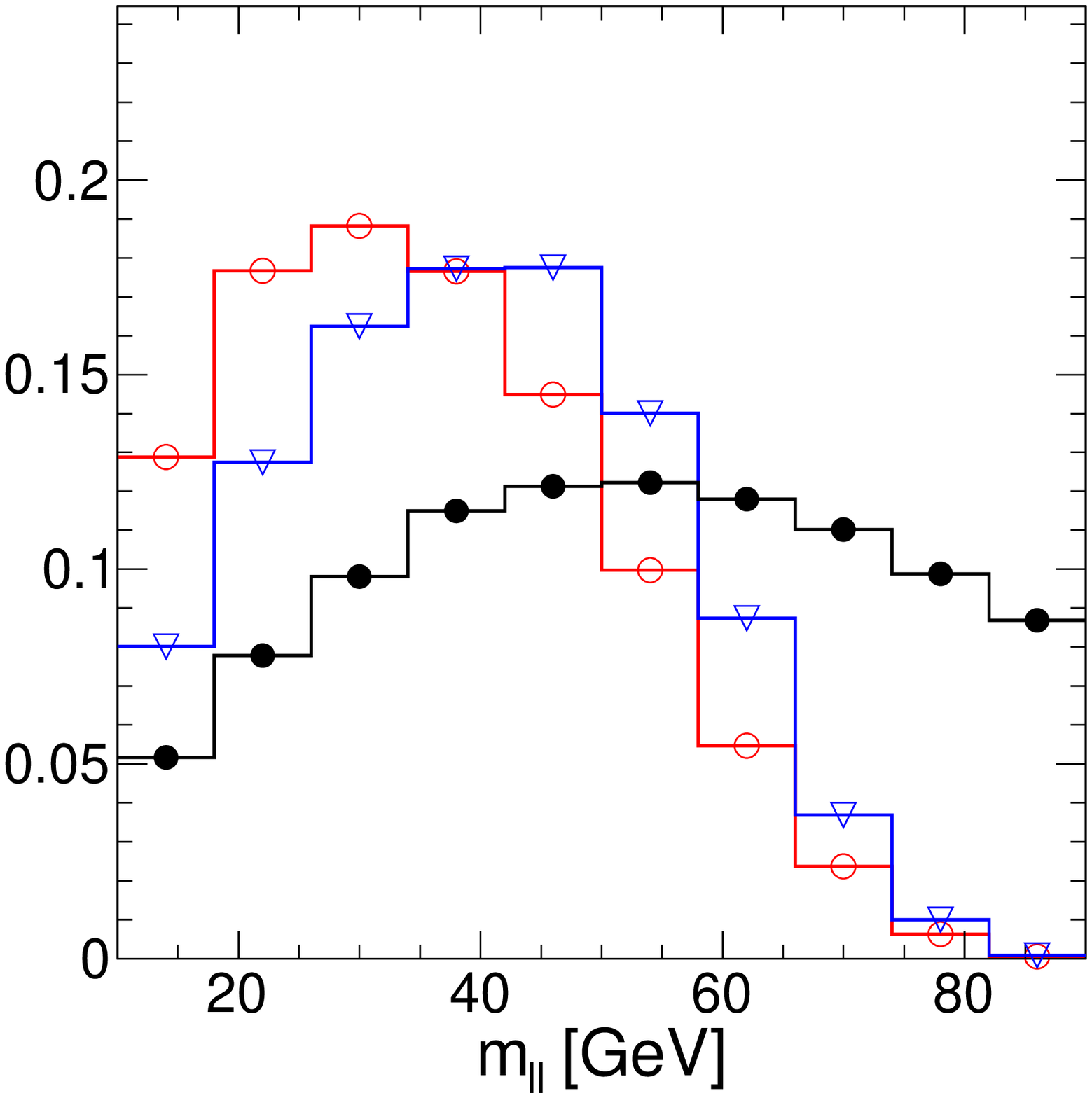,width=0.33\linewidth}
\epsfig{figure=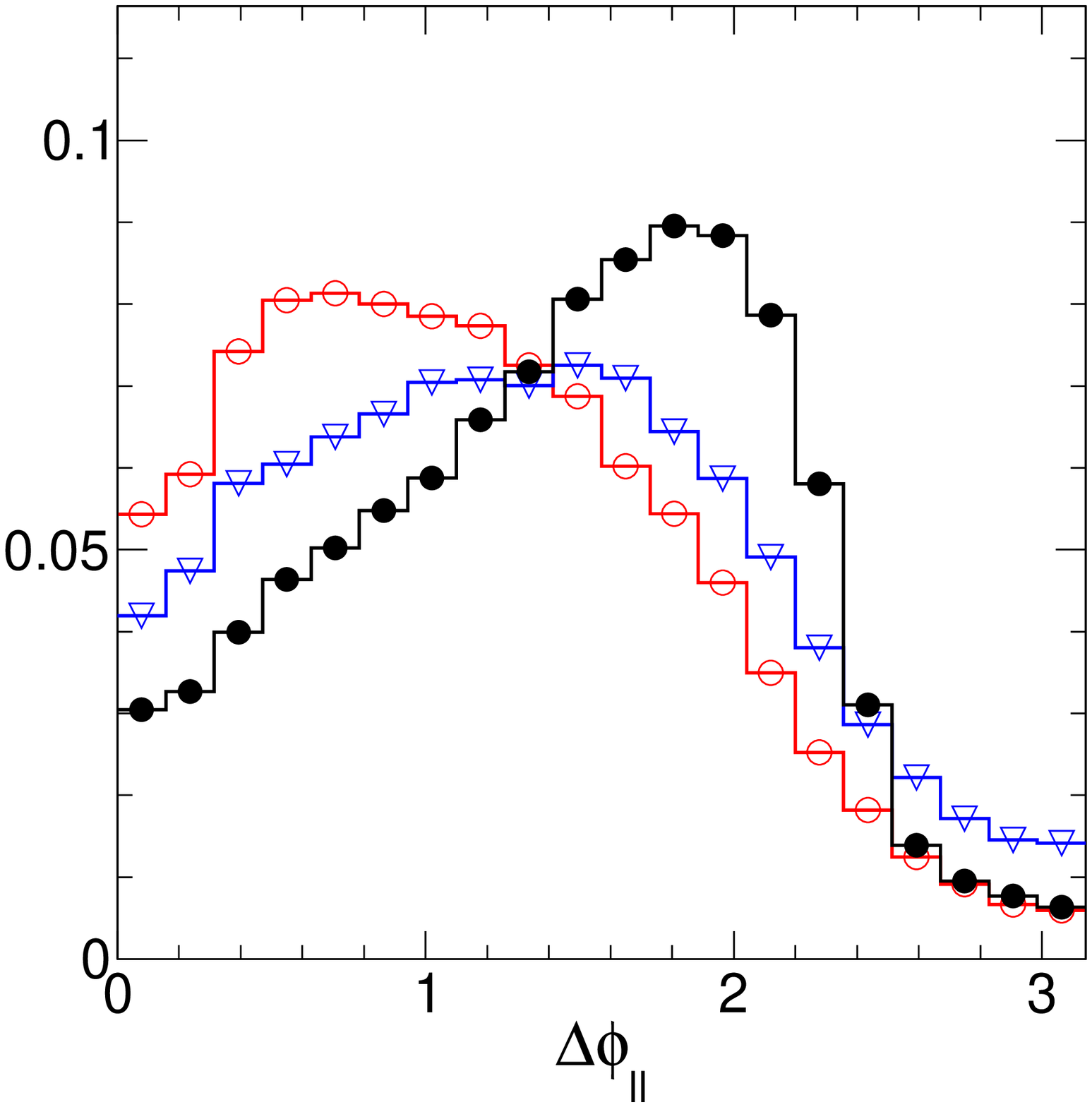,width=0.33\linewidth}
}
\caption{
Distributions of $m_T$  (left), $m_{\ell\ell}$  (middle), and $\Delta\phi_{\ell\ell}$ (right)
in the $X\to WW$ analysis for the non-resonant 
$WW$ background (black solid circles), SM Higgs boson (red open circles), and 
a spin-two resonance in the $2^+_m$ model (blue triangles).
}
\label{fig:wwobs}
\centerline{
\epsfig{figure=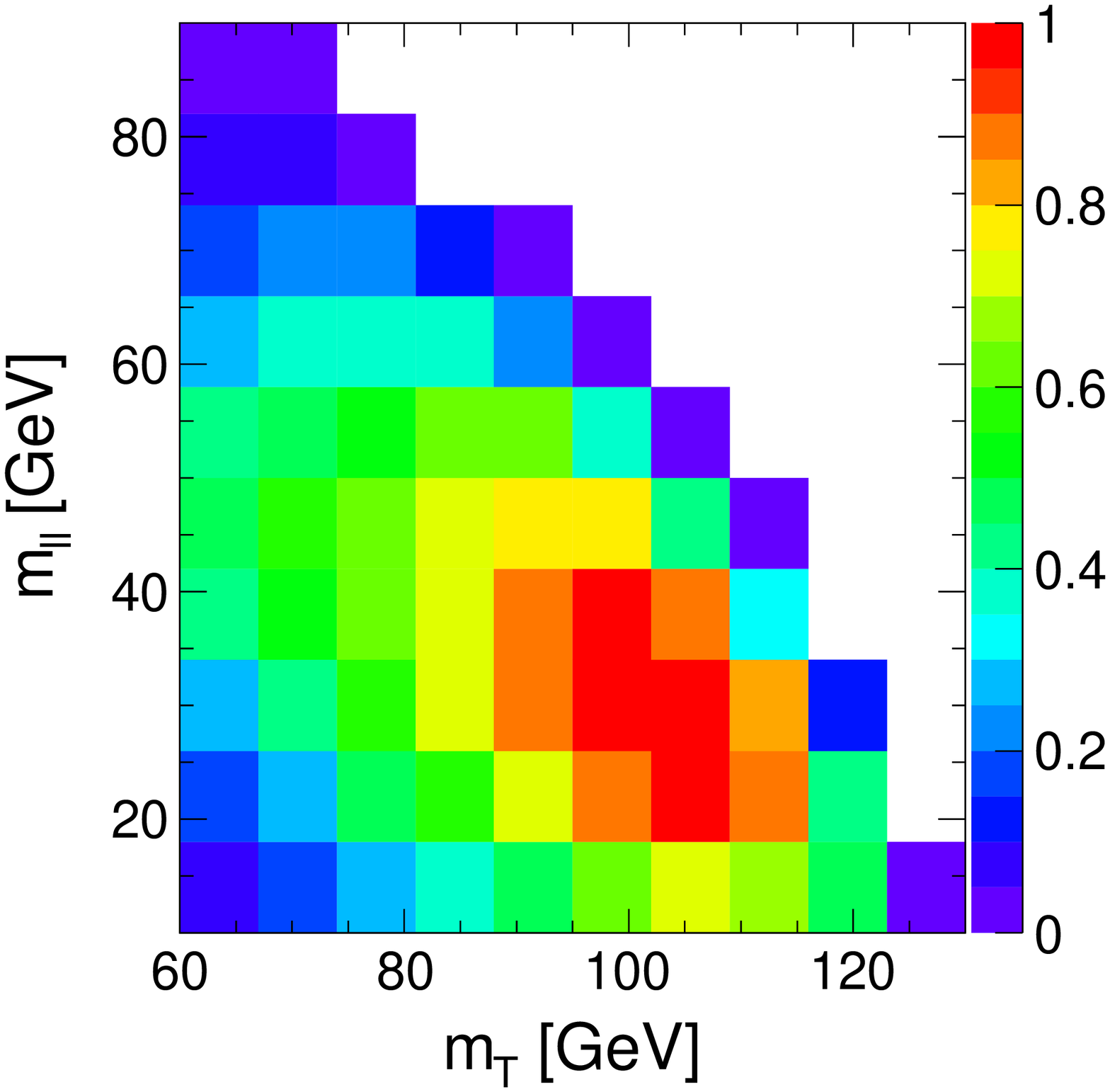,width=0.33\linewidth}
\epsfig{figure=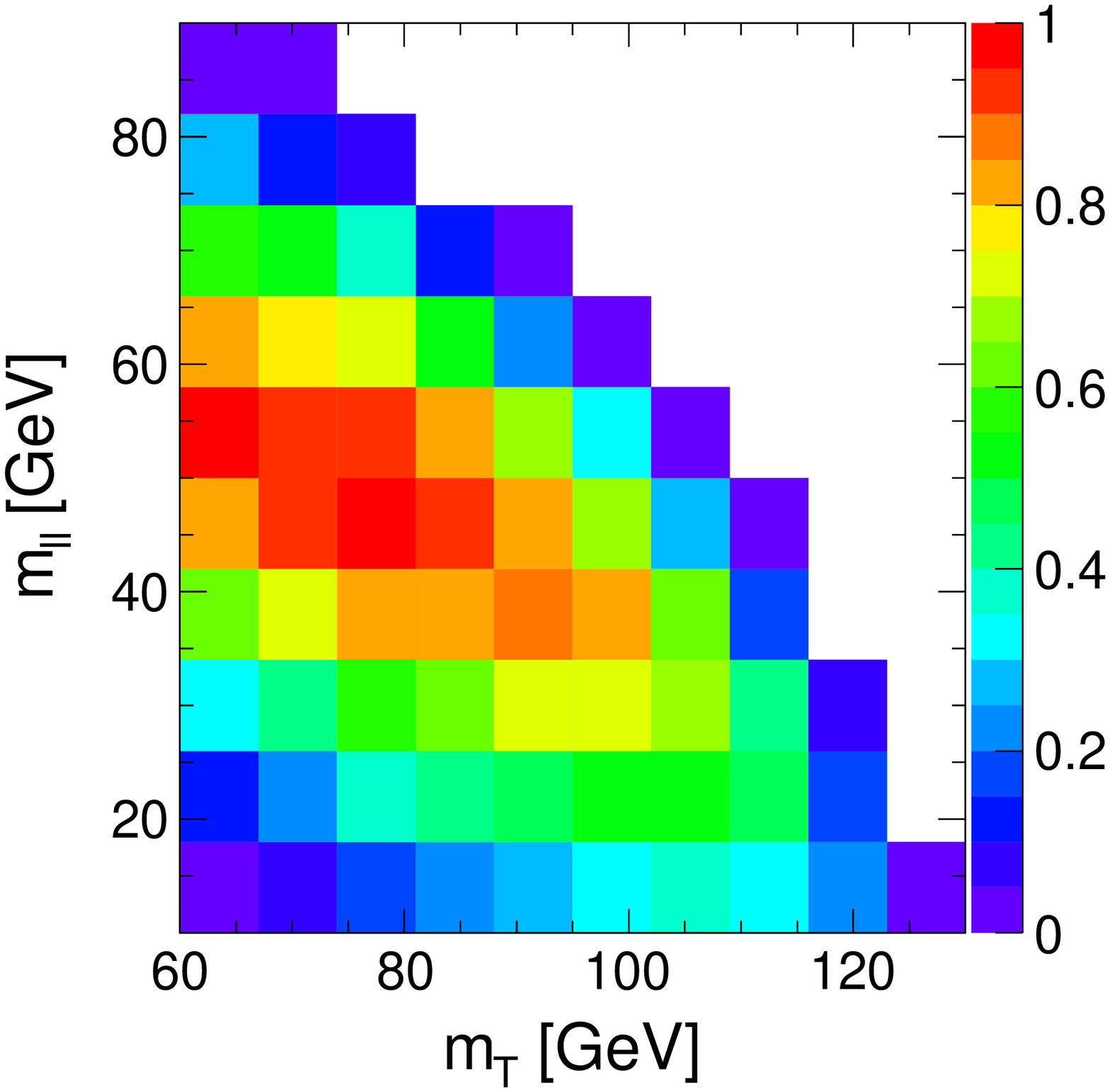,width=0.33\linewidth}
\epsfig{figure=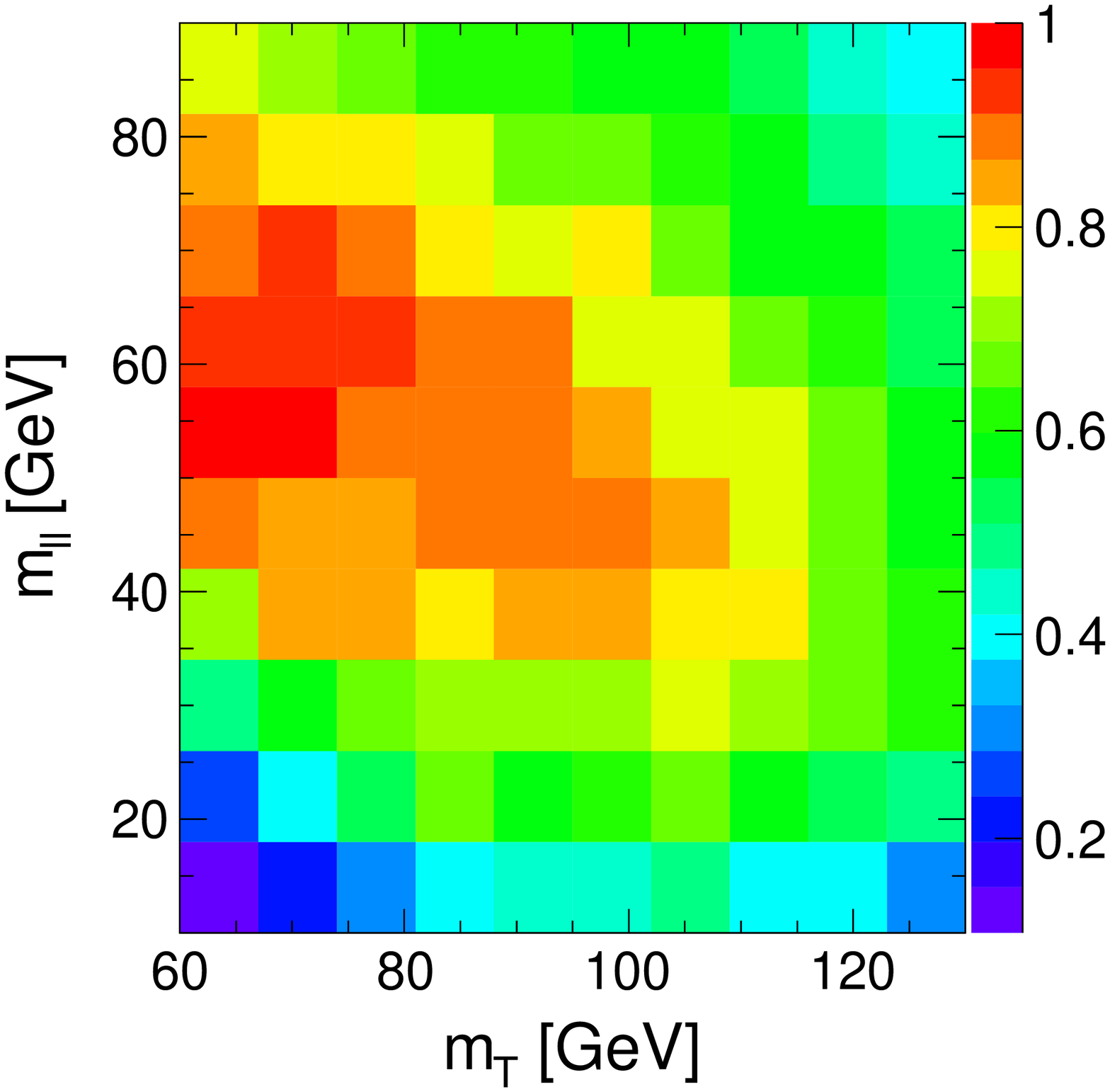,width=0.33\linewidth}
}
\caption{
Template distribution of  $m_{\ell\ell}$ vs $m_T$ in the $X\to WW$ analysis for the 
SM Higgs boson (left), spin-two resonance hypothesis $2^+_m$ (middle), and
non-resonant $WW$ background (right).
}
\label{fig:wwobs_2D}
\end{figure}

We estimate the expected number of signal and $WW$ background events after this selection 
by extrapolating the expected yields in the signal regions used in Ref.~\cite{discovery-cms} to 
the signal region defined above  using simulation. 
The estimated number of SM Higgs boson events  is 13 per $\ifb$.
The number of  non-resonant $WW$ background events is estimated to be  104
per $\ifb$.  We also assume that  continuum $WW$ production gives two-thirds  
of the total background and that kinematic distributions of the non-$WW$ backgrounds 
are the same as the ones of the $WW$ background. 
We cross-check this estimation using the  signal region used in Ref.~\cite{discovery-atlas}
and find consistent results. To extract the expected significance for separating different signal 
hypotheses ${\cal S}$, we construct a two-dimensional template based on 
two observables  $(m,D)=(m_T, m_{\ell\ell})$; this is illustrated in 
Figs.~\ref{fig:wwobs} and~\ref{fig:wwobs_2D}.
We have also considered other observables, such as the azimuthal angle $\Delta\phi_{\ell\ell}$
between the two leptons and found smaller separation compared to the case 
when $m_{\ell\ell}$ is used. On the other hand, since  there is large
correlation between $\Delta\phi_{\ell\ell}$ and  $m_{\ell\ell}$, using 
three observables in the fit is not expected to increase the significance of the separation much.

Using this simplified background model, we estimate the expected significance 
for distinguishing  the SM Higgs boson hypothesis from the background with $10~\ifb$  
using either the single observable $m_{\ell\ell}$ or the two  observables $m_T$ and $m_{\ell\ell}$. 
The former approach gives 2.6\,$\sigma$ separation from the background, similar to results 
of the  LHC~\cite{discovery-atlas,discovery-cms}, 
while the latter gives  3.5\,$\sigma$ which is an improvement of 35\%.
We follow the procedure outlined for the $X\to ZZ$ analysis above and
present the results in Fig.~\ref{fig:separation} and Table~\ref{table-separation}.
We find good separation between the SM Higgs boson and the $2_m^+$ hypotheses 
in particular, where this channel may have an advantage over the $X\to ZZ$ channel.
The reason for better performance of the $WW$ channel for $2_m^+$ separation
is the larger value of parameter $A_{f}$ defined in the Appendix~\ref{sec:appendix-a}, 
which enters the angular distributions in Eq.~(\ref{eq:mixedJtotal}).
As a consequence, there are larger azimuthal angular variations which are illustrated in 
Figs.~\ref{fig:simulated}, \ref{fig:simulated-zz-angles}, and~\ref{fig:simulated-ww}.


\subsection{\boldmath $X\to \gamma\gamma$ }
\label{sec:analysis-gg}

In the inclusive $X\to \gamma\gamma$ decay analysis, all information about the couplings is contained in 
the $\cos\theta^*$ distribution. The distribution is flat for a 
spin-zero resonance, while for a spin-two it is a normalized 
second degree polynomial 
in $\cos^2 \theta^*$ which requires  two independent parameters. 
Non-zero values of either parameter would be an unambiguous sign of a spin-two 
(or in principle higher spin) resonance.
However, relating these coefficients to general
couplings will have many ambiguities which are not generally present 
in the $ZZ$ and $WW$ channels. 
Indeed, the spin-two $X\to\gamma\gamma$ angular distribution reads

\begin{figure}[b]
\centerline{
\epsfig{figure=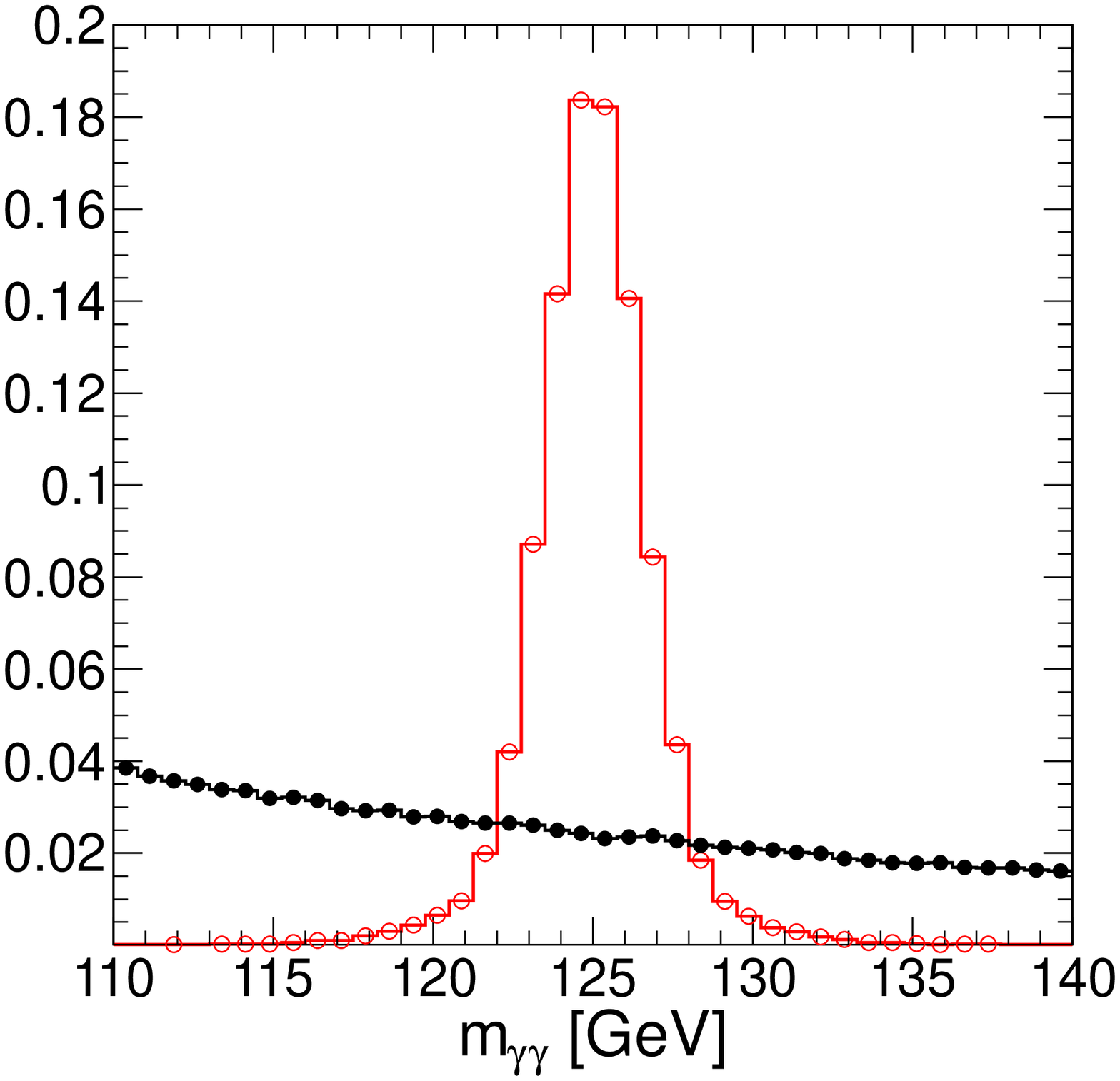,width=0.33\linewidth,height=0.33\linewidth}
\epsfig{figure=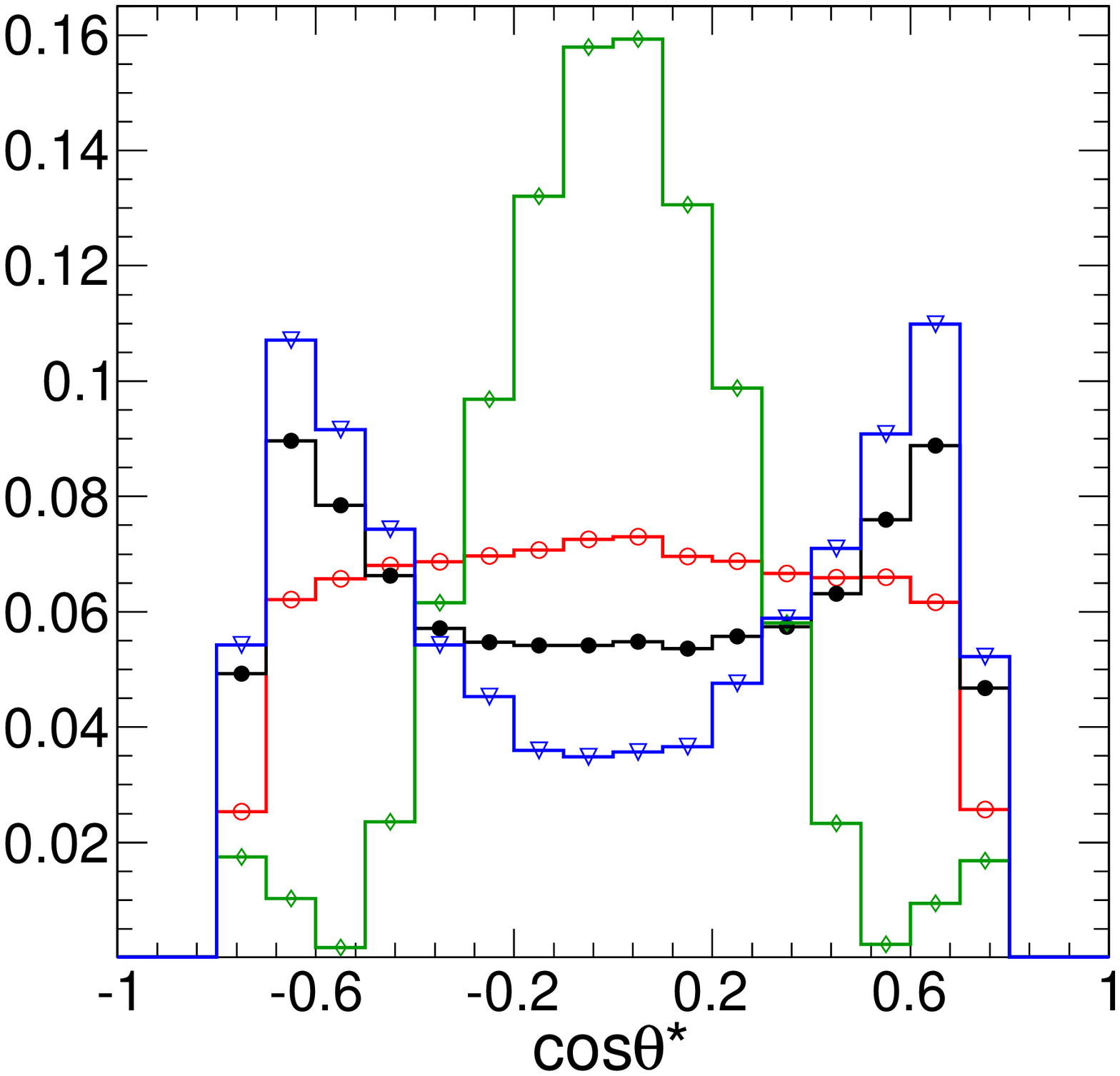,width=0.33\linewidth,height=0.33\linewidth}
}
\caption{
Distributions of $m_{\gamma\gamma}$ (left) and $\cos\theta^*$ (right) 
in the $X\to\gamma\gamma$ analysis
for the non-resonant $\gamma\gamma$ background (black solid circles), 
SM Higgs boson (red open circles), 
the spin-two resonance in the $2^+_m$ model (blue triangles)
and $2^+_h$ or $2^-_h$ models (green diamonds).
The mass range $120<m_{\gamma\gamma}<130$ GeV is shown
in the $\cos\theta^*$ plot for background.
}
\label{fig:ggobs}
\centerline{
\epsfig{figure=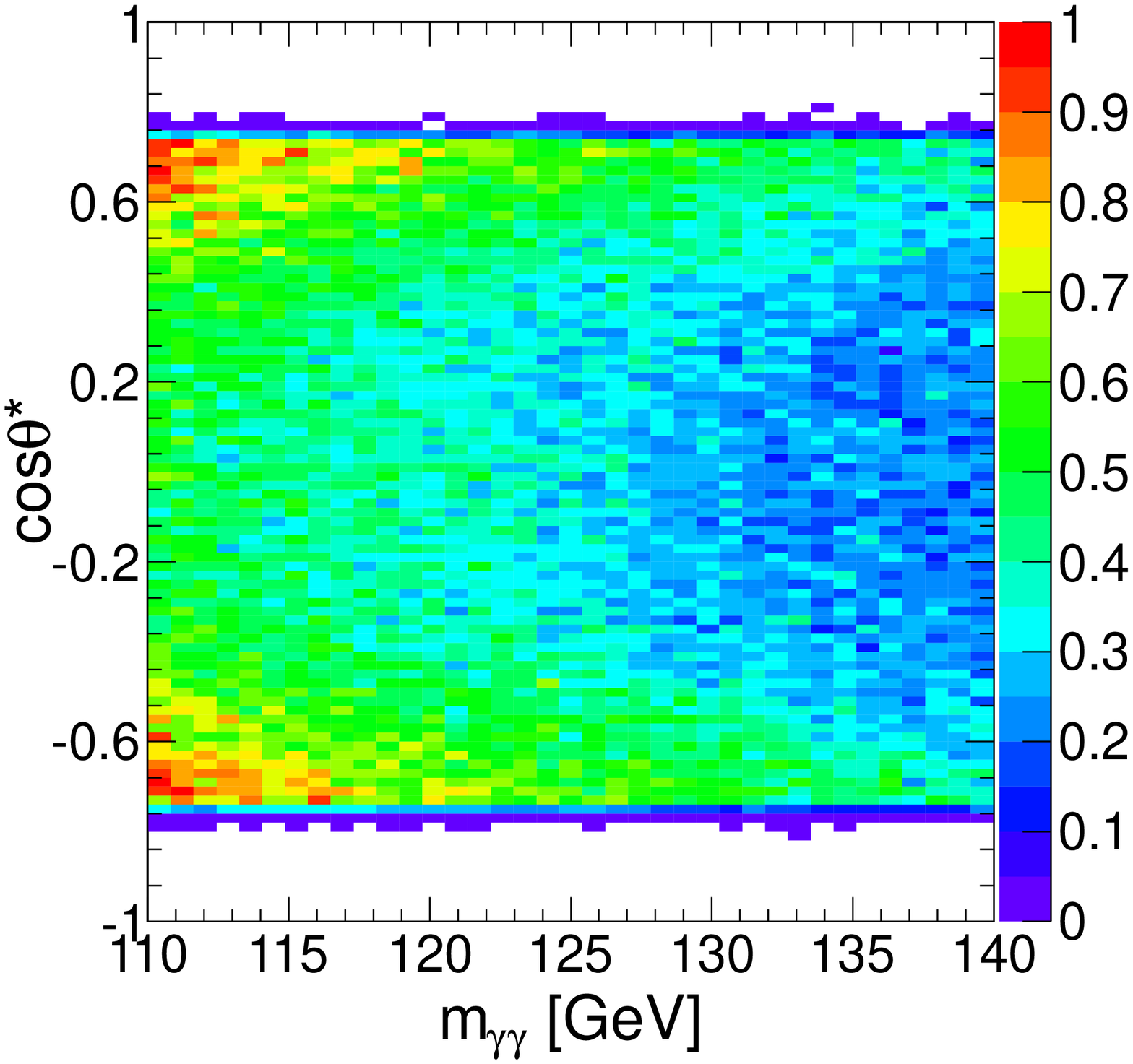,width=0.33\linewidth,height=0.33\linewidth}
\epsfig{figure=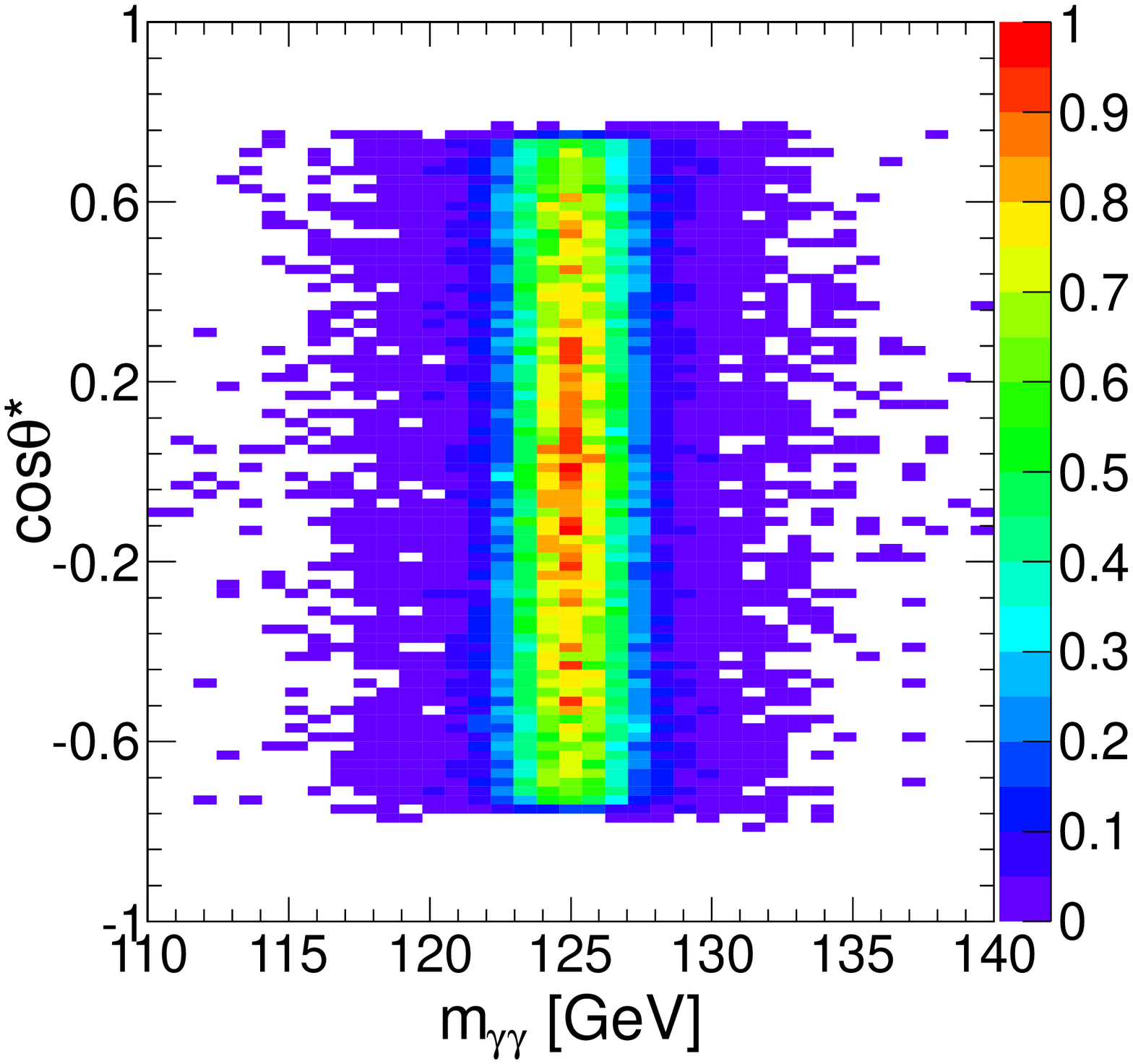,width=0.33\linewidth,height=0.33\linewidth}
\epsfig{figure=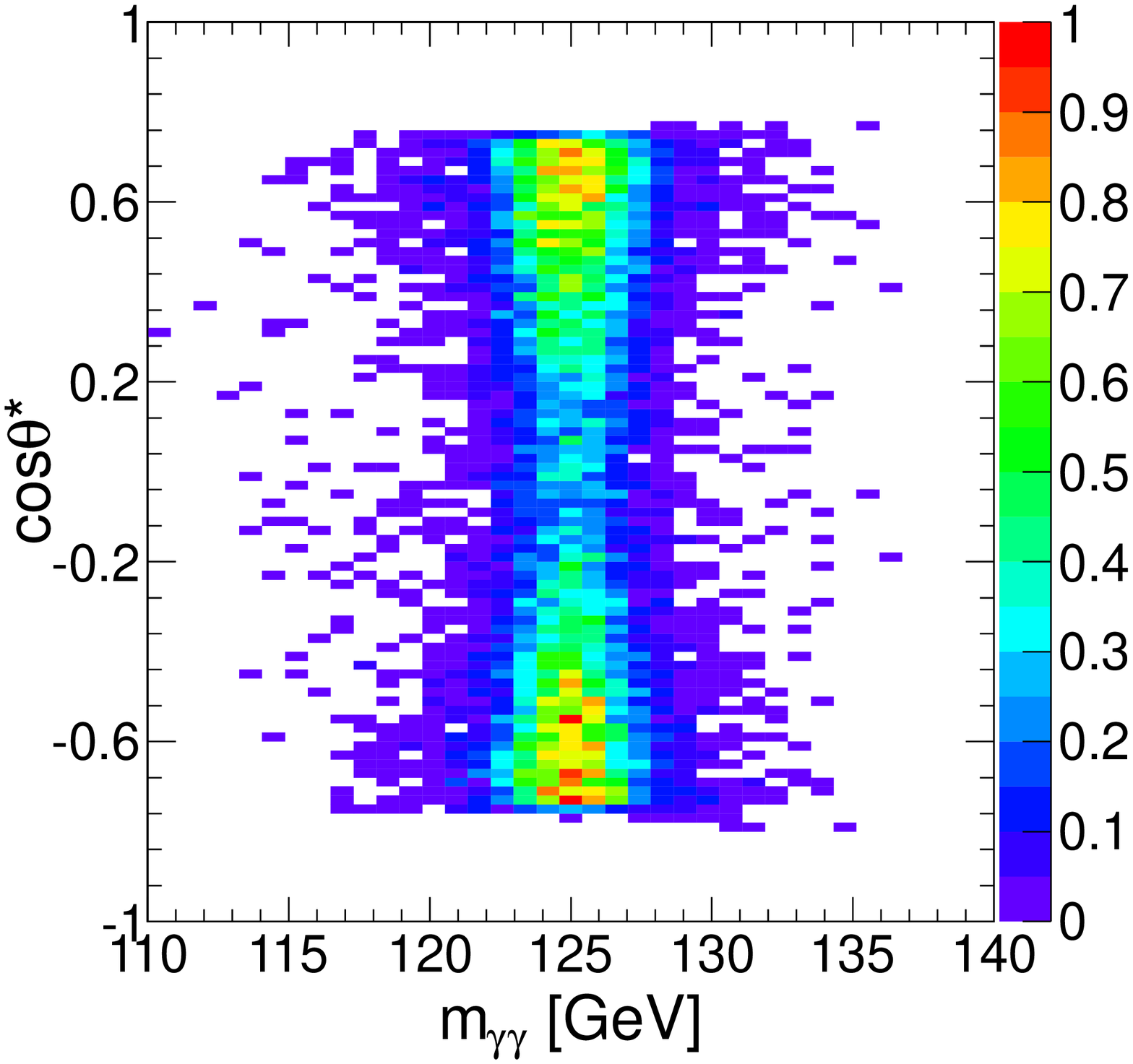,width=0.33\linewidth,height=0.33\linewidth}
}
\caption{
Template 
distributions of  $\cos\theta^*$ vs $m_{\gamma\gamma}$ in the $X\to\gamma\gamma$ analysis for the 
background (left), SM Higgs boson (middle), and spin-two resonance hypothesis $2^+_m$ (right). 
}
\label{fig:ggobs_2D}
\end{figure}

\vspace{1cm}

\begin{eqnarray}
\frac{16\, d\Gamma(X_{J=2}\to\gamma\gamma)}{5\, \Gamma d\cos\theta^\ast} 
&&   \!\!\!\!  = (2-2f_{z1}+f_{z2}) -6(2 -4f_{z1} -f_{z2})\cos^2\theta^\ast +3(6 -10f_{z1}-5f_{z2})\cos^4\theta^\ast \nonumber \\
&& +f_{+-}\,\left\{(2+2f_{z1}-7f_{z2})+6(2-6f_{z1}+f_{z2})\cos^2\theta^\ast-5(6-10f_{z1}-5f_{z2})\cos^4\theta^\ast \right\}
\nonumber \\
&& \propto 1+A\times\cos^2\theta^*+B\times\cos^4\theta^*
\,,
\label{eq:grav-twophoton}
\end{eqnarray}
where $f_{++}, f_{--}$, and $f_{-+} = f_{+-}$ are fractions of 
transverse amplitudes  in the decay, 
and $f_{z1}$ and $f_{z2}$ are polarization fractions in production,  
see Appendix A of Ref.~\cite{Gao:2010qx} for more details. 
The special case of the minimal coupling in both production and decay corresponds to $f_{z1}+f_{z2}=1$ and 
$f_{+-}=f_{-+}=1/2$. In this case, one obtains $(1+6\cos^2\theta^*+\cos^4\theta^*)$ for the $gg$ production 
mechanism with $f_{z2}=1$ and  $(1-\cos^4\theta^*)$ for the $q\bar{q}$ production mechanism with $f_{z1}=1$.
The ideal distributions in Eq.~(\ref{eq:grav-twophoton}) are shown together with generated events in  Fig.~\ref{fig:simulated}. 
These distributions 
are identical for the $2^+_h$ and $2^-_h$ hypotheses.

For illustration purposes, we proceed with the 
discussion of a simplified analysis.
The acceptance thresholds, chosen to be similar to those used in LHC analyses,  
are $E_T^1>m_{\gamma\gamma}/3$ and $E_T^2>m_{\gamma\gamma}/4$ 
for the first and second photons, respectively.
We apply $\eta$-dependent Gaussian random smearing to photon cluster energy 
which varies between between 1\% in the central pseudorapidity region and 6\% 
in the forward region.
Photons must be inside the calorimeter acceptance $|\eta|<2.5$
and outside the crack region $1.44<|\eta|<1.57$, similarly to the CMS experiment.
Considering the mass window $110 < m_{\gamma\gamma} < 140$~GeV,
we estimate the expected number of signal and background events based on Refs.~\cite{discovery-atlas, discovery-cms}
to be 22 and 3515 per $\ifb$, respectively. To extract the expected separation significance between different signal 
spin-parity hypotheses ${\cal S}$, we construct a two-dimensional template based on 
$(m,D)=(m_{\gamma\gamma}, \cos\theta^*)$ shown in Figs.~\ref{fig:ggobs} and \ref{fig:ggobs_2D}.
The loss of events at large values of $|\cos\theta^*|$ limits the precision of polarization measurements and 
is due to $p_T$ and $\eta$ selection requirements. 
Similar effects appear in the analysis of the Drell-Yan process, as discussed for example in Ref.~\cite{drellyan-cms}.
We rely on the shapes of the distributions after the above kinematic selection,
and the normalization is taken from data.
Using two-dimensional $(m_{\gamma\gamma}, \cos\theta^*)$ templates,
we obtain 2.7\,$\sigma$ significance with $10~\ifb$, 
which is similar to the LHC results expected for the SM Higgs boson~\cite{discovery-atlas,discovery-cms}.

We find good separation between 
the SM Higgs boson hypothesis and the spin-two models considered,
as can be seen in Fig.~\ref{fig:separation} and in Table~\ref{table-separation}. 
However, since just one angle is available in the analysis, the separation power may be weak 
or absent for other models where the $\cos\theta^*$ distribution is close to flat.


\section{Summary and Conclusion}
\label{sec:summary}

We have described a framework to determine the spin, parity, and general 
tensor structure of interactions of the new boson observed at the LHC. 
We consider a variety of Lorentz structures for spin-parity 
hypothesis testing that go beyond the minimal couplings expected for the SM Higgs boson or the 
graviton-like interactions of a spin-two boson. 
The full analytical calculation  of  angular and mass dependence 
of the decay amplitude $X \to V^*V^*$  
allows the most general analysis of a resonance with any integer spin $J$. 
A Monte Carlo simulation of the process $pp \to X \to V^*V^*$, 
with $V=Z$, $W$, and $\gamma$,  with off-shell electroweak gauge bosons, 
all spin correlations, and general couplings enables experimental 
investigation  of the properties of the new resonance.  
Both the analytic formulas and the event generator are publicly available, 
see Ref.~\cite{support}. 

We have illustrated how the spin and parity of  the new boson can be tested 
in the processes $pp \to X \to ZZ$, $WW$, 
and $\gamma\gamma$, using simplified simulation 
of the background and of the detector effects at the LHC experiments.
We have presented  the expected significance of spin-parity hypothesis 
separation for several  scenarios in Table~\ref{table-separation}, where 
it is assumed that the $5\sigma$ signal-to-background separation is achieved 
in each channel. 
The linearity of the relation between the signal-to-background 
significance and the spin-parity signal hypothesis separation significance allows 
us to extrapolate expectations to different luminosity scenarios, as shown in 
Fig.~\ref{fig:separation1} for $0^-$ and $2_m^+$ models.
We rely on the expected signal-to-background significances reported by the LHC experiments for the
integrated luminosity of about 10~fb$^{-1}$, which we take as 3.8, 2.4, and 2.8\,$\sigma$ 
in the $X\to ZZ$, $WW$, and $\gamma\gamma$ channels, respectively~\cite{discovery-cms}. 
In Table~\ref{table-separation-projected} we show examples of hypothesis separation expectations, 
per each LHC experiment, by the end of the 8~TeV LHC run, assuming 35~fb$^{-1}$
of integrated luminosity.

\begin{table}[t]
\caption{
Expected separation significance ${\cal S}$ (Gaussian $\sigma$)
between the SM Higgs boson scenario ($0_m^+$) and $0^-$ or $2_m^+$ 
hypotheses in the analyzed channels and combined, for the scenario corresponding 
approximately to 35~fb$^{-1}$ of integrated luminosity at one LHC experiment.
}
\begin{tabular}{lcccc}
\hline\hline
\vspace{0.1cm}
scenario & ~~$X\to ZZ$~~ &  ~~$X\to WW$~~ & ~~$X\to\gamma\gamma$~~ & ~~combined~~  \\
\hline
\vspace{0.1cm}
$0_m^+$ vs background & 7.1 & 4.5 &  5.2 &  9.9 \\
\hline
$0_m^+$ vs $0^-$   &  4.1 &  1.1 &  0.0  &  4.2   \\
\vspace{0.1cm}
$0_m^+$ vs $2_m^+$  &  2.2  &  2.5 &  2.5 &  4.2  \\
\hline\hline
\end{tabular}
\label{table-separation-projected}
\end{table}

\begin{figure}[t]
\centerline{
\epsfig{figure=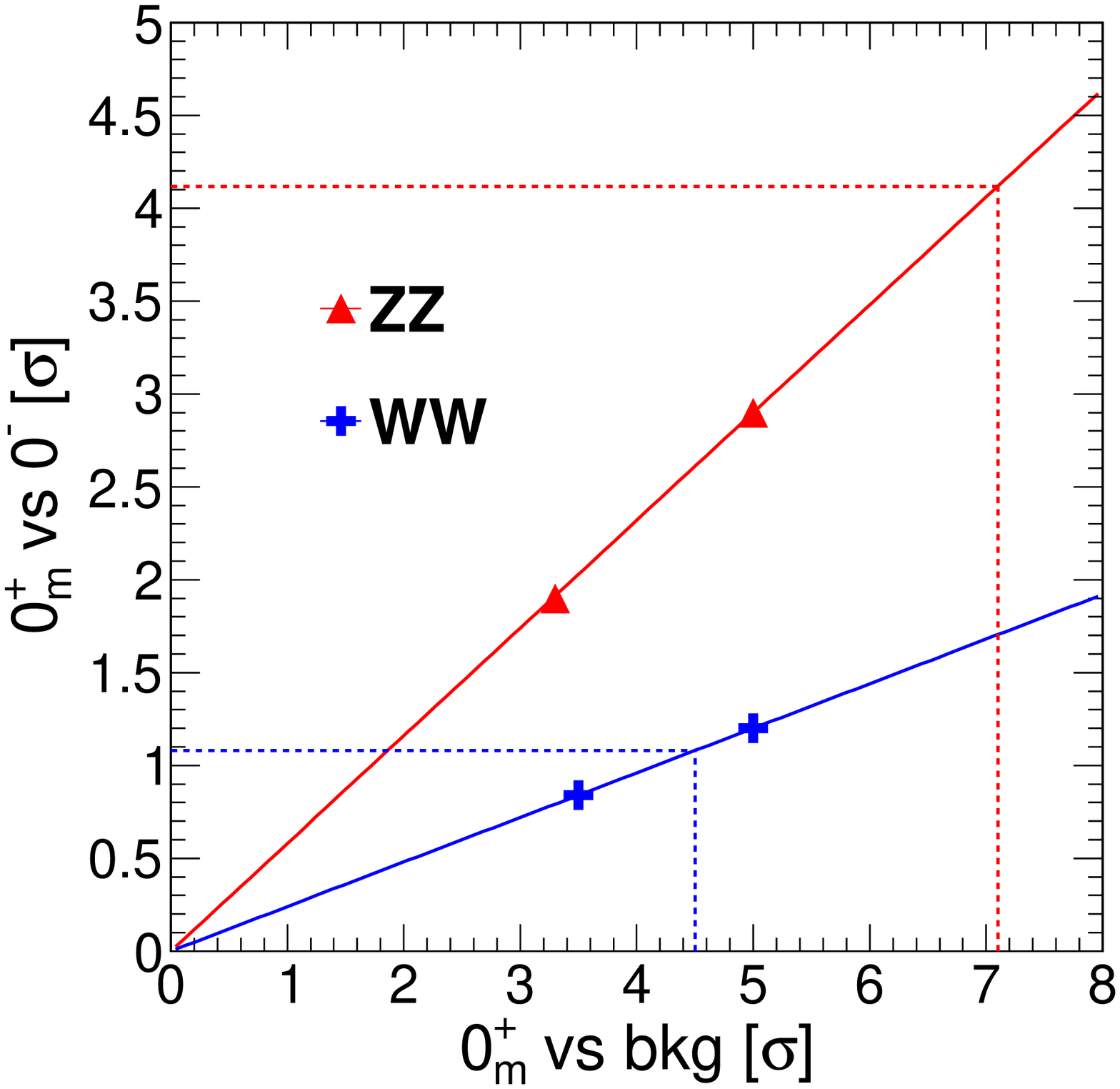,width=0.45\linewidth}
\epsfig{figure=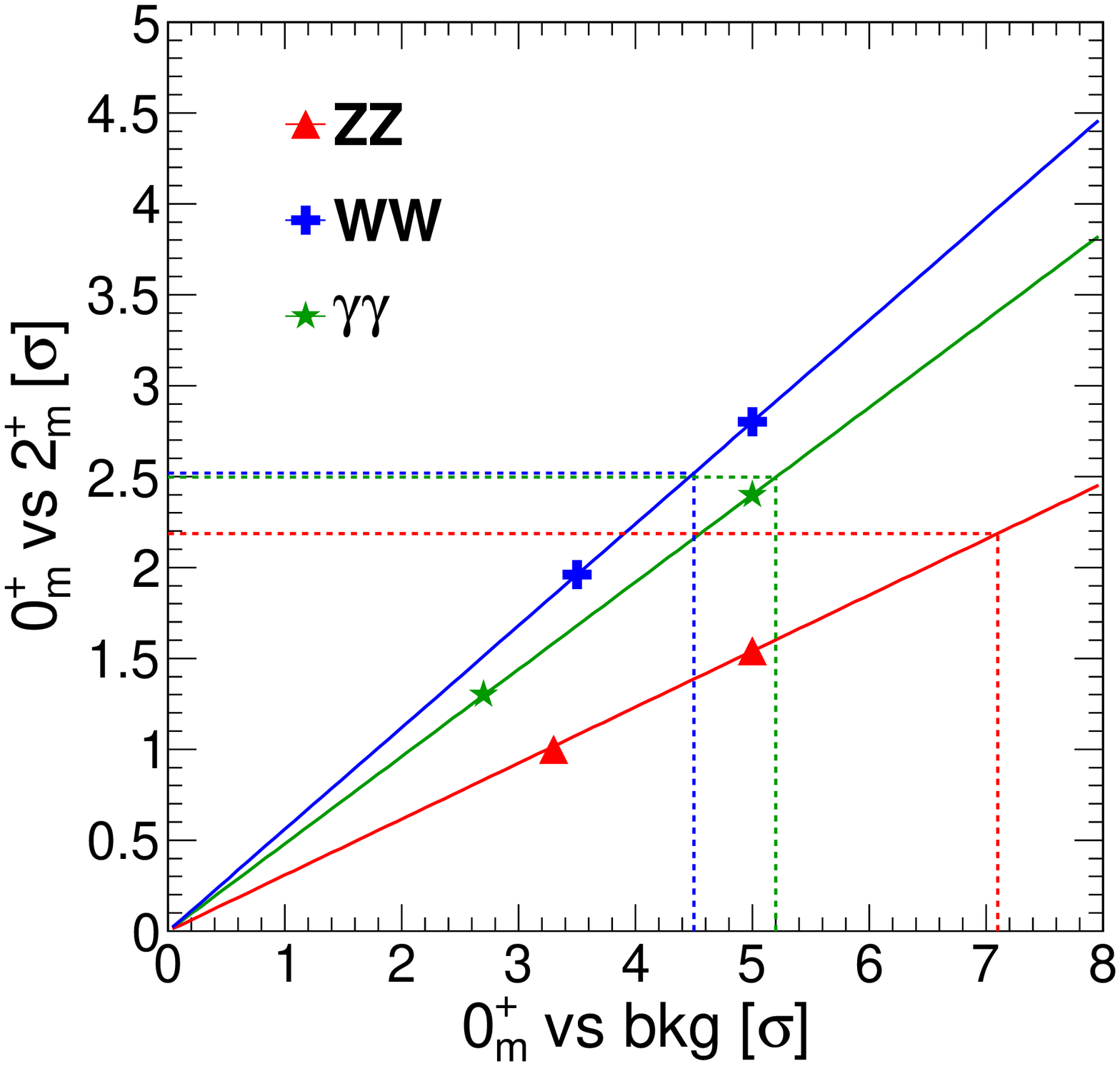,width=0.45\linewidth}
}
\caption{
Expected hypotheses separation significance vs signal observation significance
for the SM Higgs boson vs $0^-$ (left) and $2^+_m$ (right) hypotheses.
Points show two luminosity scenarios tested with generated experiments
and expectations are extrapolated linearly to other significance scenarios. 
Dashed lines indicate what might be expected with 35~fb$^{-1}$ of data 
at one LHC experiment.
}
\label{fig:separation1}
\end{figure}

We would also like to comment on some other potential final states in the decay
of the new boson, such as $Z\gamma$ and fermion-antifermion final states. 
Since no significant excess of events in these final states has been observed, we
leave detailed discussion of these final states to later work. However, 
the techniques discussed in this paper are applicable to them as well.
For example, analysis of associated  production 
$q\bar{q}\to Z^*\to ZX\to (\ell^-\ell^+) (b\bar{b})$ or $(\ell^-\ell^+) (\tau^-\tau^+)$,
and  similarly $q \bar q' \to W^*\to WX$, would follow the same formalism as discussed above.
In the above processes, the angular distributions of decay products should allow discrimination 
between the spin and coupling hypotheses for both $XVV$ and $Xf\bar{f}$. 
For a spin-zero $X$ decay, the fermion angular distributions are flat;
for a spin-one $X$, the angular distributions are similar to those in Eq.~(\ref{eq:mixedJtotal}) with $J=1$;
and for a spin-two $X$, the angular distributions can be obtained in a similar manner. 
The angular distributions for general couplings in the decay $X\to f\bar{f}$ 
can be obtained from Ref.~\cite{Gao:2010qx}.

In view of the importance of the discovery of the new boson for particle physics, 
it is important to confront all theoretical assumptions about its properties against 
experimental facts. The results presented in this paper point towards a realistic 
possibility that by the end of the 8 TeV run of the LHC,  extreme hypotheses about 
spin and parity of the new boson can be experimentally excluded.
However, it will be much harder to exclude contributions of anomalous 
couplings of the Higgs bosons to gauge bosons if they are smaller than  
ten percent of the SM couplings.  
For that, a significantly larger dataset will be required and multivariate 
fitting techniques~\cite{Gao:2010qx} will help to achieve this goal.
Nonetheless, if the nature of the new boson discovered at the LHC is exotic, 
there is a good chance to determine this already in the coming year.


\bigskip
\noindent
{\bf Acknowledgments}:
Several of us would like to thank CMS collaboration colleagues for feedback during the working group 
presentations of this analysis, and in particular Serguei Ganjour and Chia Ming Kuo for discussion 
of the two-photon analysis performance. 
We thank Rostislav Konoplich for pointing to an inconsistency with indices in Eqs.~(\ref{eq:fullampl-spin0}) and~(\ref{eq13}).
This research is partially supported by US NSF under grants PHY-1100862 and PHY-1214000, 
and by US DOE under grants DE-AC02-06CD11357 and DE-AC02-07CH11359.
We also acknowledge support from the LPC-CMS Fellows program operated through FNAL.
Calculations reported in this paper were performed on the Homewood High Performance Cluster 
of the Johns Hopkins University. 
%



\clearpage
\appendix

\section{Analytical angular distributions}
\label{sec:appendix-a}

In this appendix we present the general angular distribution in the production and decay of a particle $X$, 
with any integer spin $J$, in parton collisions
$ab\to X\to V_1(q_1) V_2(q_2)$, $V_1 \to  f(q_{11}) \bar{f}(q_{12})$, $V_2 \to f(q_{21}) \bar{f}(q_{22})$,
as derived in Ref.~\cite{Gao:2010qx} and generalized here to remove the constraint between the
$A_{\alpha\beta}$ and $A_{\beta\alpha}$ amplitudes.
Helicity amplitudes $A_{\alpha\beta}$ depend on the vector boson resonance masses $m_1$ and $m_2$,
as described in Eqs.~(\ref{eq:relate-spin0}), (\ref{eq:relate-spin1}), and (\ref{eq:relate-spin2}),
and related formulas incorporating the couplings. 
We work in the rest frame of the resonance 
$X$ and all angles that we use below are defined in Section~II.

The amplitudes $A_{\alpha\beta}$ are, in general, complex and the angular distribution 
is parameterized by the magnitude of the amplitude $|A_{\alpha\beta}|$ and the phase 
$\phi_{\alpha\beta} =\arg(A_{\alpha\beta}/A_{00})$.
The angular distribution is
%
\begin{eqnarray}
&&\frac{{\cal{N}}_J \; d \Gamma_{J}(m_1,m_2,\cos\theta^\ast,\Psi,\cos\theta_1,\cos\theta_2,\Phi)}
             { \; d\cos\theta^\ast d\Psi d\cos\theta_1d\cos\theta_2 d\Phi } = \nonumber \\
&& F^J_{0,0} (\theta^\ast) \times  \Bigl[  4\, |A_{00}|^2\,\sin^2\theta_1 \sin^2\theta_2 \nonumber \\ 
&& ~~~~~ ~~~~~ ~~~ + |A_{++}|^2 \left(1+2 A_{f_1} \cos\theta_1+\cos^2\theta_1 \right) \left(1+2 A_{f_2} \cos\theta_2+\cos^2\theta_2 \right) \nonumber \\ 
&& ~~~~~ ~~~~~ ~~~ + |A_{--}|^2 \left(1-2 A_{f_1} \cos\theta_1+\cos^2\theta_1\right) \left(1-2 A_{f_2} \cos\theta_2+\cos^2\theta_2\right) \nonumber \\ 
&& ~~~~~ ~~~~~ ~~~ + 4|A_{00}| |A_{++}| (A_{f_1} + \cos\theta_1) \sin\theta_1 (A_{f_2} + \cos\theta_2) \sin\theta_2 \cos(\Phi + \phi_{++}) \nonumber \\ 
&& ~~~~~ ~~~~~ ~~~ + 4|A_{00}| |A_{--}| (A_{f_1} - \cos\theta_1) \sin\theta_1 (A_{f_2} - \cos\theta_2) \sin\theta_2 \cos(\Phi - \phi_{--})  \nonumber \\ 
&& ~~~~~ ~~~~~ ~~~ + 2|A_{++}| |A_{--}| \sin^2\theta_1 \sin^2\theta_2 \cos(2\Phi - \phi_{--} + \phi_{++})\Bigr] \nonumber \\ 
&& + F^J_{1,1} (\theta^\ast) \times \Bigl[ 2|A_{+0}|^2 (1 + 2 A_{f_1} \cos\theta_1 + \cos^2\theta_1) \sin^2\theta_2 \nonumber \\ 
&& ~~~~~ ~~~~~ ~~~~ + 2|A_{0-}|^2 \sin^2\theta_1 (1 - 2 A_{f_2} \cos\theta_2 + \cos^2\theta_2) \nonumber \\ 
&& ~~~~~ ~~~~~ ~~~~ + 2|A_{-0}|^2 (1 - 2 A_{f_1} \cos\theta_1 + \cos^2\theta_1) \sin^2\theta_2  \nonumber \\ 
&& ~~~~~ ~~~~~ ~~~~ + 2|A_{0+}|^2 \sin^2\theta_1 (1 + 2 A_{f_2} \cos\theta_2 + \cos^2\theta_2) \nonumber \\ 
&& ~~~~~ ~~~~~ ~~~~ + 4 |A_{+0}||A_{0-}| (A_{f_1} + \cos\theta_1) \sin\theta_1 (A_{f_2} - \cos\theta_2) \sin\theta_2 \cos(\Phi + \phi_{+0} - \phi_{0-}) \nonumber \\ 
&& ~~~~~ ~~~~~ ~~~~ + 4 |A_{0+}||A_{-0}| (A_{f_1} - \cos\theta_1) \sin\theta_1 (A_{f_2} + \cos\theta_2) \sin\theta_2 \cos(\Phi + \phi_{0+} - \phi_{-0}) \Bigr] \nonumber \\ 
&& + F^J_{1,-1} (\theta^\ast) \times \Bigl[ 4 |A_{+0}||A_{0+}| (A_{f_1} + \cos\theta_1) \sin\theta_1 (A_{f_2} + \cos\theta_2) \sin\theta_2 \cos(2\Psi - \phi_{+0} + \phi_{0+}) \nonumber \\ 
&& ~~~~~ ~~~~~ ~~~~~ ~ + 4 |A_{+0}||A_{-0}| \sin^2\theta_1 \sin^2\theta_2 \cos(2\Psi - \Phi - \phi_{+0} + \phi_{-0}) \nonumber \\ 
&& ~~~~~ ~~~~~ ~~~~~ ~ + 4 |A_{0-}||A_{0+}| \sin^2\theta_1 \sin^2\theta_2 \cos(2\Psi + \Phi -\phi_{0-} + \phi_{0+}) \nonumber \\ 
&& ~~~~~ ~~~~~ ~~~~~ ~ + 4 |A_{0-}||A_{-0}| (A_{f_1} - \cos\theta_1) \sin\theta_1 (A_{f_2} - \cos\theta_2) \sin\theta_2 \cos(2\Psi -\phi_{0-} + \phi_{-0}) \Bigr] \nonumber \\ 
&& + F^J_{2,2} (\theta^\ast) \times \Bigl[ |A_{+-}|^2 (1 + 2 A_{f_1} \cos\theta_1 + \cos^2\theta_1) (1 - 2 A_{f_2} \cos\theta_2 + \cos^2\theta_2) \nonumber \\ 
&& ~~~~~ ~~~~~ ~~~~ + |A_{-+}|^2 (1 - 2 A_{f_1} \cos\theta_1 + \cos^2\theta_1) (1 + 2 A_{f_2} \cos\theta_2 + \cos^2\theta_2) \Bigr] \nonumber\\ 
&& + F^J_{2,-2} (\theta^\ast) \times \Bigl[ 2 |A_{+-}||A_{-+}| \sin^2\theta_1 \sin^2\theta_2 \cos(4\Psi -\phi_{+-} + \phi_{-+})\Bigr] \nonumber \\ 
&& + F^J_{0,1} (\theta^\ast) \times \Bigl[ 4\sqrt{2} |A_{00}| |A_{+0}| (A_{f_1} + \cos\theta_1) \sin\theta_1 \sin^2\theta_2 \cos(\Psi - \Phi/2 - \phi_{+0}) \nonumber \\ 
&& ~~~~~ ~~~~~ ~~~~ + 4\sqrt{2} |A_{00}| |A_{0-}| \sin^2\theta_1 (A_{f_2} - \cos\theta_2) \sin\theta_2 \cos(\Psi + \Phi/2 - \phi_{0-}) \nonumber \\ 
&& ~~~~~ ~~~~~ ~~~~ + 2\sqrt{2} |A_{--}| |A_{+0}| \sin^2\theta_1 (A_{f_2} - \cos\theta_2) \sin\theta_2 \cos(-\Psi + 3\Phi/2 + \phi_{+0} - \phi_{--}) \nonumber \\ 
&& ~~~~~ ~~~~~ ~~~~ + 2\sqrt{2} |A_{--}| |A_{0-}| (A_{f_1} - \cos\theta_1) \sin\theta_1 (1 - 2 A_{f_2} \cos\theta_2 + \cos^2\theta_2) \cos(- \Psi + \Phi/2 + \phi_{0-} - \phi_{--}) \nonumber \\ 
&& ~~~~~ ~~~~~ ~~~~ + 2\sqrt{2} |A_{++}| |A_{+0}| (1 + 2 A_{f_1} \cos\theta_1 + \cos^2\theta_1) (A_{f_2} + \cos\theta_2) \sin\theta_2 \cos(\Psi + \Phi/2 - \phi_{+0} + \phi_{++}) \nonumber \\ 
&& ~~~~~ ~~~~~ ~~~~ + 2\sqrt{2} |A_{++}| |A_{0-}| (A_{f_1} + \cos\theta_1) \sin\theta_1 \sin^2\theta_2 \cos(\Psi + 3\Phi/2 - \phi_{0-} + \phi_{++}) \Bigr] \nonumber \\ 
&& + F^J_{0,-1} (\theta^\ast) \times \Bigl[ 4\sqrt{2} |A_{00}| |A_{0+}| \sin^2\theta_1 (A_{f_2} + \cos\theta_2) \sin\theta_2 \cos(\Psi + \Phi/2 + \phi_{+0}) \nonumber \\ 
&& ~~~~~ ~~~~~ ~~~~~ + 4\sqrt{2} |A_{00}| |A_{-0}| (A_{f_1} -  \cos\theta_1) \sin\theta_1 \sin^2\theta_2 \cos(\Psi - \Phi/2 + \phi_{-0}) \nonumber \\ 
&& ~~~~~ ~~~~~ ~~~~~ + 2\sqrt{2} |A_{--}| |A_{0+}| (A_{f_1} - \cos\theta_1) \sin\theta_1 \sin^2\theta_2 \cos(\Psi + 3\Phi/2 + \phi_{0+} - \phi_{--}) \nonumber \\ 
&& ~~~~~ ~~~~~ ~~~~~ + 2\sqrt{2} |A_{--}| |A_{-0}| (1 - 2 A_{f_1} \cos\theta_1 + \cos^2\theta_1) (A_{f_2} - \cos\theta_2) \sin\theta_2 \cos(\Psi + \Phi/2 + \phi_{-0} - \phi_{--})  \nonumber \\ 
&& ~~~~~ ~~~~~ ~~~~~ + 2\sqrt{2} |A_{++}| |A_{0+}| (A_{f_1} + \cos\theta_1) \sin\theta_1 (1 + 2 A_{f_2} \cos\theta_2 + \cos^2\theta_2) \cos(\Psi - \Phi/2 + \phi_{0+} - \phi_{++}) \nonumber \\ 
&& ~~~~~ ~~~~~ ~~~~~ + 2\sqrt{2} |A_{++}| |A_{-0}| \sin^2\theta_1 (A_{f_2} + \cos\theta_2) \sin\theta_2 \cos(-\Psi + 3\Phi/2 - \phi_{-0} + \phi_{++}) \Bigr] \nonumber \\ 
&& + F^J_{0,2} (\theta^\ast) \times \Bigl[ 4 |A_{00}| |A_{+-}| (A_{f_1} + \cos\theta_1) \sin\theta_1 (A_{f_2} - \cos\theta_2) \sin\theta_2 \cos( 2\Psi - \phi_{+-}) \nonumber \\ 
&& ~~~~~ ~~~~~ ~~~~ + 2 |A_{--}| |A_{+-}| \sin^2\theta_1 (1 - 2 A_{f_2} \cos\theta_2 + \cos^2\theta_2) \cos(2\Psi - \Phi + \phi_{--} - \phi_{+-})  \nonumber \\ 
&& ~~~~~ ~~~~~ ~~~~ + 2 |A_{++}| |A_{+-}| (1 + 2 A_{f_1} \cos\theta_1 + \cos^2\theta_1) \sin^2\theta_2 \cos(2\Psi + \Phi + \phi_{++} - \phi_{+-}) \Bigr]  \nonumber \\ 
&& + F^J_{0,-2} (\theta^\ast) \times \Bigl[ 4 |A_{00}| |A_{-+}| (A_{f_1} - \cos\theta_1) \sin\theta_1 (A_{f_2} + \cos\theta_2) \sin\theta_2 \cos( 2\Psi + \phi_{-+}) \nonumber \\ 
&& ~~~~~ ~~~~~ ~~~~~ + 2 |A_{--}| |A_{-+}| (1 - 2 A_{f_1} \cos\theta_1 + \cos^2\theta_1) \sin^2\theta_2 \cos(2\Psi + \Phi - \phi_{--} + \phi_{-+})  \nonumber \\ 
&& ~~~~~ ~~~~~ ~~~~~ + 2 |A_{++}| |A_{-+}| \sin^2\theta_1 (1 + 2 A_{f_2} \cos\theta_2 + \cos^2\theta_2) \cos(2\Psi - \Phi - \phi_{++} + \phi_{-+})  \Bigr] \nonumber \\ 
&& + F^J_{1,2} (\theta^\ast)  \times \Bigl[ 2\sqrt{2} |A_{+0}| |A_{+-}| (1 + 2 A_{f_1} \cos\theta_1 + \cos^2\theta_1) (A_{f_2} - \cos\theta_2) \sin\theta_2 \cos(\Psi + \Phi/2+\phi_{+0}-\phi_{+-}) \nonumber \\ 
&& ~~~~~ ~~~~~ ~~~~ + 2\sqrt{2} |A_{0-}| |A_{+-}| (A_{f_1} + \cos\theta_1) \sin\theta_1 (1 - 2 A_{f_2} \cos\theta_2 + \cos^2\theta_2) \cos(\Psi - \Phi/2 + \phi_{0-} - \phi_{+-}) \nonumber \\ 
&& ~~~~~ ~~~~~ ~~~~ - 2\sqrt{2} |A_{0+}||A_{-+}| (A_{f_1} - \cos\theta_1) \sin\theta_1 (1 + 2 A_{f_2} \cos\theta_2 + \cos^2\theta_2) \cos(-\Psi + \Phi/2 + \phi_{0+} - \phi_{-+}) \nonumber \\ 
&& ~~~~~ ~~~~~ ~~~~ - 2\sqrt{2} |A_{-0}||A_{-+}| (1 - 2 A_{f_1} \cos\theta_1 + \cos^2\theta_1) (A_{f_2} + \cos\theta_2) \sin\theta_2 \cos(\Psi + \Phi/2 + \phi_{-0} - \phi_{-+}) \Bigr] \nonumber \\ 
&& + F^J_{1,-2} (\theta^\ast) \times \Bigl[ 2\sqrt{2} |A_{+0}| |A_{-+}| \sin^2\theta_1 (A_{f_2} + \cos\theta_2) \sin\theta_2 \cos(3\Psi - \Phi/2 -\phi_{+0} + \phi_{-+})   \nonumber \\ 
&& ~~~~~ ~~~~~ ~~~~~ + 2\sqrt{2} |A_{0-}||A_{-+}| (A_{f_1} - \cos\theta_1) \sin\theta_1 \sin^2\theta_2 \cos(3\Psi + \Phi/2 - \phi_{0-} + \phi_{-+}) \nonumber \\ 
&& ~~~~~ ~~~~~ ~~~~~ - 2\sqrt{2} |A_{0+}||A_{+-}| (A_{f_1} + \cos\theta_1) \sin\theta_1 \sin^2\theta_2 \cos(3\Psi + \Phi/2 + \phi_{0+} - \phi_{+-}) \nonumber \\ 
&& ~~~~~ ~~~~~ ~~~~~ - 2\sqrt{2} |A_{-0}||A_{+-}| \sin^2\theta_1 (A_{f_2} - \cos\theta_2) \sin\theta_2 \cos(3\Psi - \Phi/2 + \phi_{-0} - \phi_{+-}) \Bigr] 
\,, 
\label{eq:mixedJtotal}
\end{eqnarray}
%
where ${\cal N}_J$ is the normalization constant which does not affect the angular and mass distributions. 
Because decays of vector bosons $V_i\to f_i\bar{f}_i$ are involved, the 
angular distributions depend on the parameter $A_{f_i}$ characterizing their decay,  
defined as $A_{f}=2\bar{g}_V^f\bar{g}_A^f/(\bar{g}_V^{f2}+\bar{g}_A^{f2})$~\cite{pdg}.
This parameter is $1$ for $W$ decays and approximately 0.15 for $Z\to\ell^-\ell^+$.
Equation~(\ref{eq:mixedJtotal}) represents a more general version of Eq.~(B1) from Ref.~\cite{Gao:2010qx},
where sign conventions are different between the two equations.  
Conventions for Eq.~(\ref{eq:mixedJtotal}) are consistent with Eqs.~(\ref{eq:angles-1})--(\ref{eq:angles-3}).
The functions $F^{J}_{i,j}(\theta^\ast)$ are defined through the Wigner $d$-functions 
as\footnote{
The convention presented here differs from that in Ref.~\cite{Gao:2010qx}. 
All probability distributions are invariant under the simultaneous transformations
$\theta^*\rightarrow(\pi-\theta^*)$ and $\Phi_1\rightarrow(\pi+\Phi_1)$.
The different convention is equivalent to either of these two transformations.
}
\begin{eqnarray}
F^J_{i,j} (\theta^\ast) = \sum_{m=0,\pm1,\pm2} f_m \, d^J_{im} (\theta^\ast) d^J_{jm} (\theta^\ast) \,,
\label{eq:formfactors}
\end{eqnarray}
where $f_m$ are fractions of the $X$ particle polarization
as defined in Ref.~\cite{Gao:2010qx}. 
In $q\bar{q}$ annihilation the resonance $X$ can only be produced by $m=\pm1$,
whereas in gluon fusion $m=\pm2$ or $0$.
The relative fractions of $m=\pm2$ and $0$ are determined
by amplitudes in Eq.~(\ref{eq:relate-spin2}) which simplify in the case of couplings
to two massless gluons and depend on production couplings in Eq.~(\ref{eq:fullampl-spin2}).
The relative fraction of $q\bar{q}\to X$ production is denoted by $f_{q\bar{q}}$ and is 
determined by the ratio of cross-sections, including effects of parton structure functions. 
This leads to 
\begin{eqnarray}
f_{+1}=f_{-1} &=& \frac{f_{z1}}{2} = \frac{f_{q\bar{q}}}{2}
\,, \nonumber \\
f_{+2}=f_{-2}
&=& \frac{f_{z2}}{2} = (1-f_{q\bar{q}})\frac{|A^{gg}_{+-}|^2}{\sum_{\alpha,\beta=\pm1} |A^{gg}_{\alpha\beta}|^2}
 =  (1-f_{q\bar{q}})\frac{|A^{gg}_{-+}|^2}{\sum_{\alpha,\beta=\pm1} |A^{gg}_{\alpha\beta}|^2}
\,, \nonumber \\
f_{0} &=& {f_{z0}} = (1-f_{q\bar{q}})\frac{|A^{gg}_{++}|^2+|A^{gg}_{--}|^2}{\sum_{\alpha,\beta=\pm1} |A^{gg}_{\alpha\beta}|^2}
\,.
\label{eq:fractions}
\end{eqnarray}
For a spin-zero resonance $f_{q\bar{q}}=0$ and $f_{0}=1$. 
For a spin-one resonance $f_{q\bar{q}}=1$.
For a spin-two resonance, generally all polarizations are possible.
The minimal couplings of a spin-two resonance correspond to $f_{0}=0$.
Specific examples of $F^J_{i,j}$ for $J = 0,1,2$ are given in Ref.~\cite{Gao:2010qx}.


\begin{figure}[b]
\centerline{
\epsfig{figure=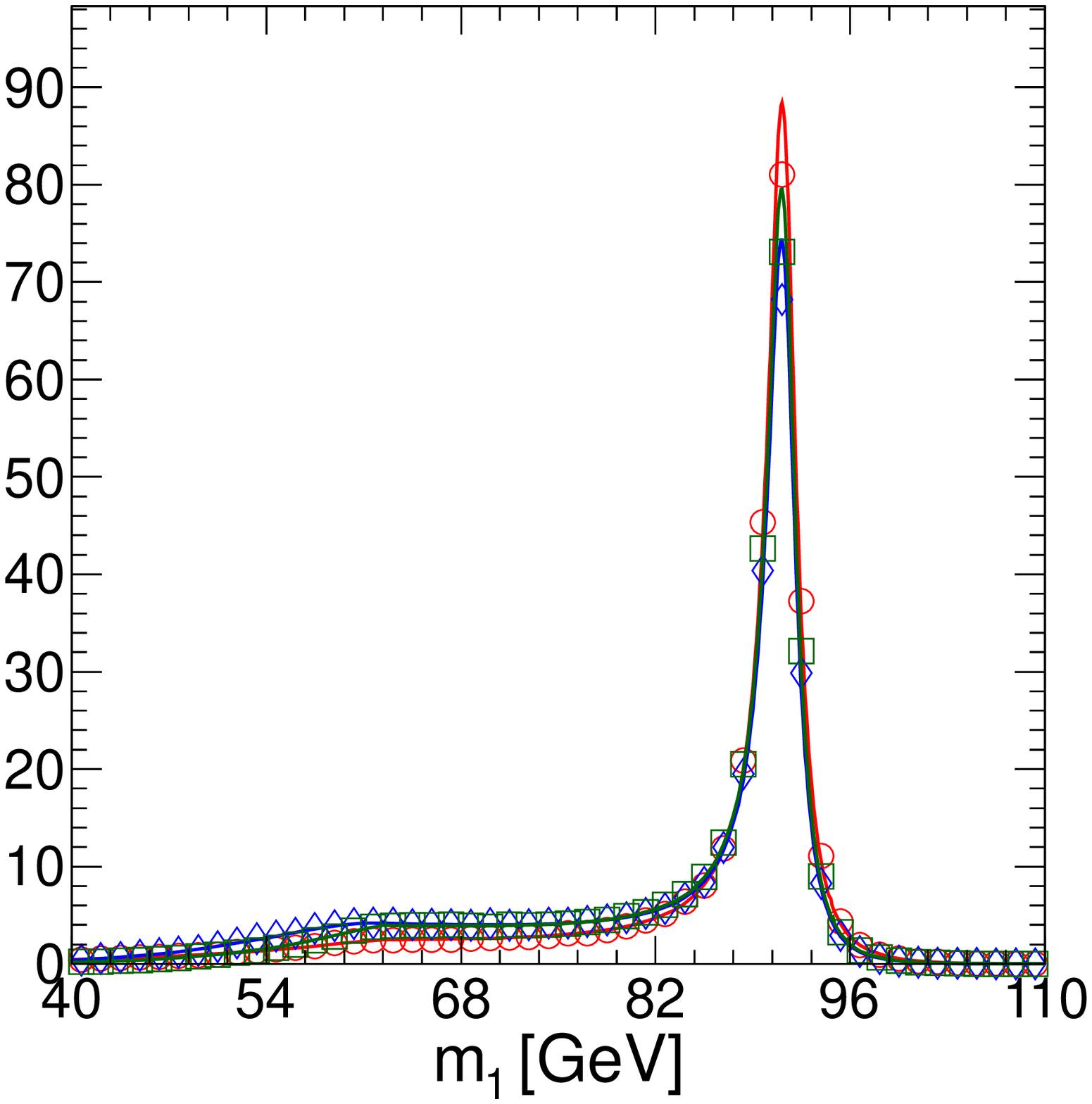,width=0.25\linewidth}
\epsfig{figure=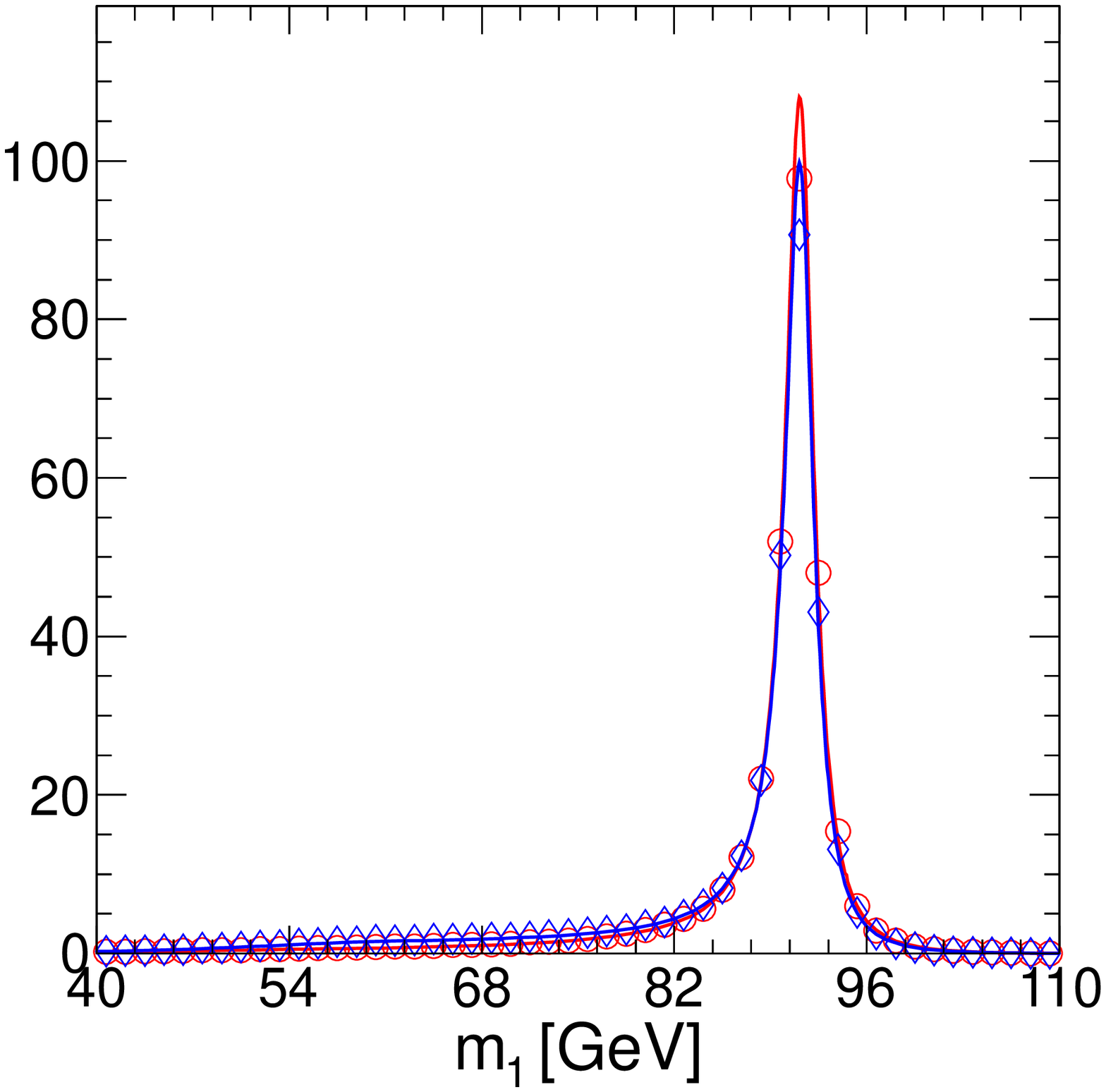,width=0.25\linewidth}
\epsfig{figure=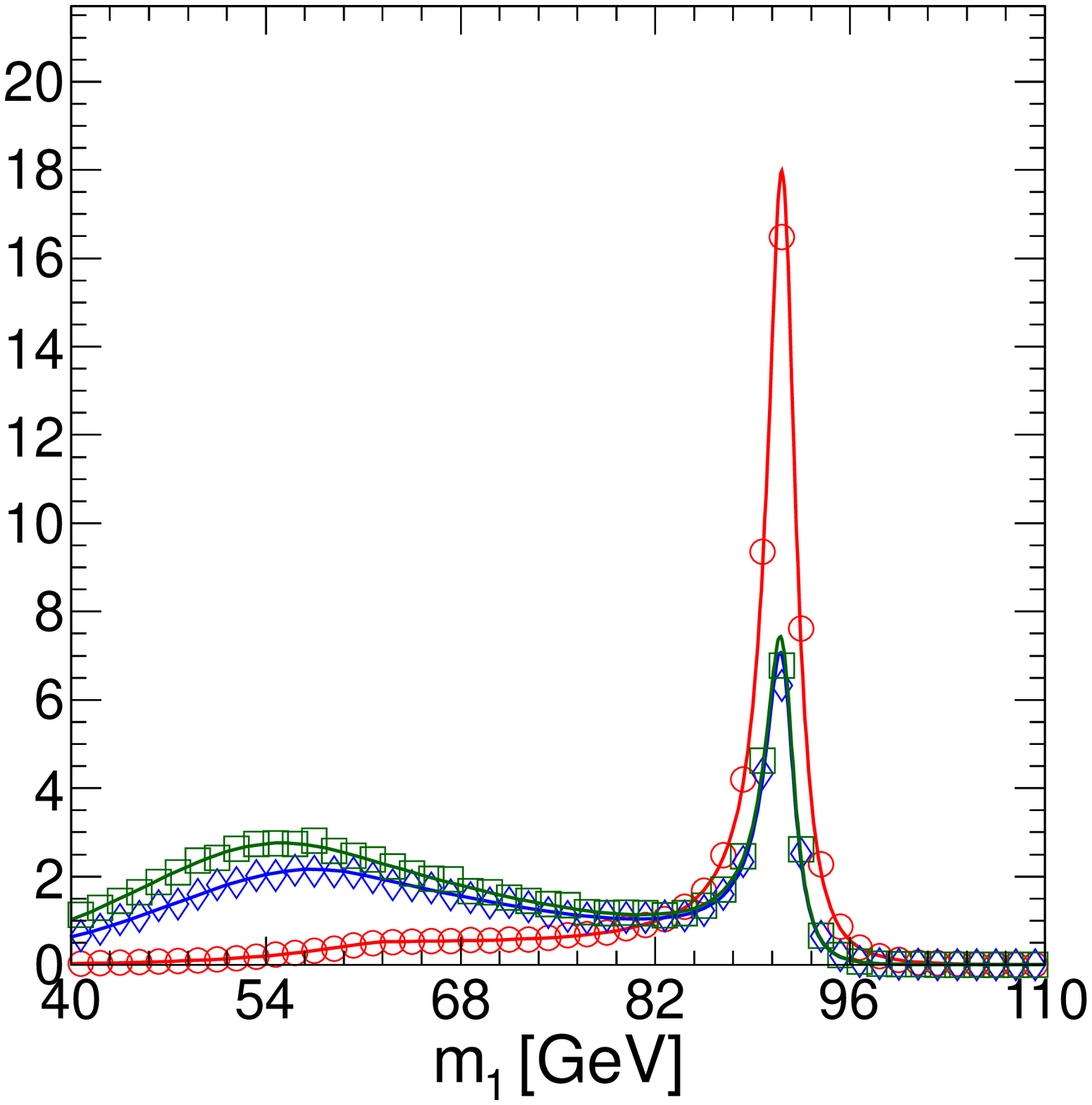,width=0.25\linewidth}
\epsfig{figure=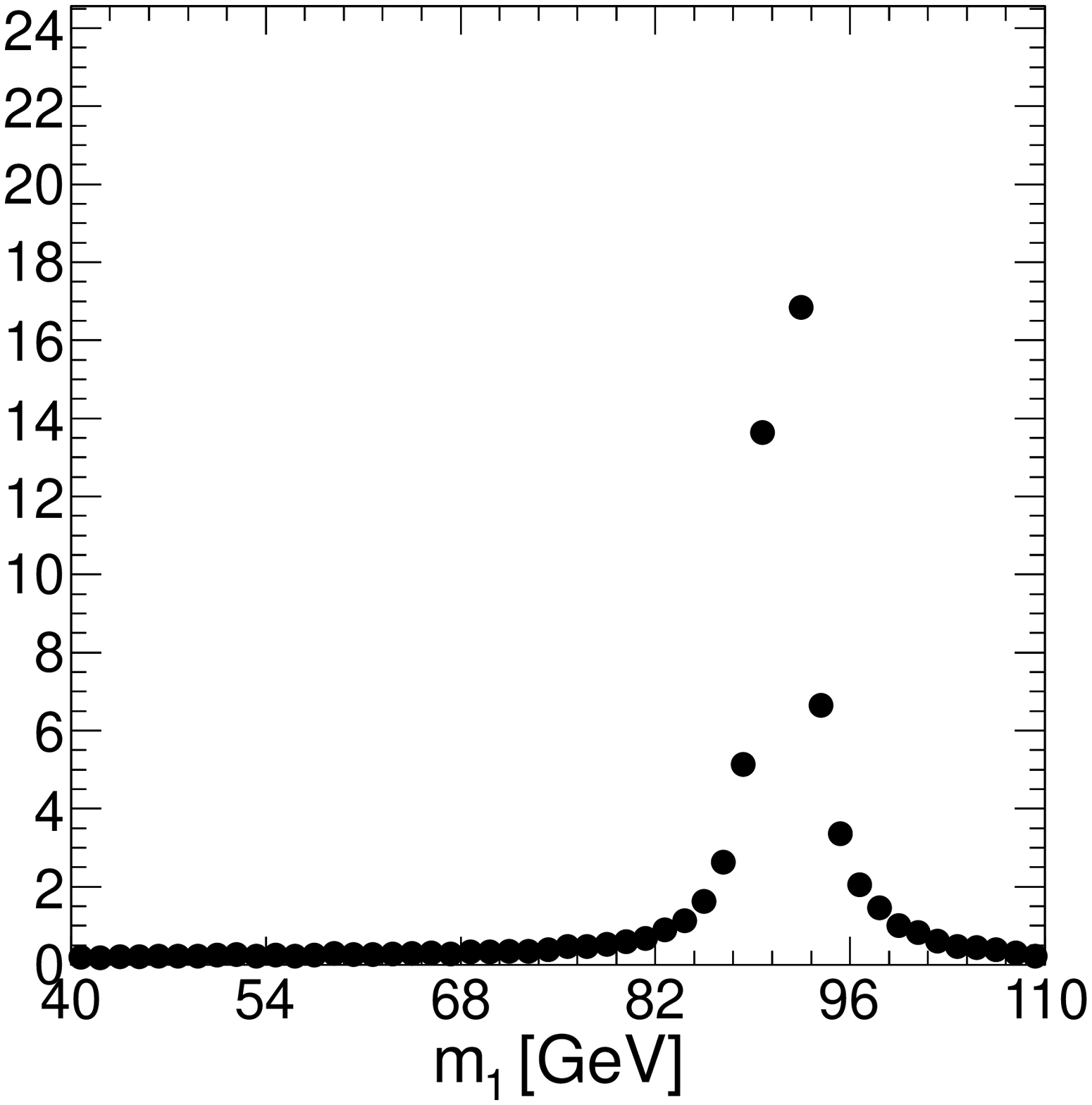,width=0.25\linewidth}
}
\centerline{
\epsfig{figure=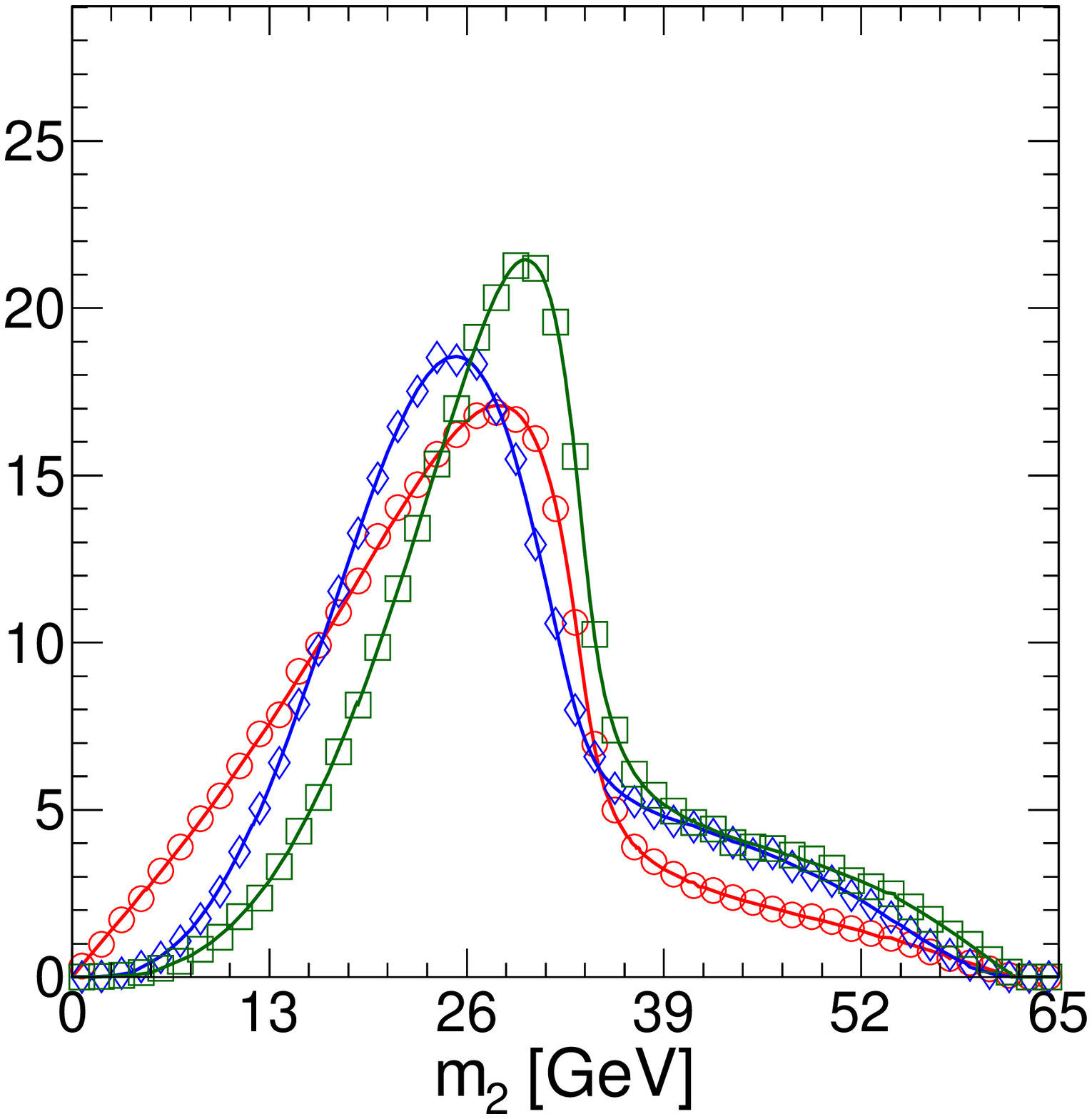,width=0.25\linewidth}
\epsfig{figure=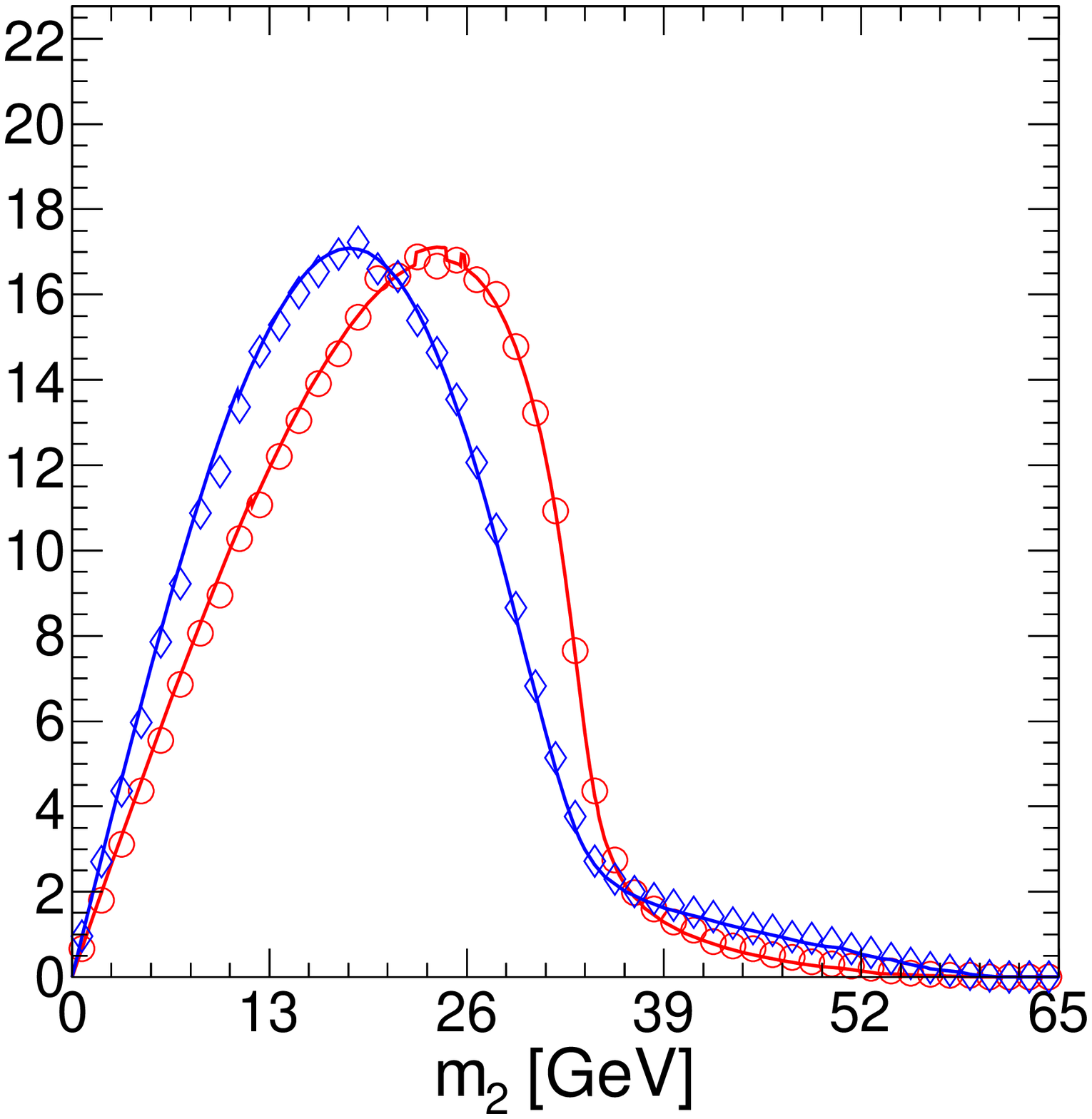,width=0.25\linewidth}
\epsfig{figure=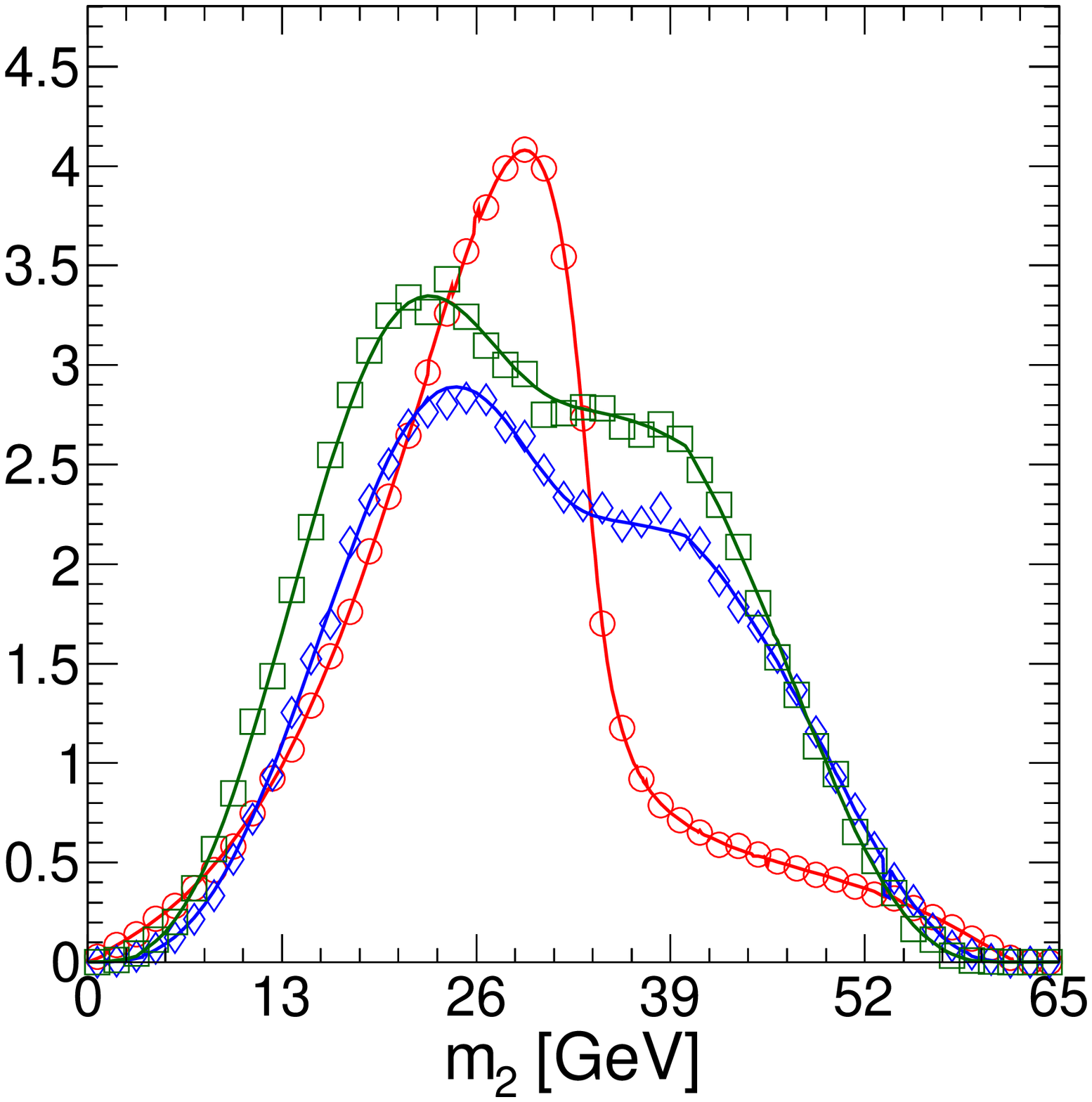,width=0.25\linewidth}
\epsfig{figure=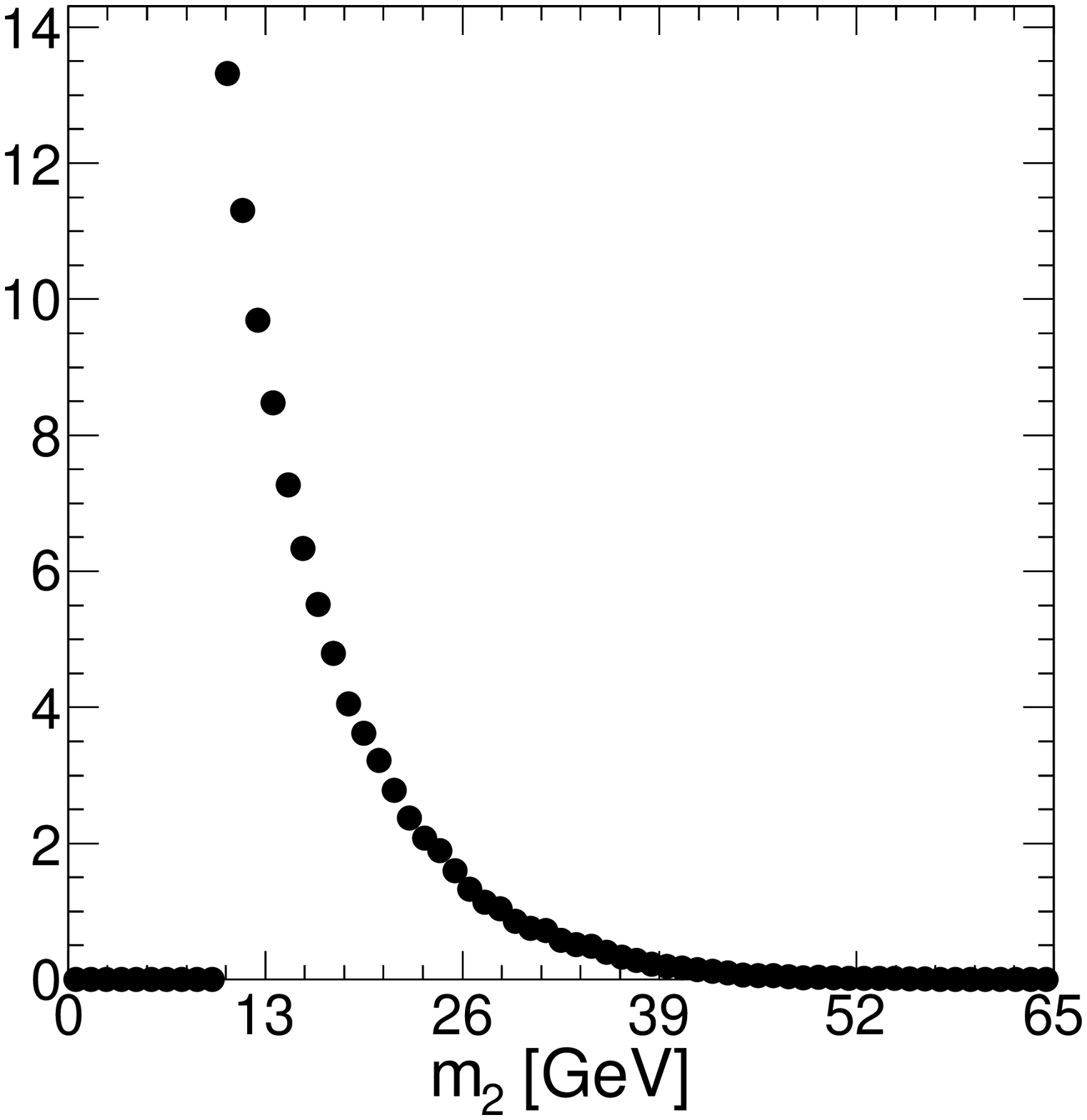,width=0.25\linewidth}
}
\caption{
Distributions of the observables in the $X\to ZZ$ analysis, from left to right:
spin-zero, spin-one, and spin-two signal, and $q\bar{q}\to ZZ$ background.
The signal hypotheses shown are $J^+_m$ (red circles), $J^+_h$ (green squares), 
$J^-_h$ (blue diamonds), as defined in Table~\ref{table-scenarios}. 
Background is shown with the requirements $m_2>10$ GeV and $120<m_{4\ell}<130$ GeV.
The observables shown from top to bottom: $m_1$ and $m_2$ (where $m_1>m_2$).
Points show simulated events and lines show projections of analytical distributions.
}
\label{fig:simulated-zz-mass}
\end{figure}

\section{Angular and mass distributions}
\label{sec:appendix-b}

We illustrate MC simulation and compare it to the derived analytical angular and mass distributions
in Figs.~\ref{fig:simulated-zz-mass} and \ref{fig:simulated-zz-angles} for the $ZZ$,
and in Fig.~\ref{fig:simulated-ww} for the $WW$ final states.
The $X\to\gamma\gamma$ distributions are shown in Fig.~\ref{fig:simulated}.
We have also validated that results presented in this paper using Eqs.~(\ref{eq:mela}) and~(\ref{eq:melaSig}) 
are nearly identical if in place of analytical parameterization of the probabilities, 
${\cal P}$, we use matrix element calculations from the vector algebra employed in the 
event generator. The two methods are conceptually independent but are mathematically 
equivalent, apart from the normalization of the probabilities which is easier to 
calculate with the analytical parameterization. We provide the necessary code for 
both methods in Ref.~\cite{support}.

\begin{figure}[t]
\centerline{
\epsfig{figure=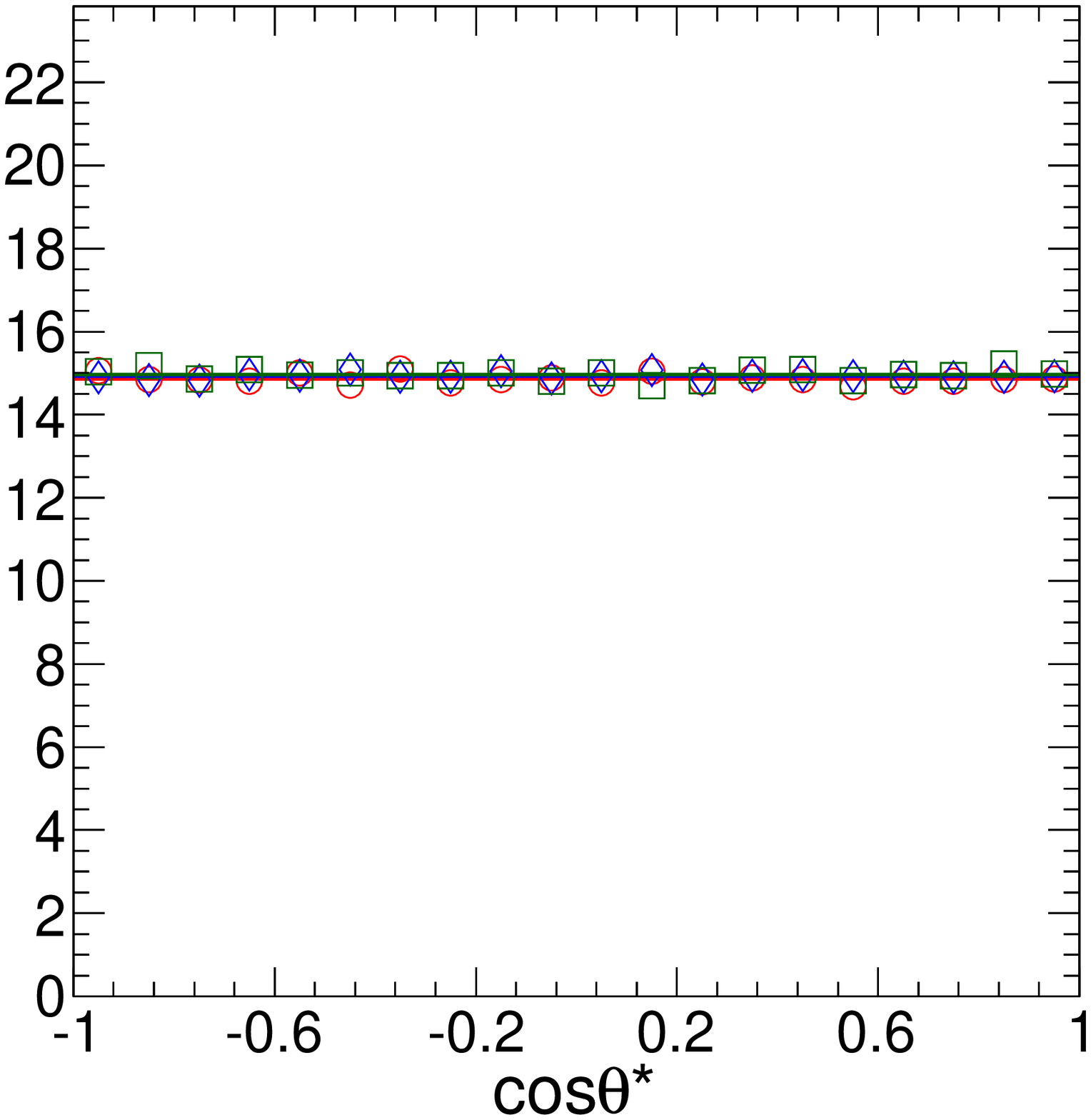,width=0.25\linewidth}
\epsfig{figure=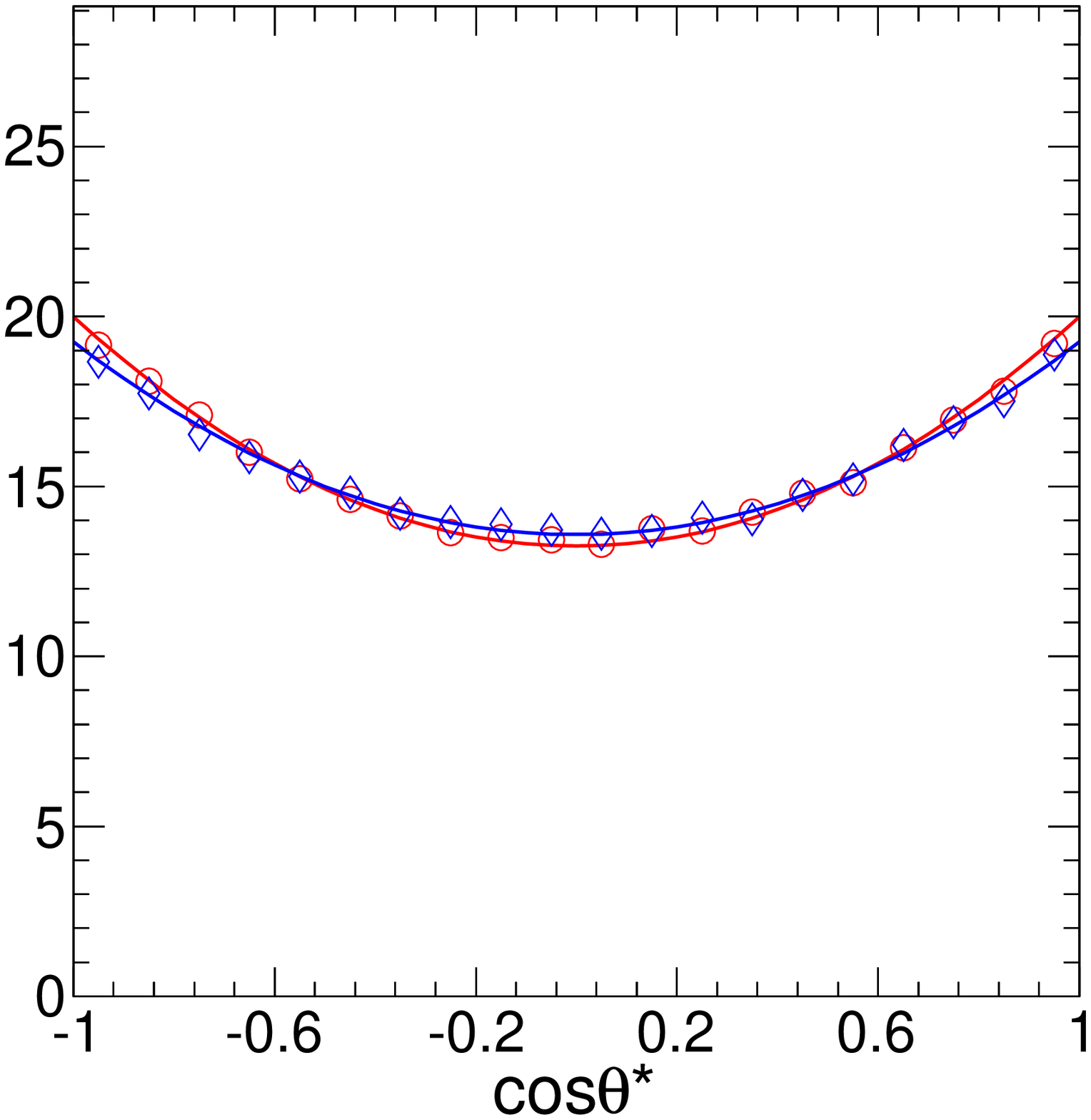,width=0.25\linewidth}
\epsfig{figure=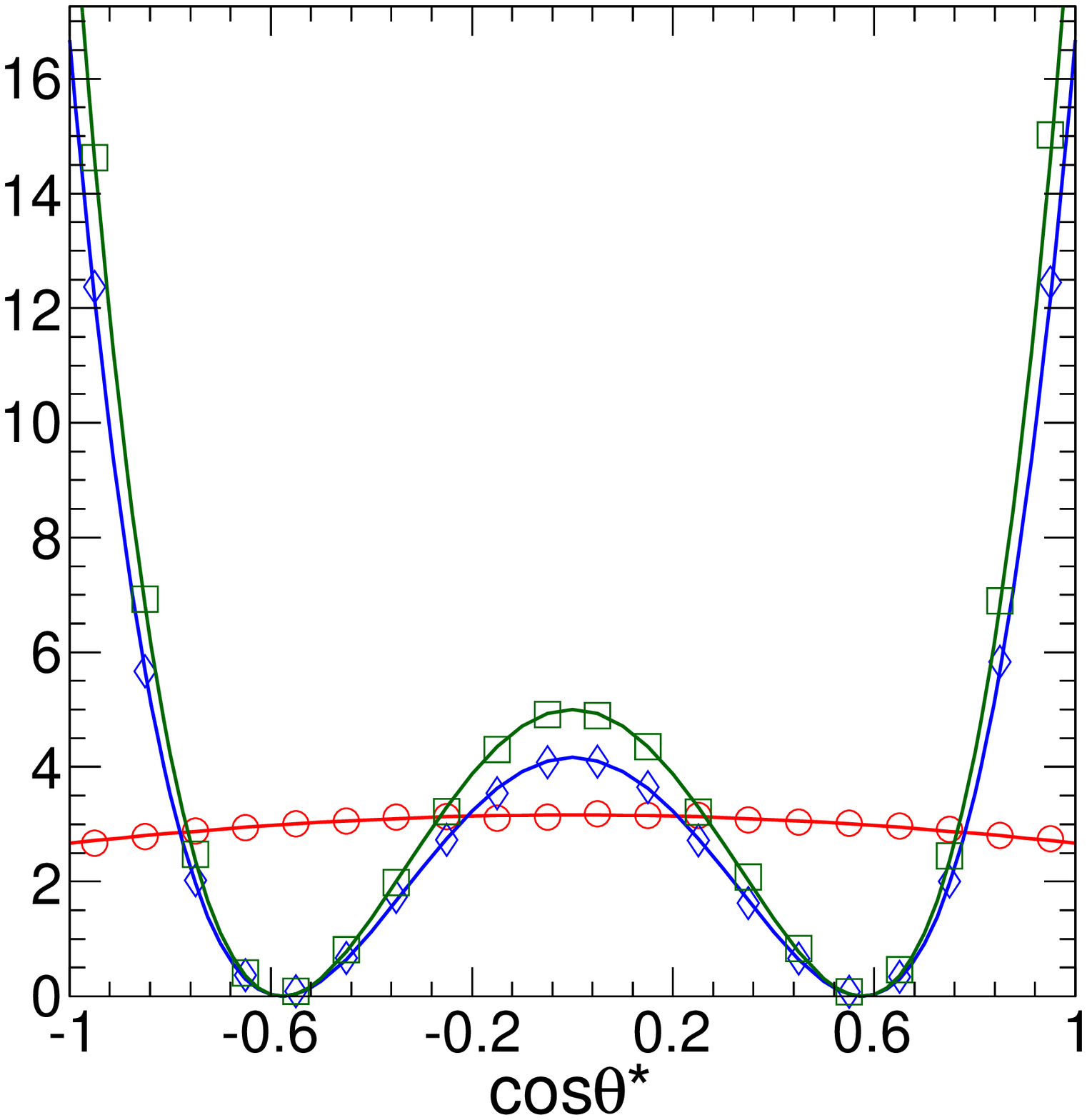,width=0.25\linewidth}
\epsfig{figure=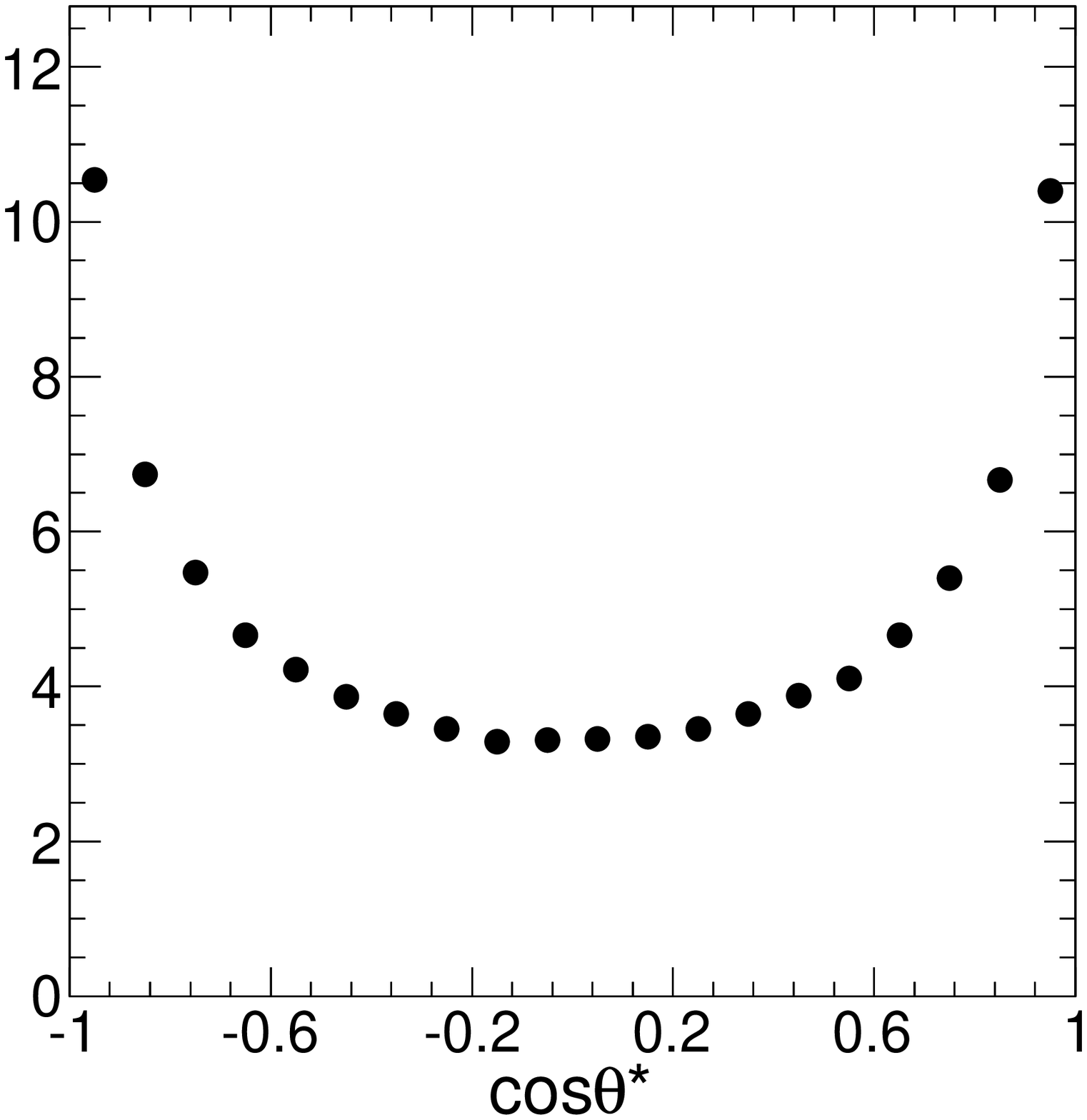,width=0.25\linewidth}
}
\centerline{
\epsfig{figure=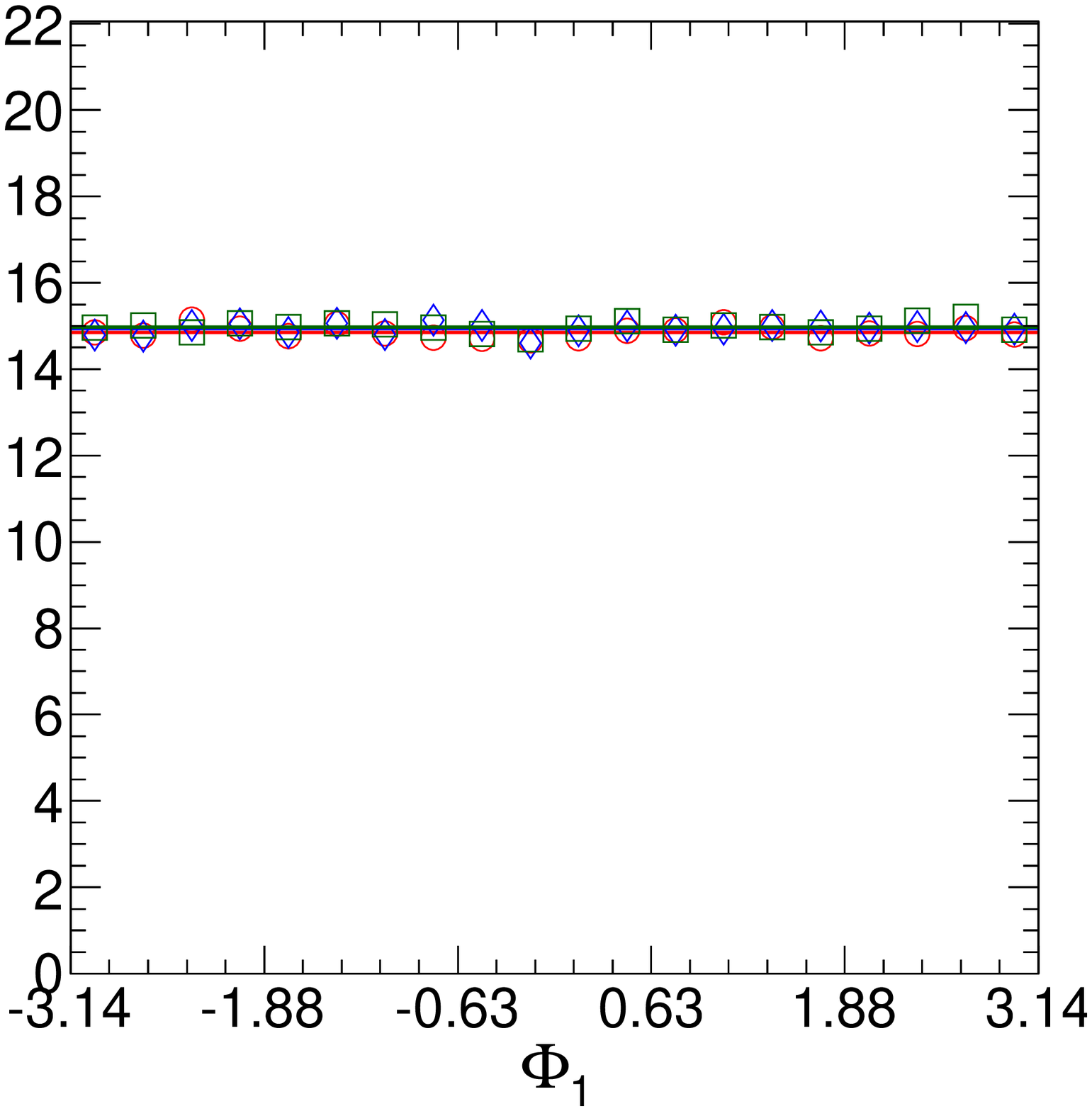,width=0.25\linewidth}
\epsfig{figure=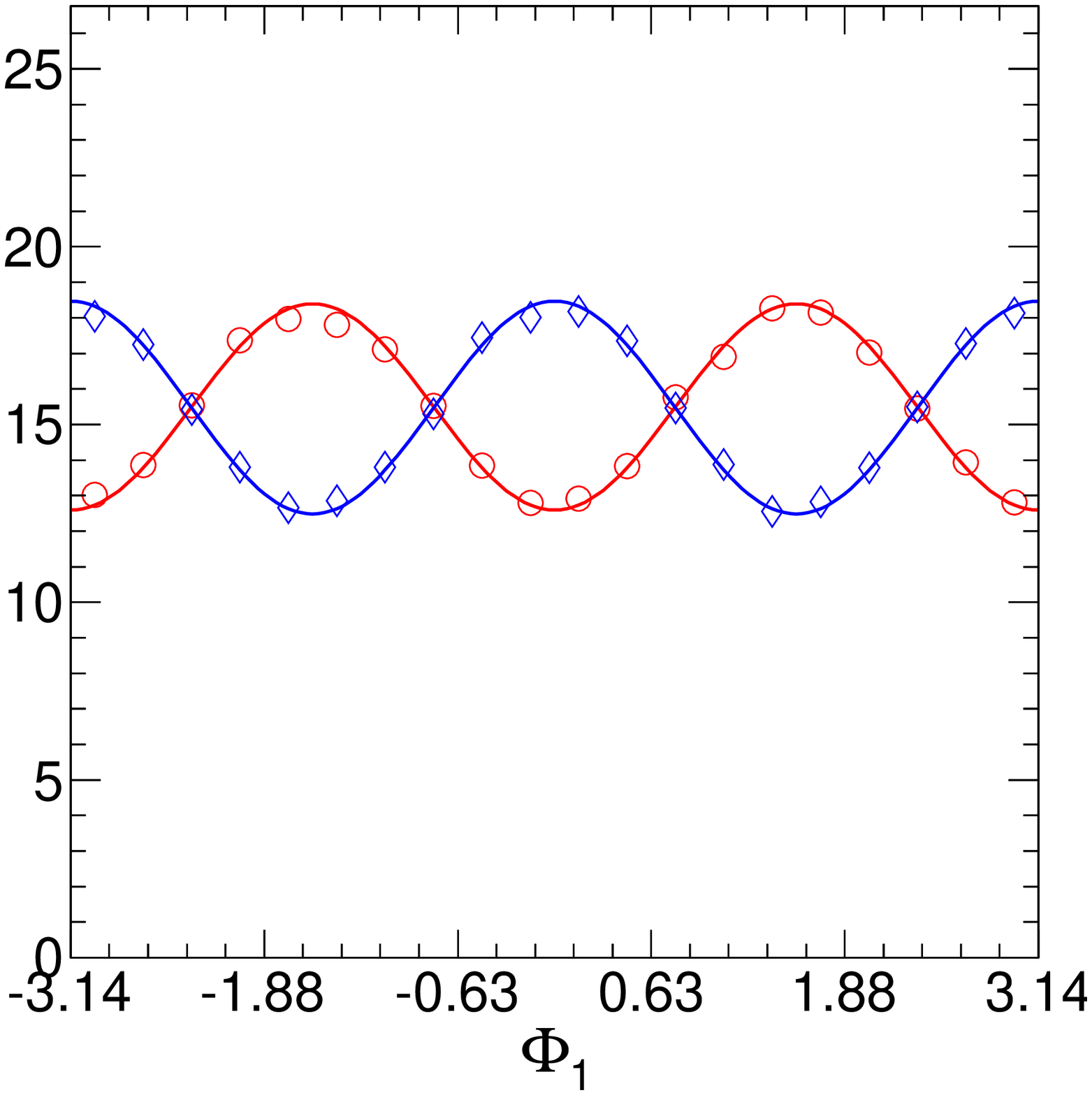,width=0.25\linewidth}
\epsfig{figure=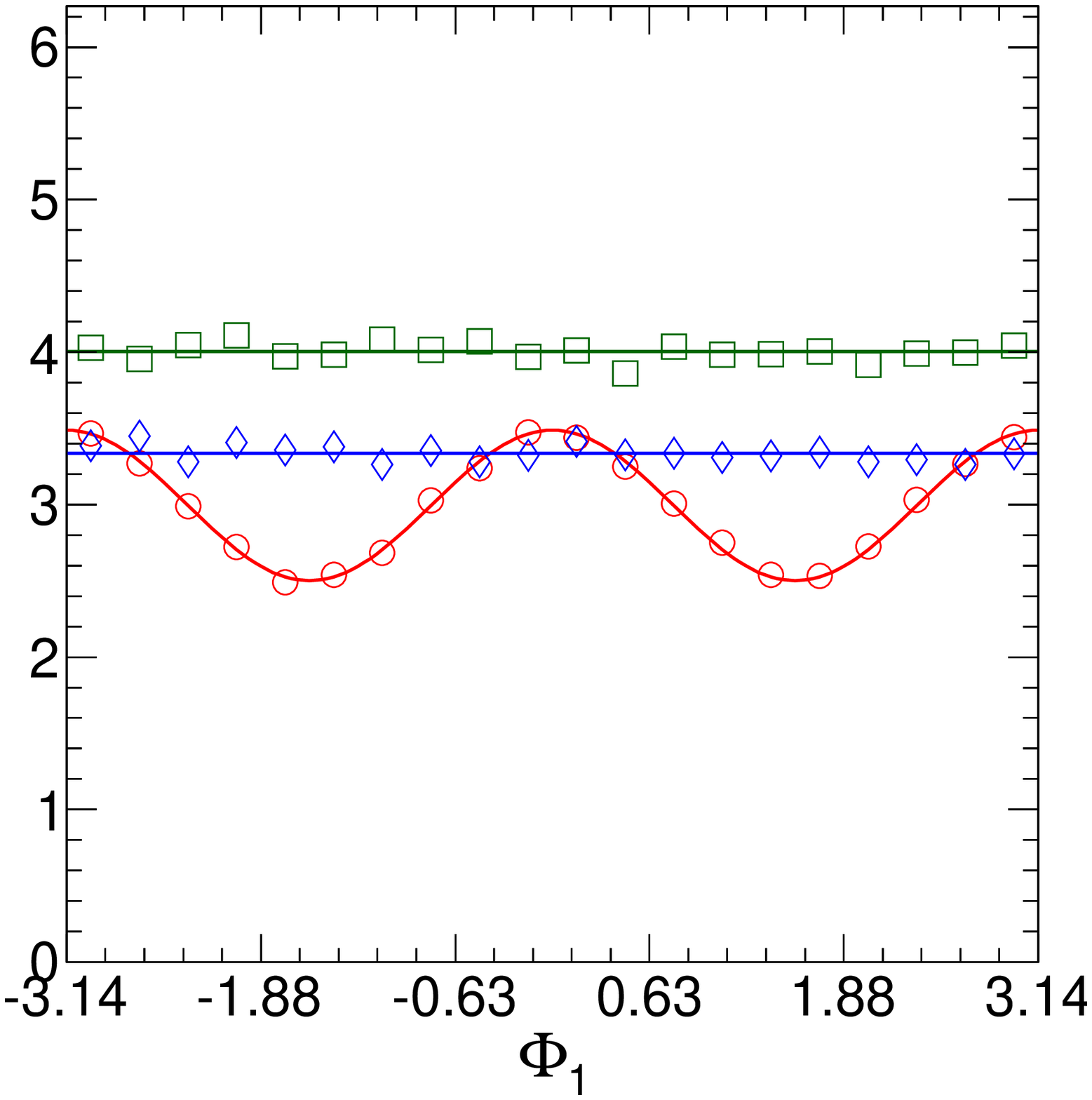,width=0.25\linewidth}
\epsfig{figure=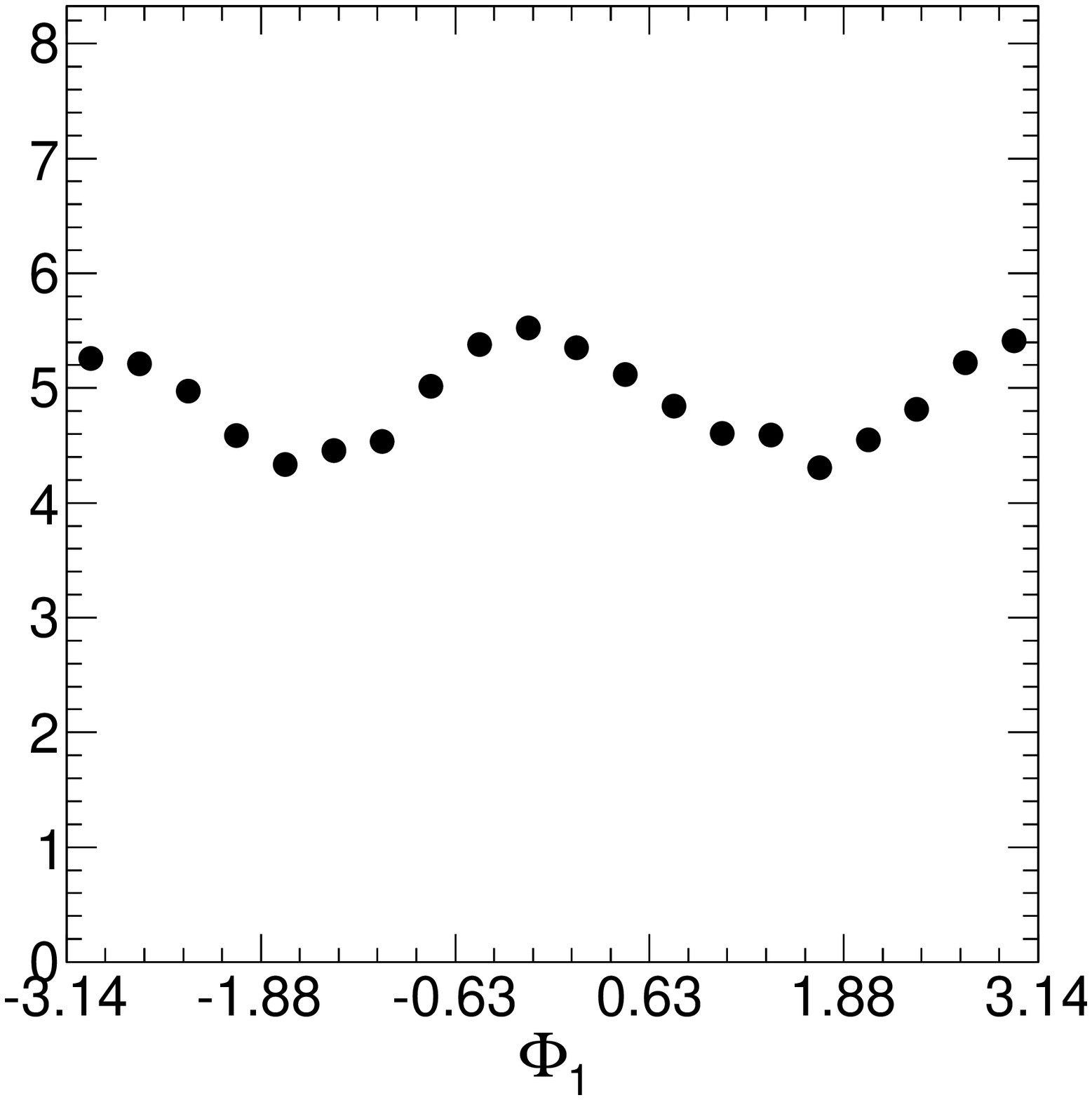,width=0.25\linewidth}
}
\centerline{
\epsfig{figure=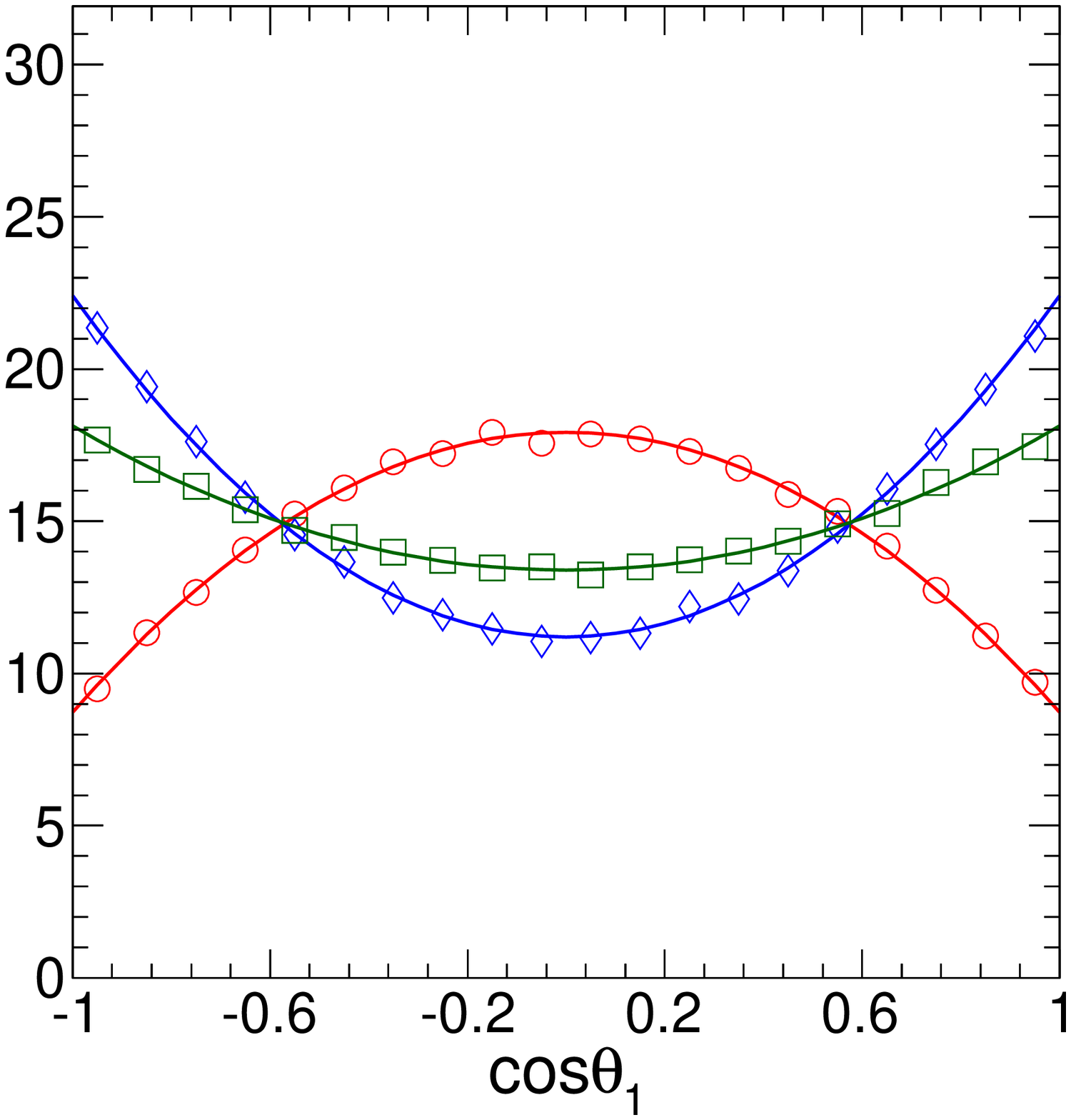,width=0.25\linewidth}
\epsfig{figure=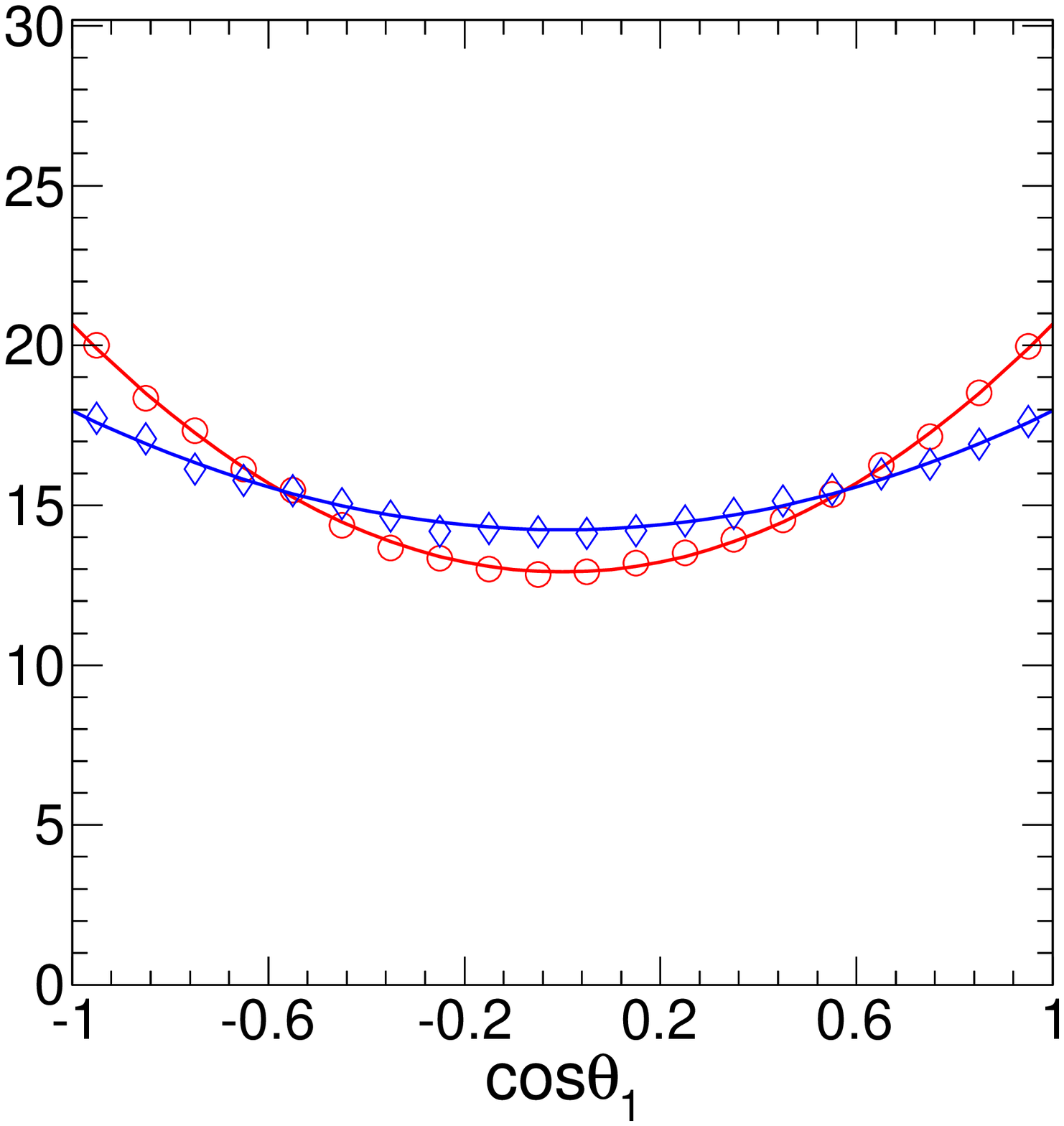,width=0.25\linewidth}
\epsfig{figure=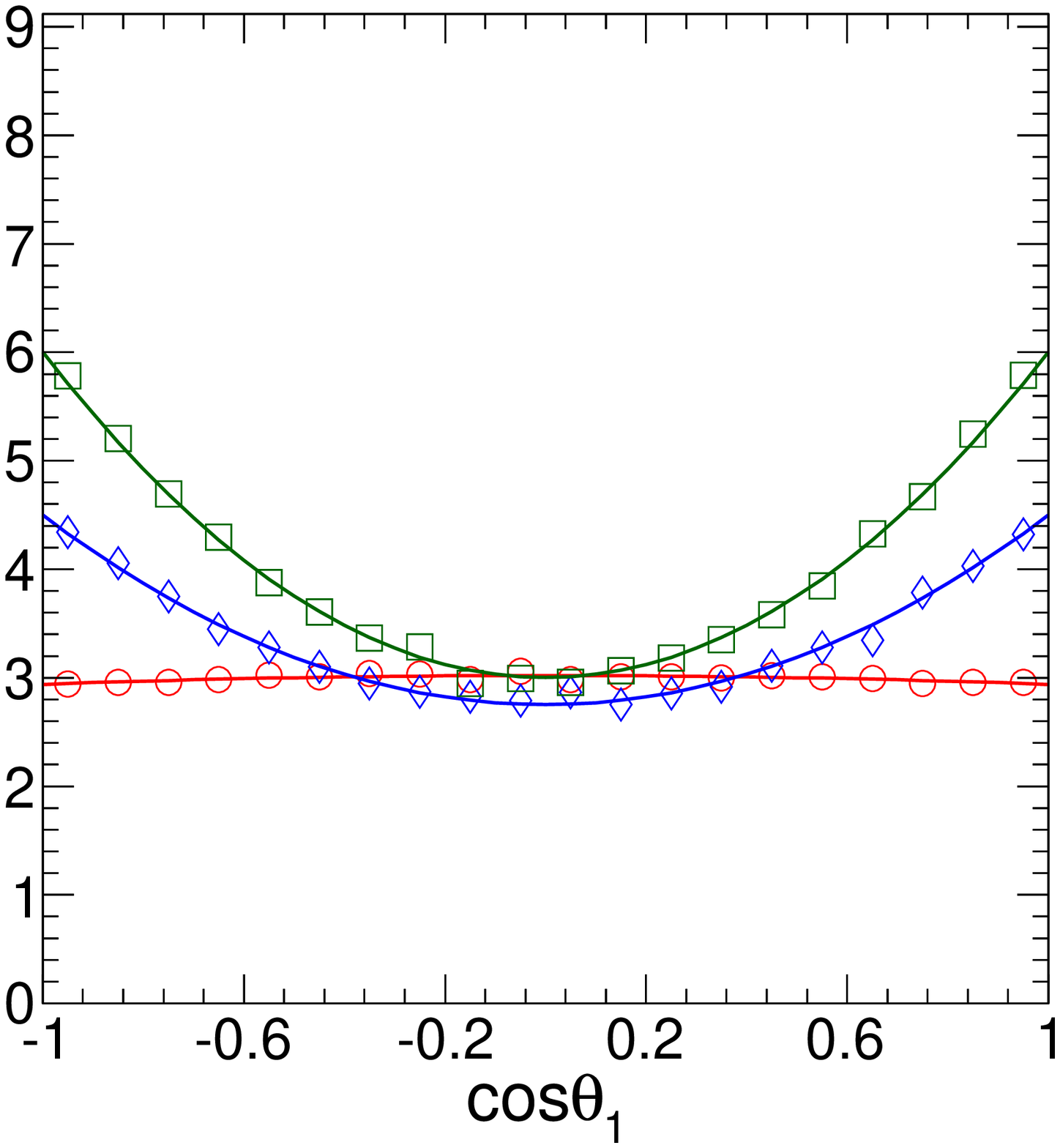,width=0.25\linewidth}
\epsfig{figure=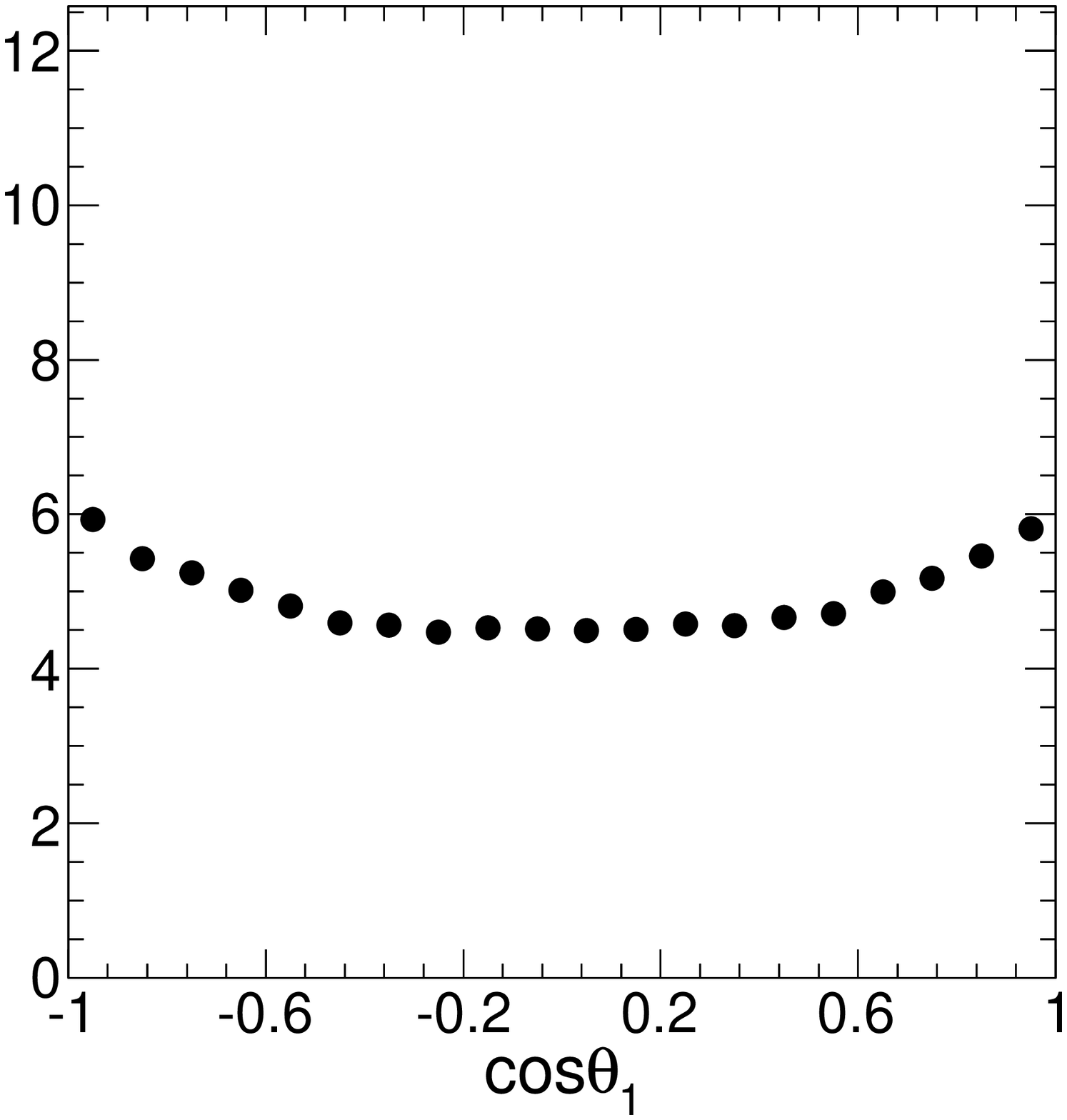,width=0.25\linewidth}
}
\centerline{
\epsfig{figure=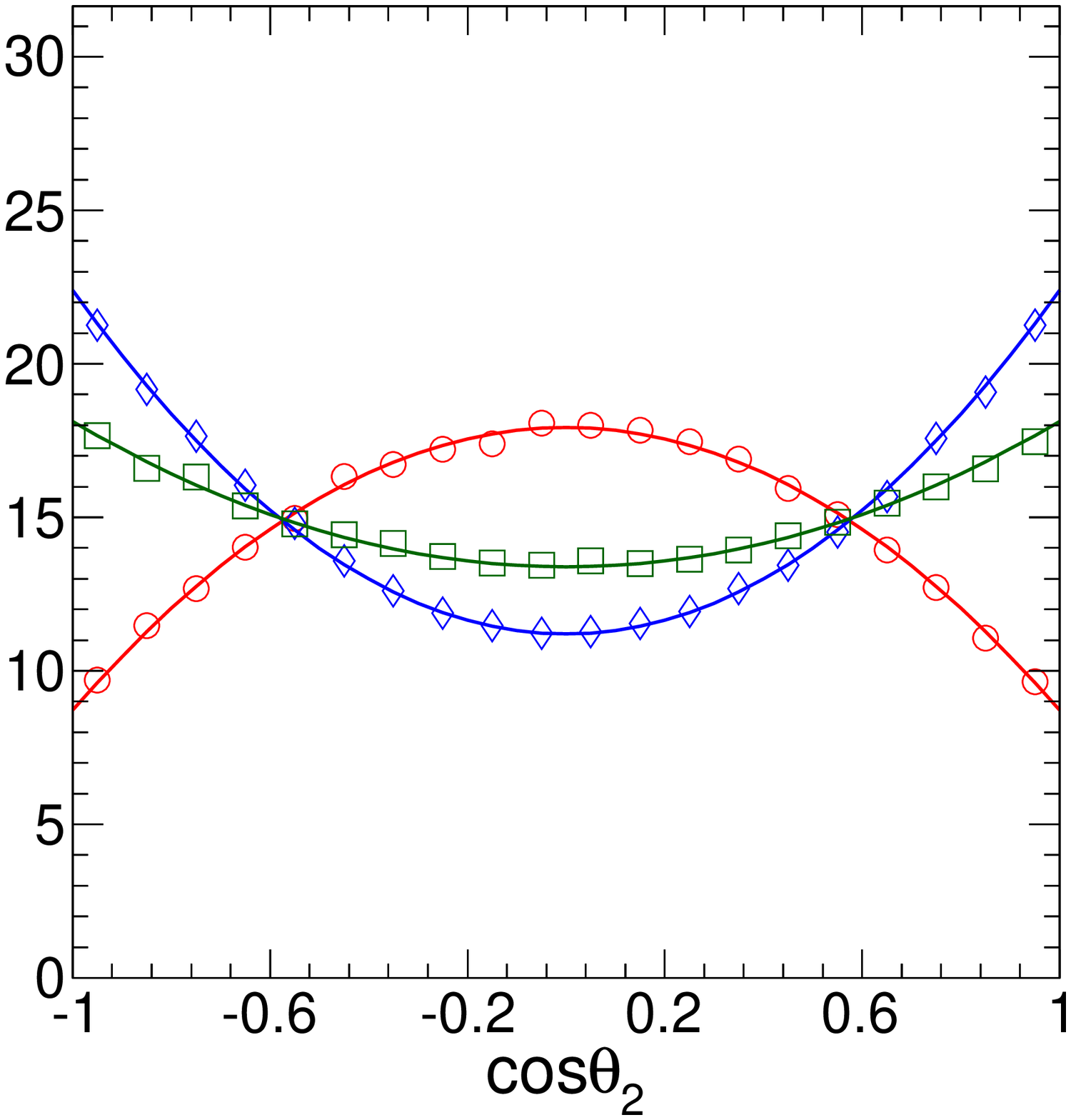,width=0.25\linewidth}
\epsfig{figure=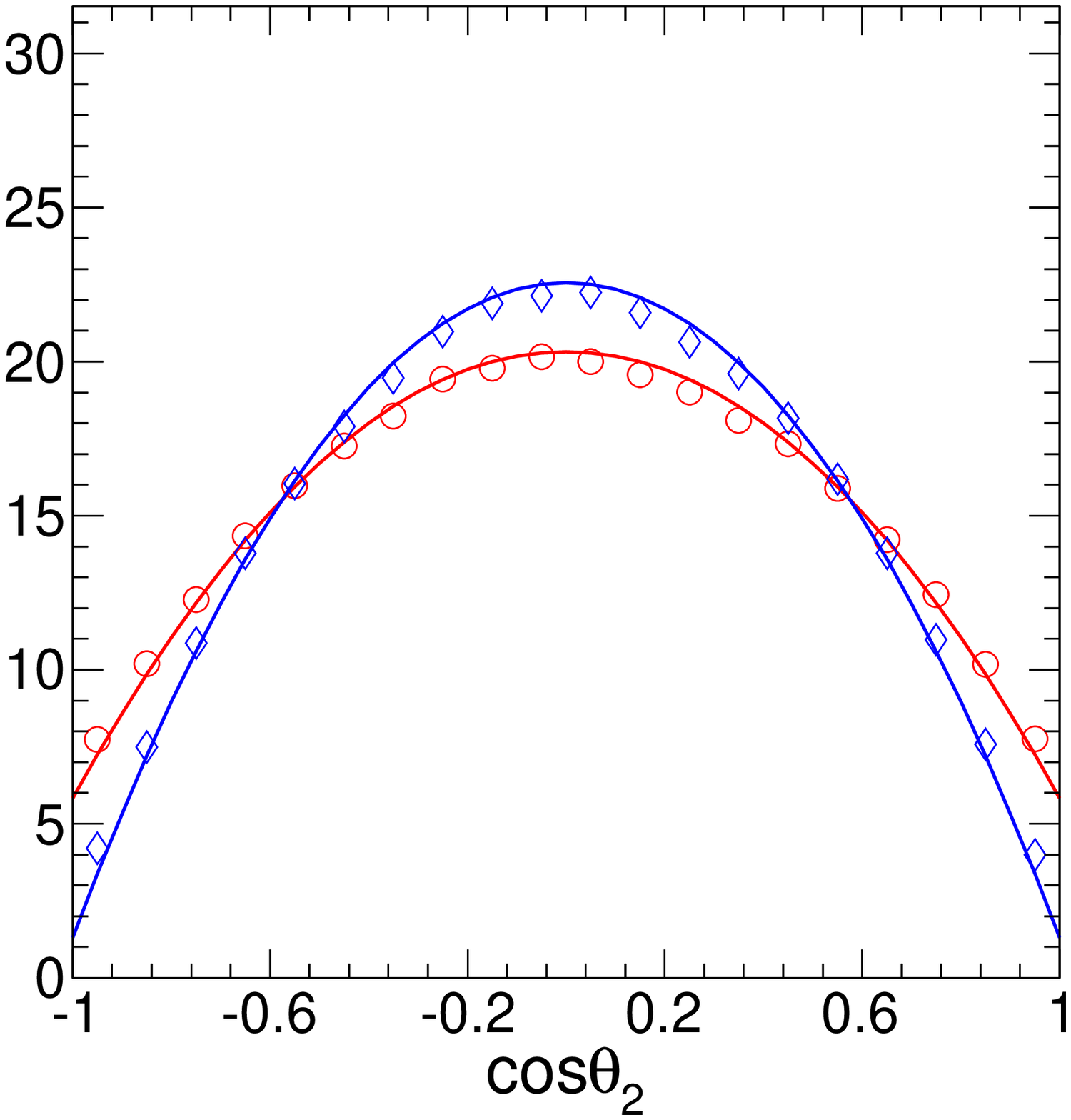,width=0.25\linewidth}
\epsfig{figure=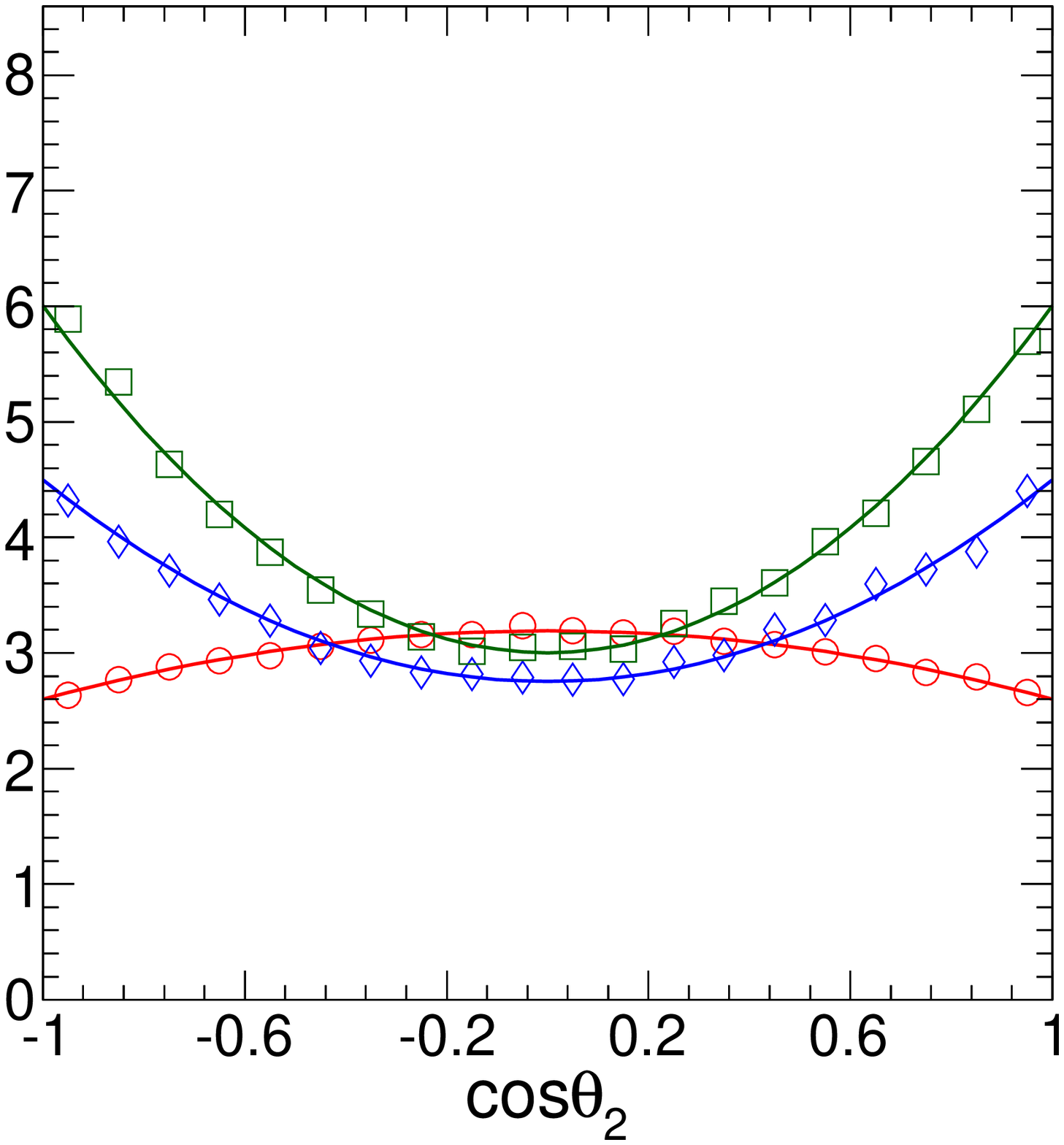,width=0.25\linewidth}
\epsfig{figure=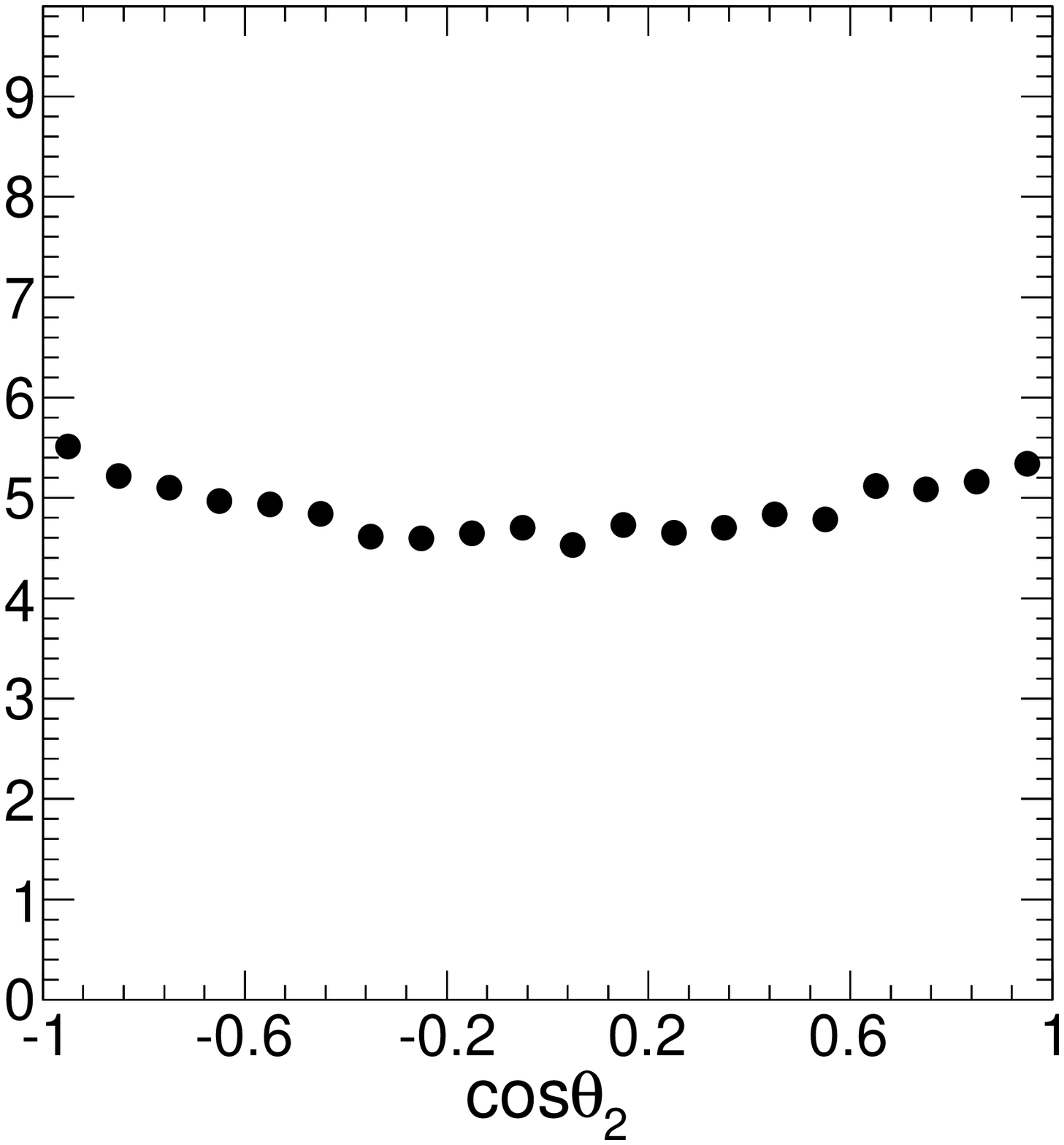,width=0.25\linewidth}
}
\centerline{
\epsfig{figure=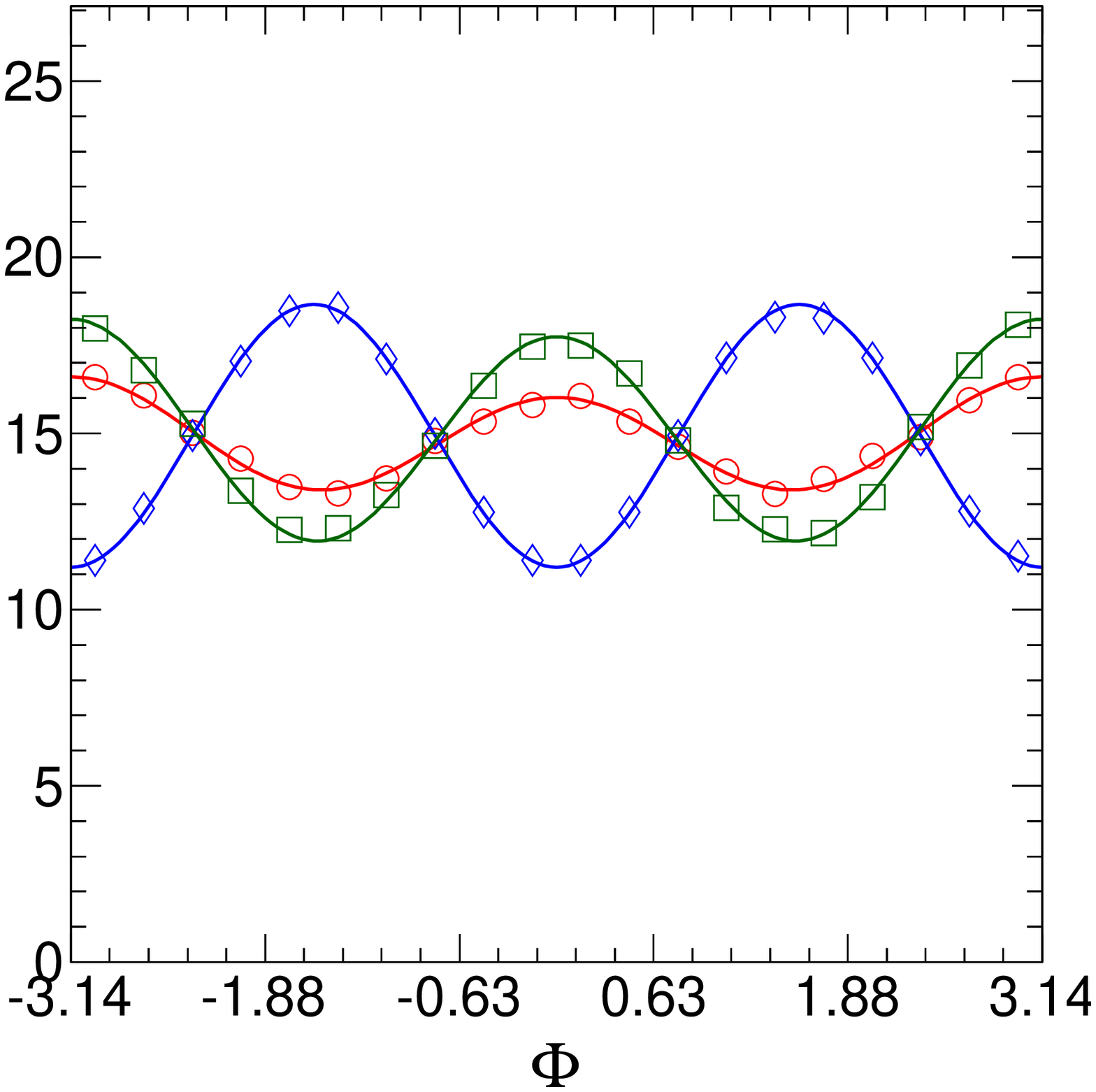,width=0.25\linewidth}
\epsfig{figure=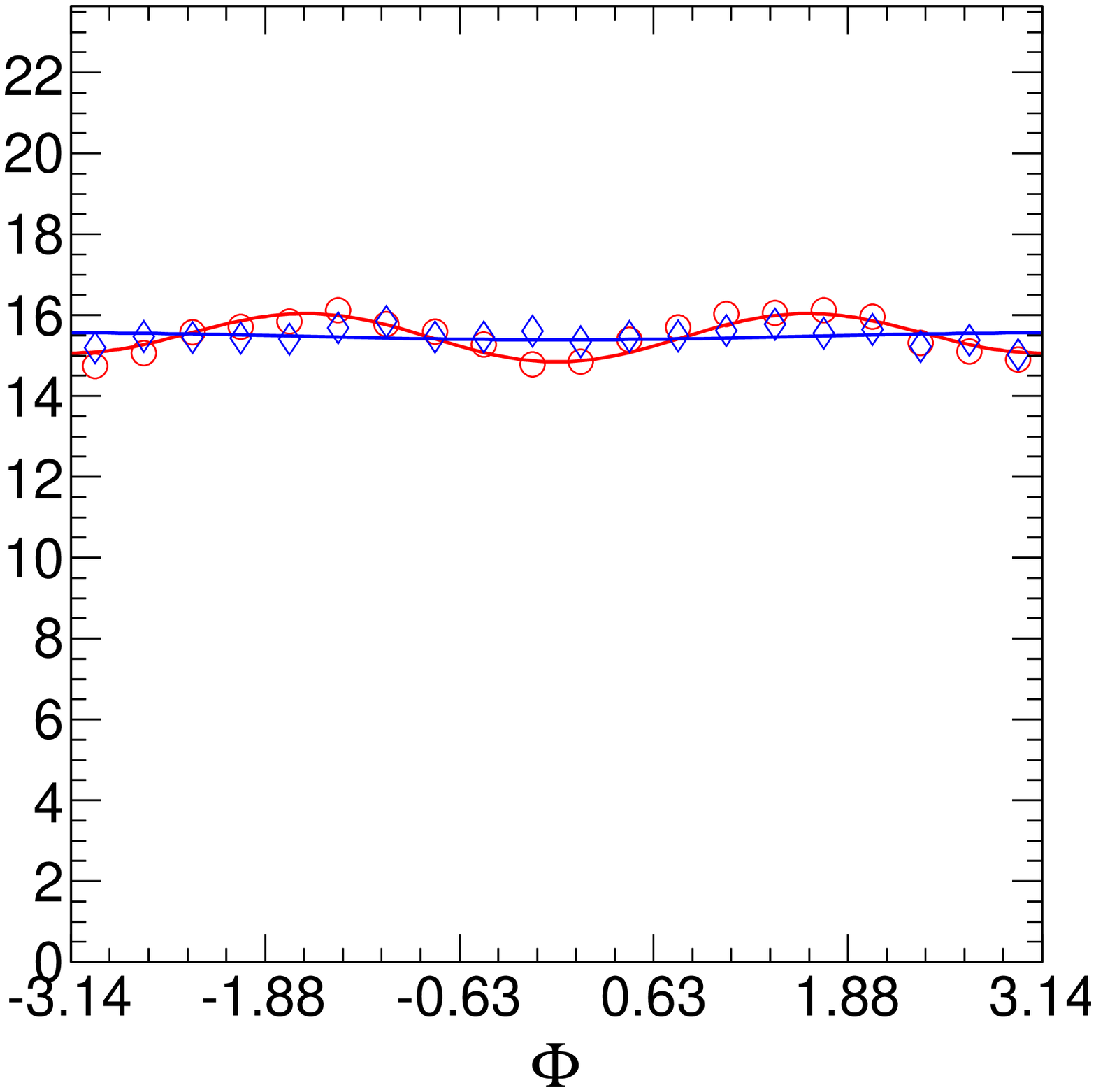,width=0.25\linewidth}
\epsfig{figure=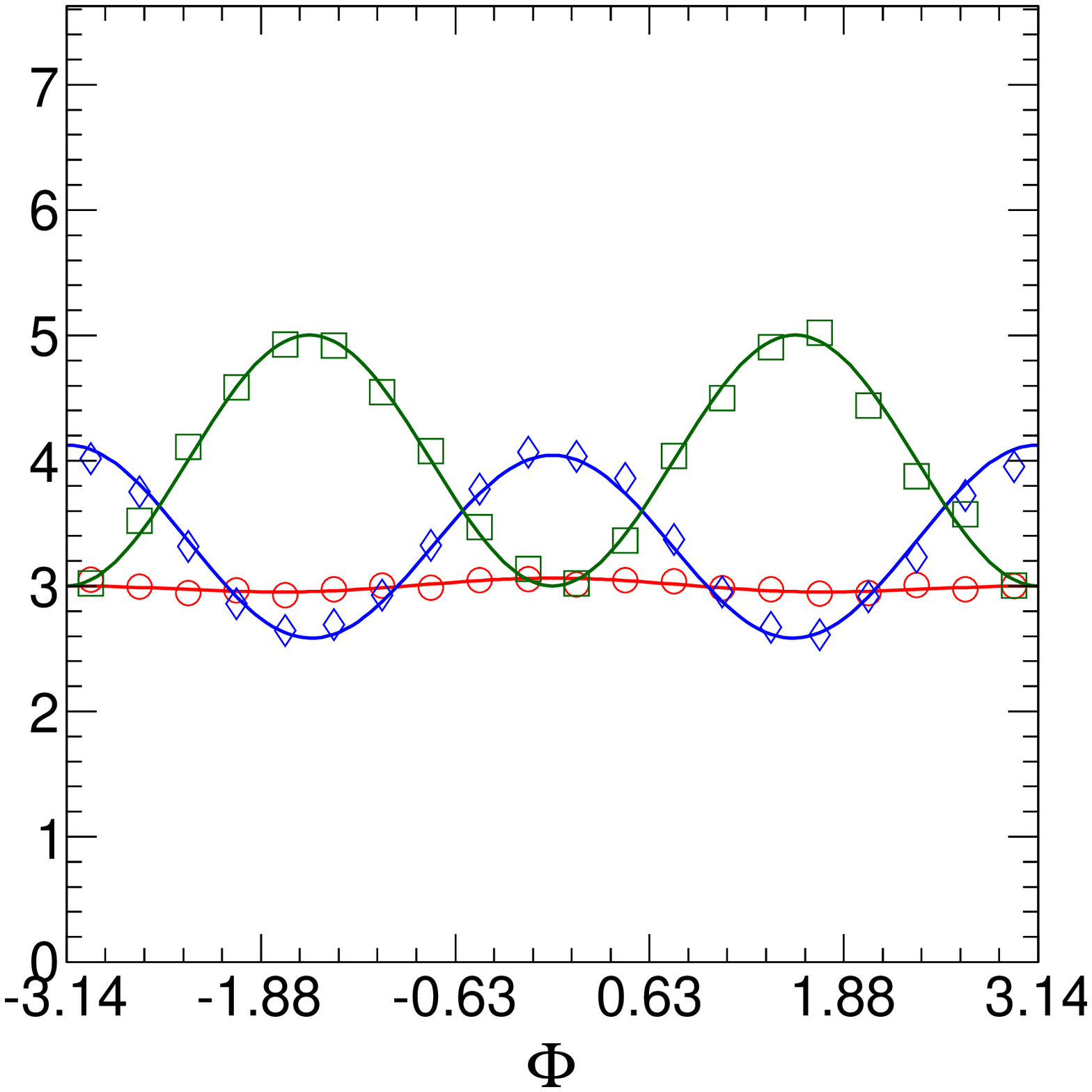,width=0.25\linewidth}
\epsfig{figure=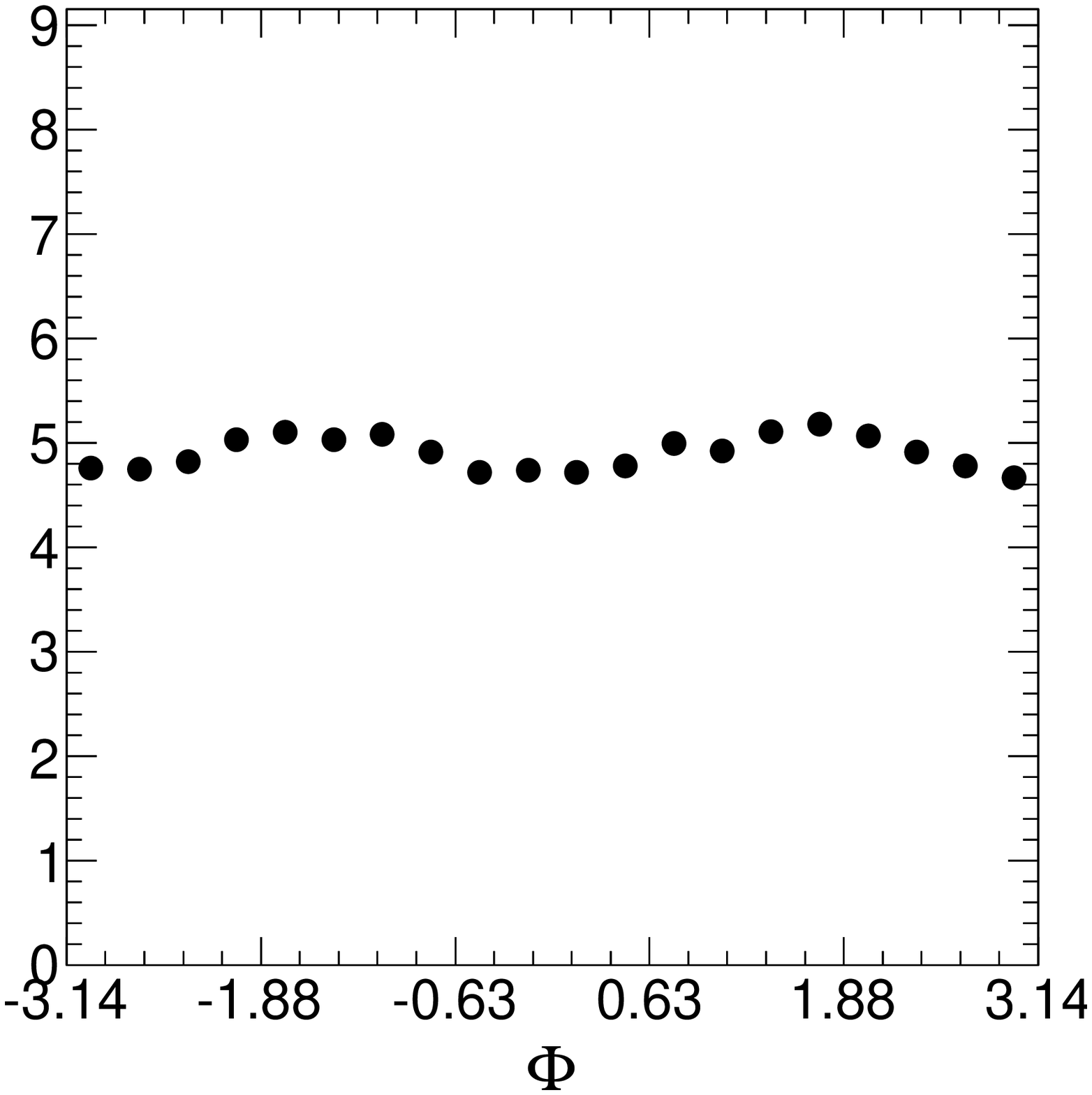,width=0.25\linewidth}
}
\caption{
Distributions of the observables in the $X\to ZZ$ analysis, from left to right:
spin-zero, spin-one, and spin-two signal, and $q\bar{q}\to ZZ$ background.
The signal hypotheses shown are $J^+_m$ (red circles), $J^+_h$ (green squares), 
$J^-_h$ (blue diamonds), as defined in Table~\ref{table-scenarios}. 
Background is shown with the requirements $m_2>10$ GeV and $110<m_{4\ell}<140$ GeV.
The observables shown from top to bottom: 
$\cos\theta^*$, $\Phi_1$, $\cos\theta_1$, $\cos\theta_2$, and $\Phi$.
Points show simulated events and lines show projections of analytical distributions.
}
\label{fig:simulated-zz-angles}
\end{figure}

\begin{figure}[t]
\centerline{
\epsfig{figure=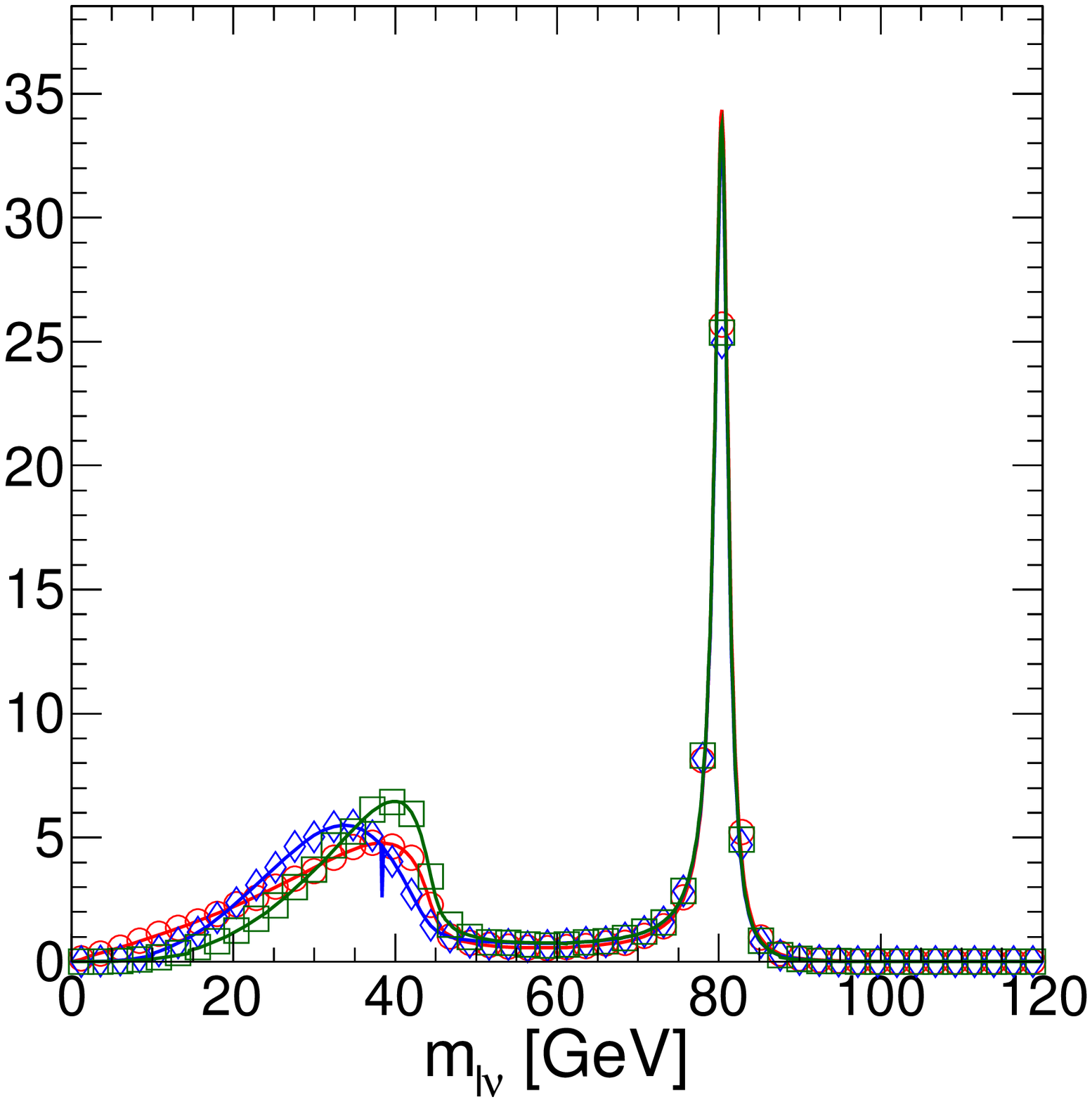,width=0.25\linewidth}
\epsfig{figure=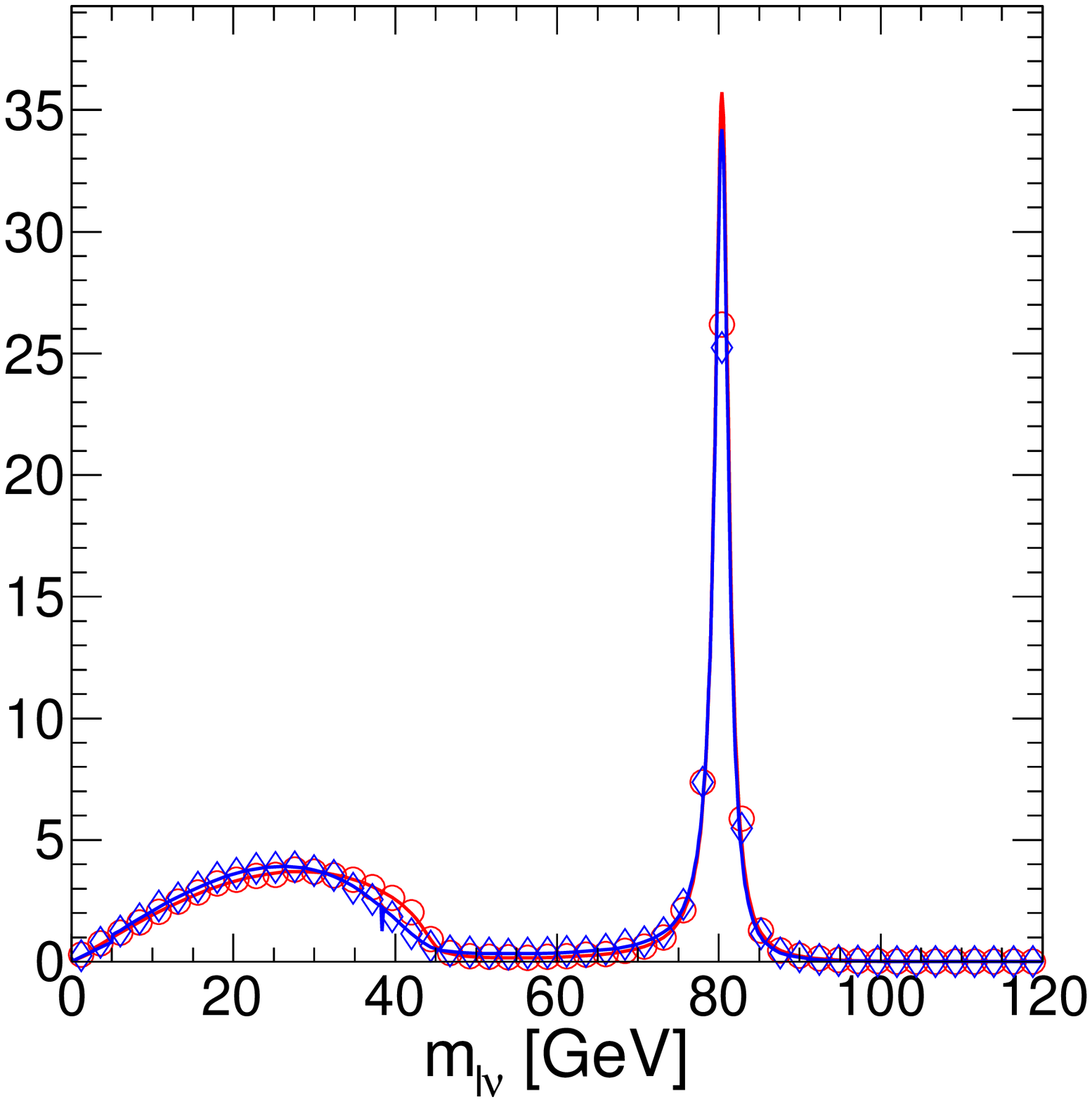,width=0.25\linewidth}
\epsfig{figure=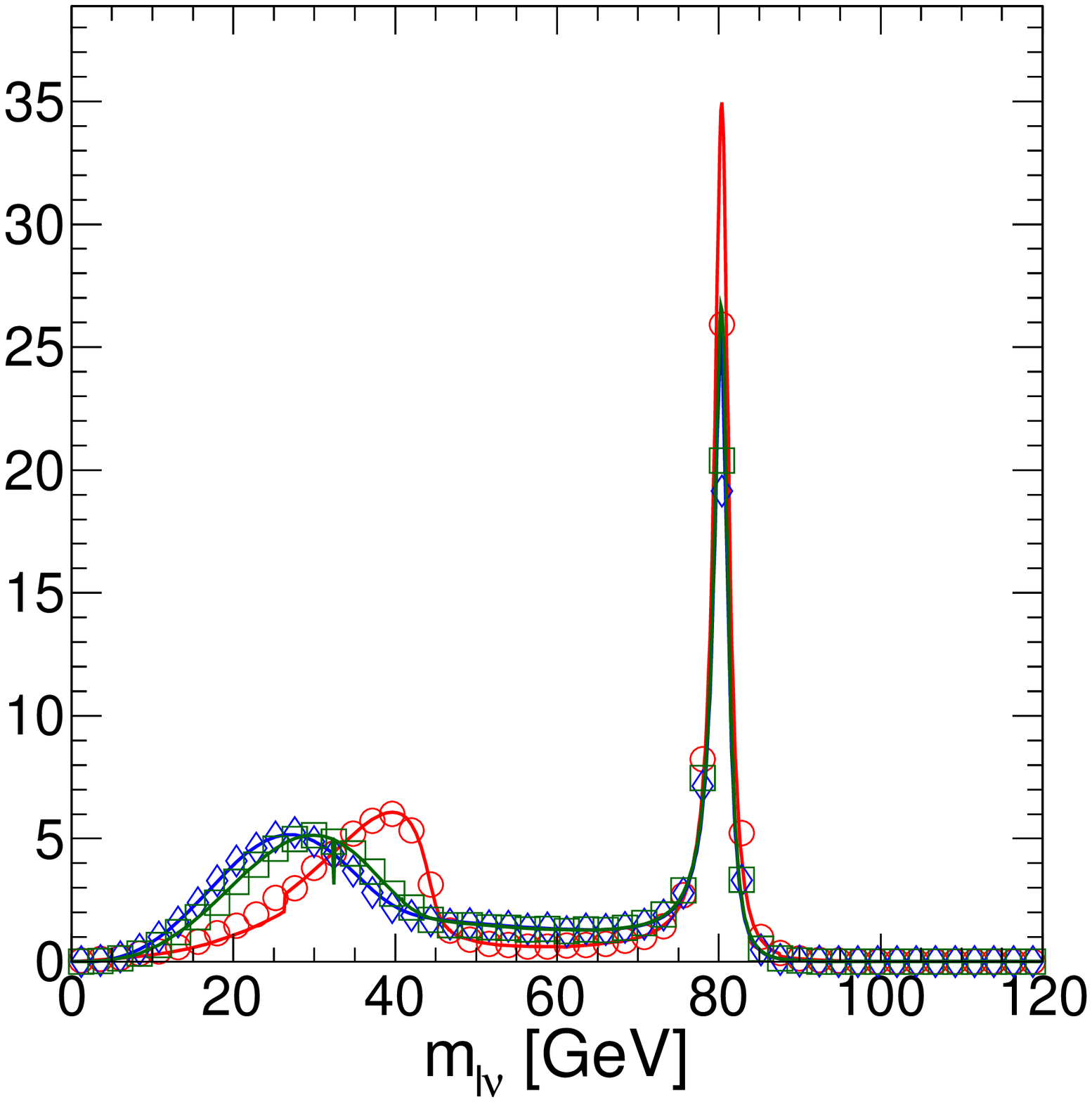,width=0.25\linewidth}
}
\centerline{
\epsfig{figure=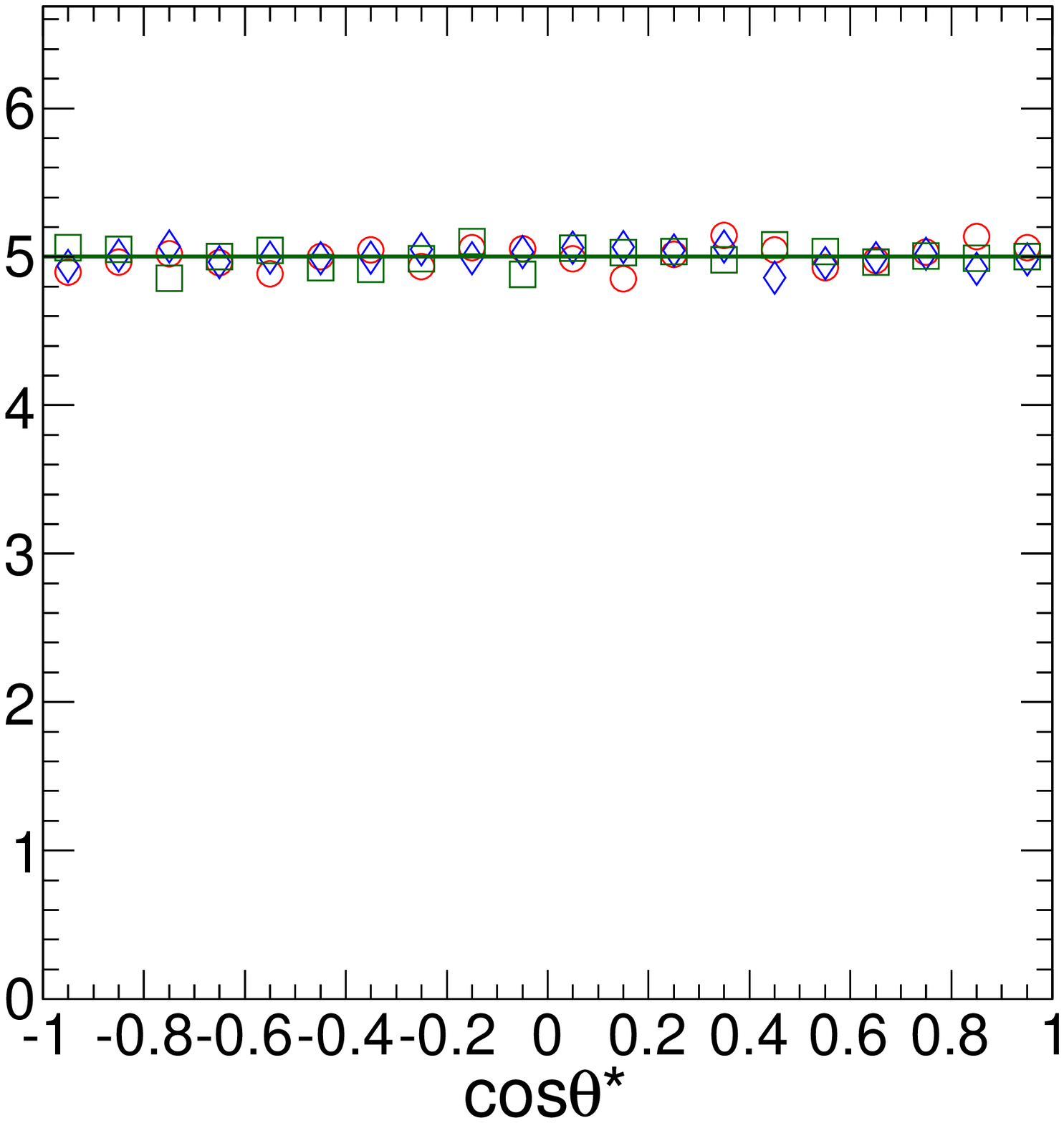,width=0.25\linewidth}
\epsfig{figure=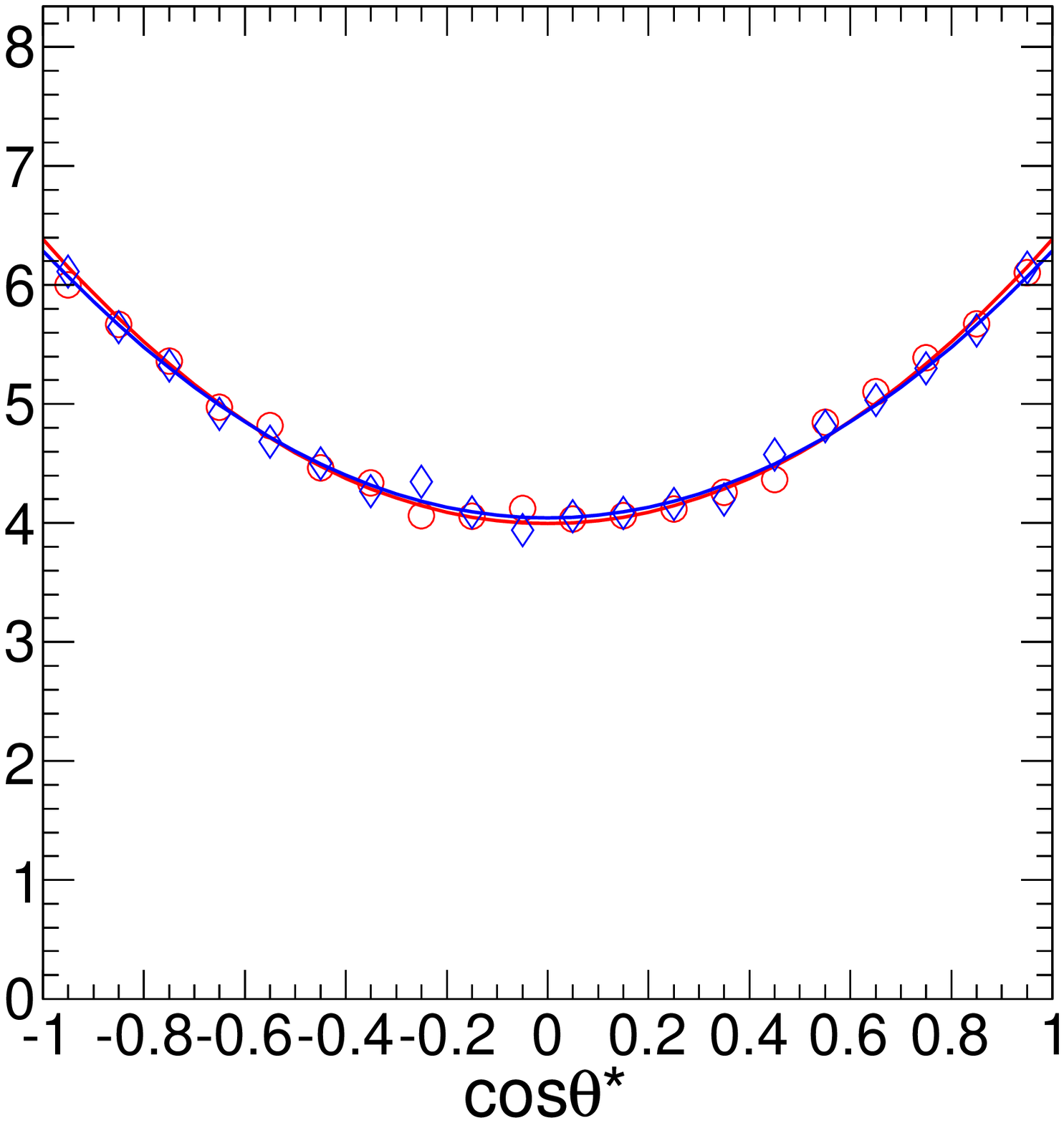,width=0.25\linewidth}
\epsfig{figure=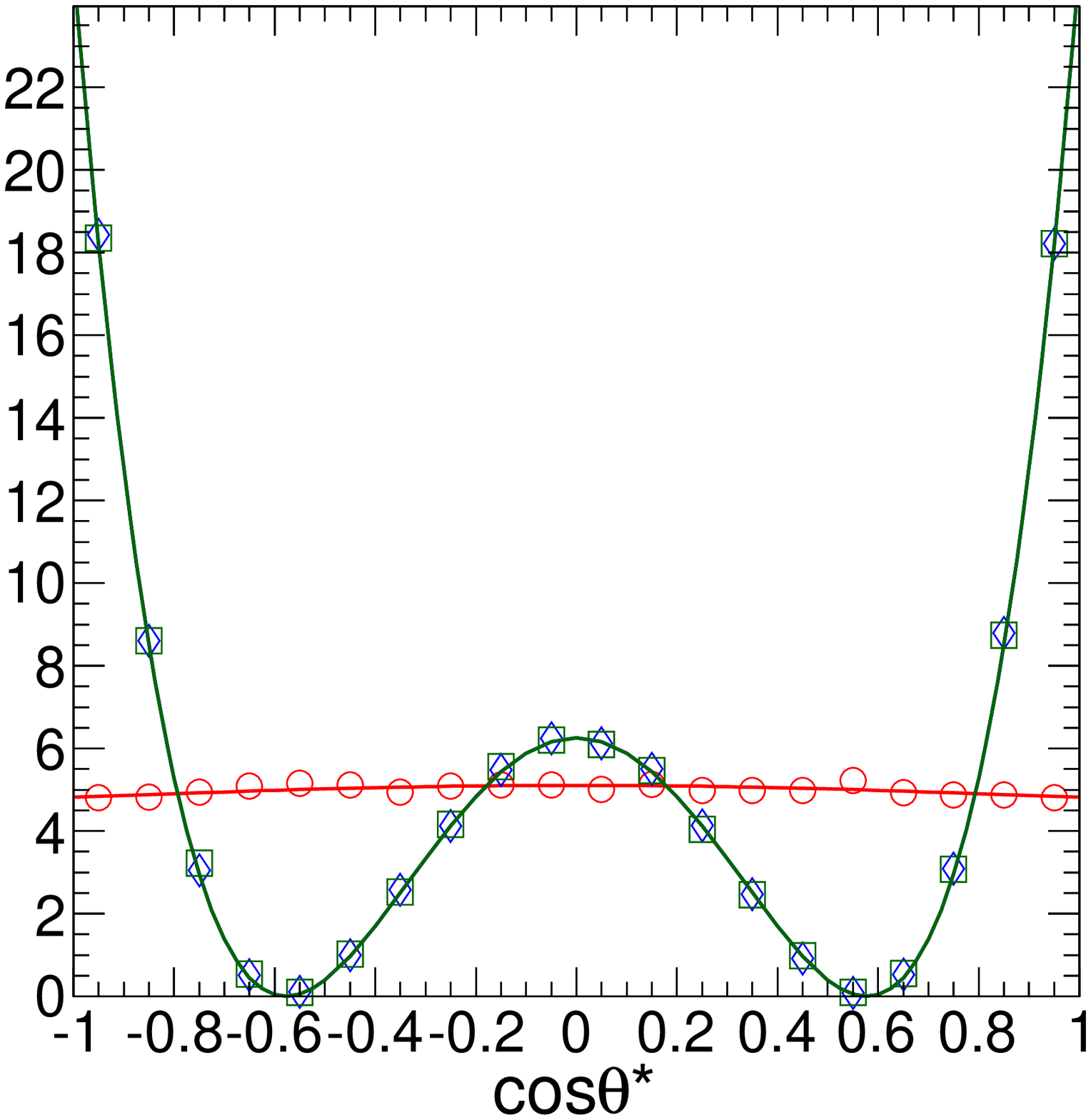,width=0.25\linewidth}
}
\centerline{
\epsfig{figure=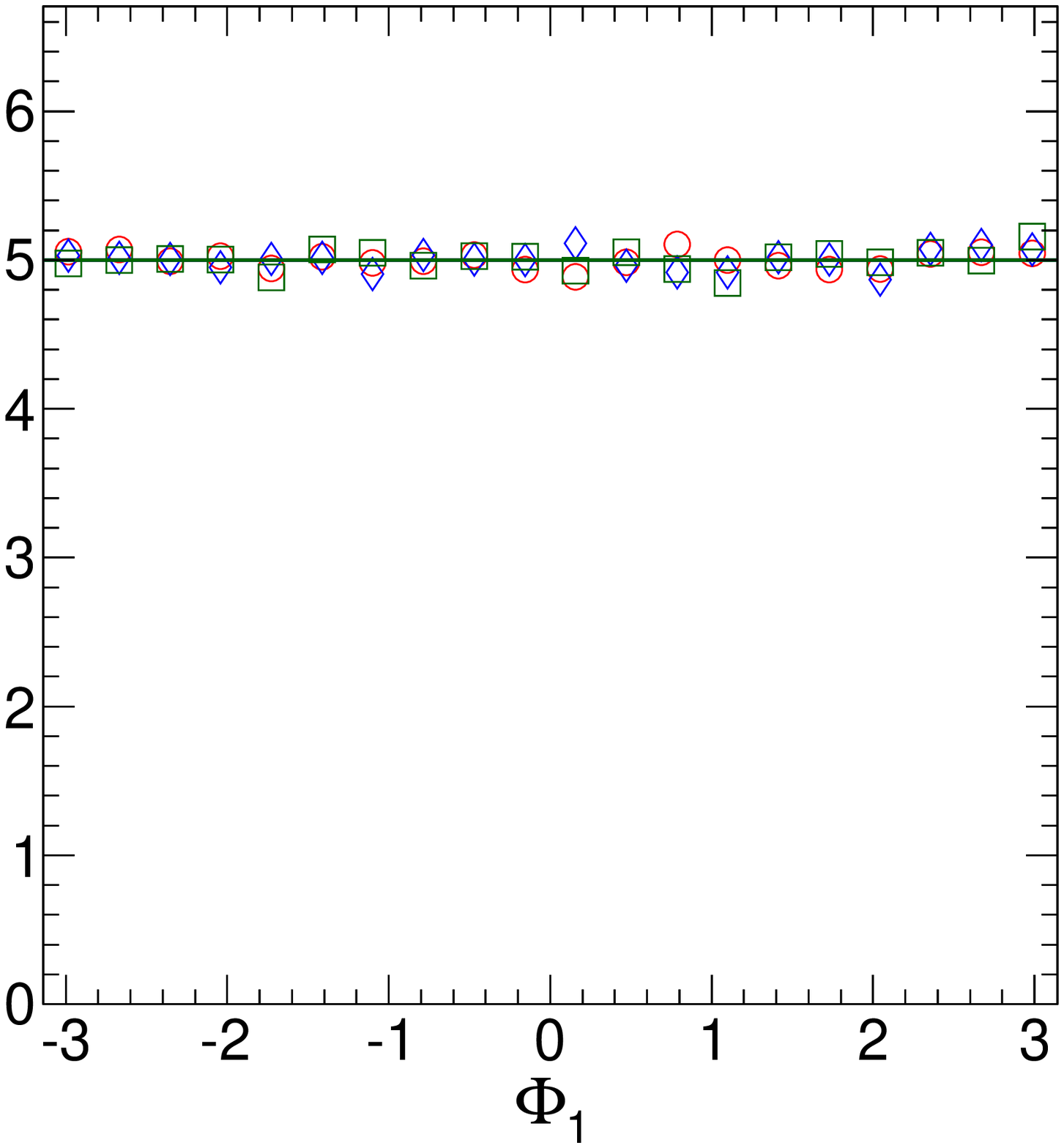,width=0.25\linewidth}
\epsfig{figure=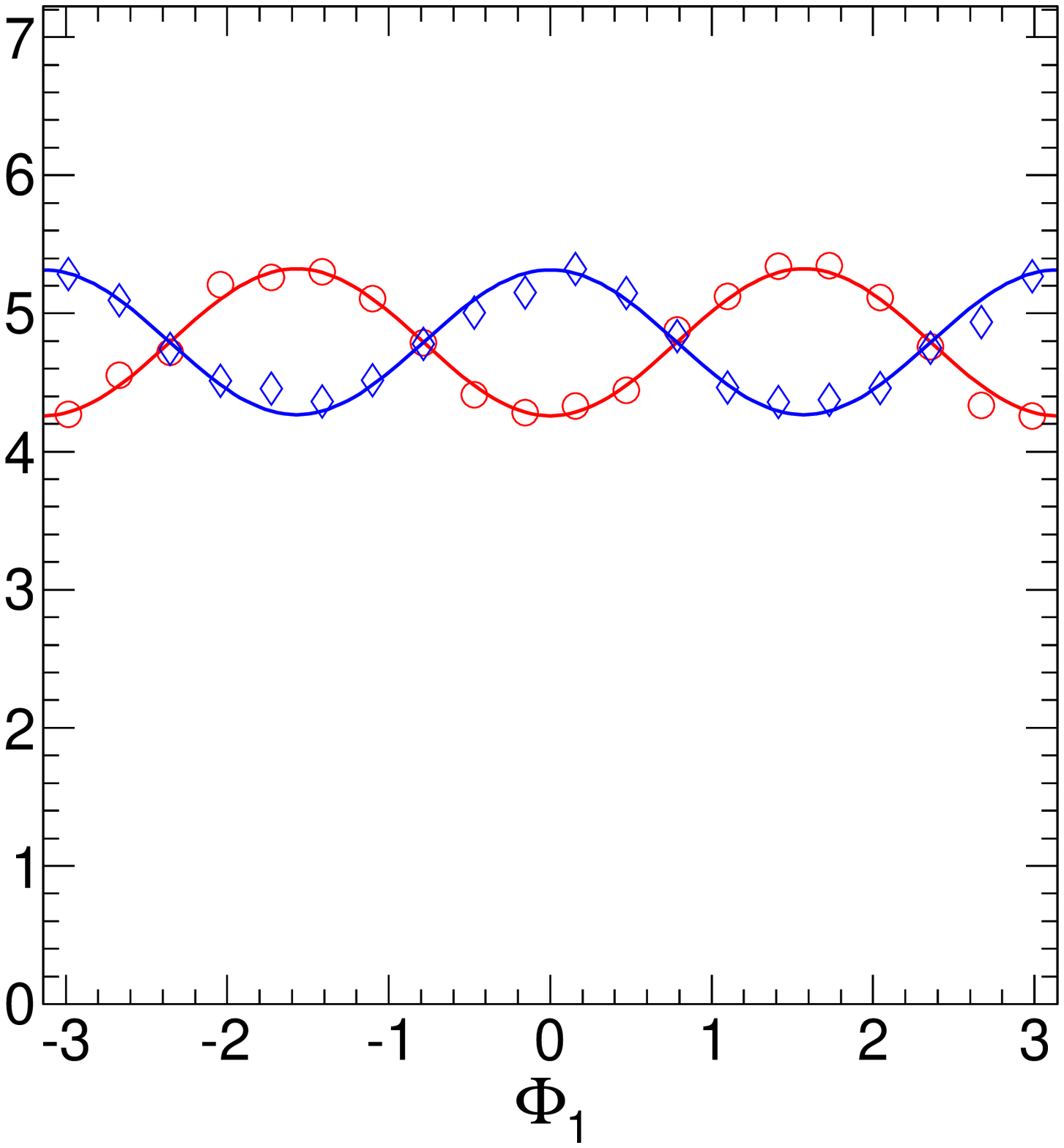,width=0.25\linewidth}
\epsfig{figure=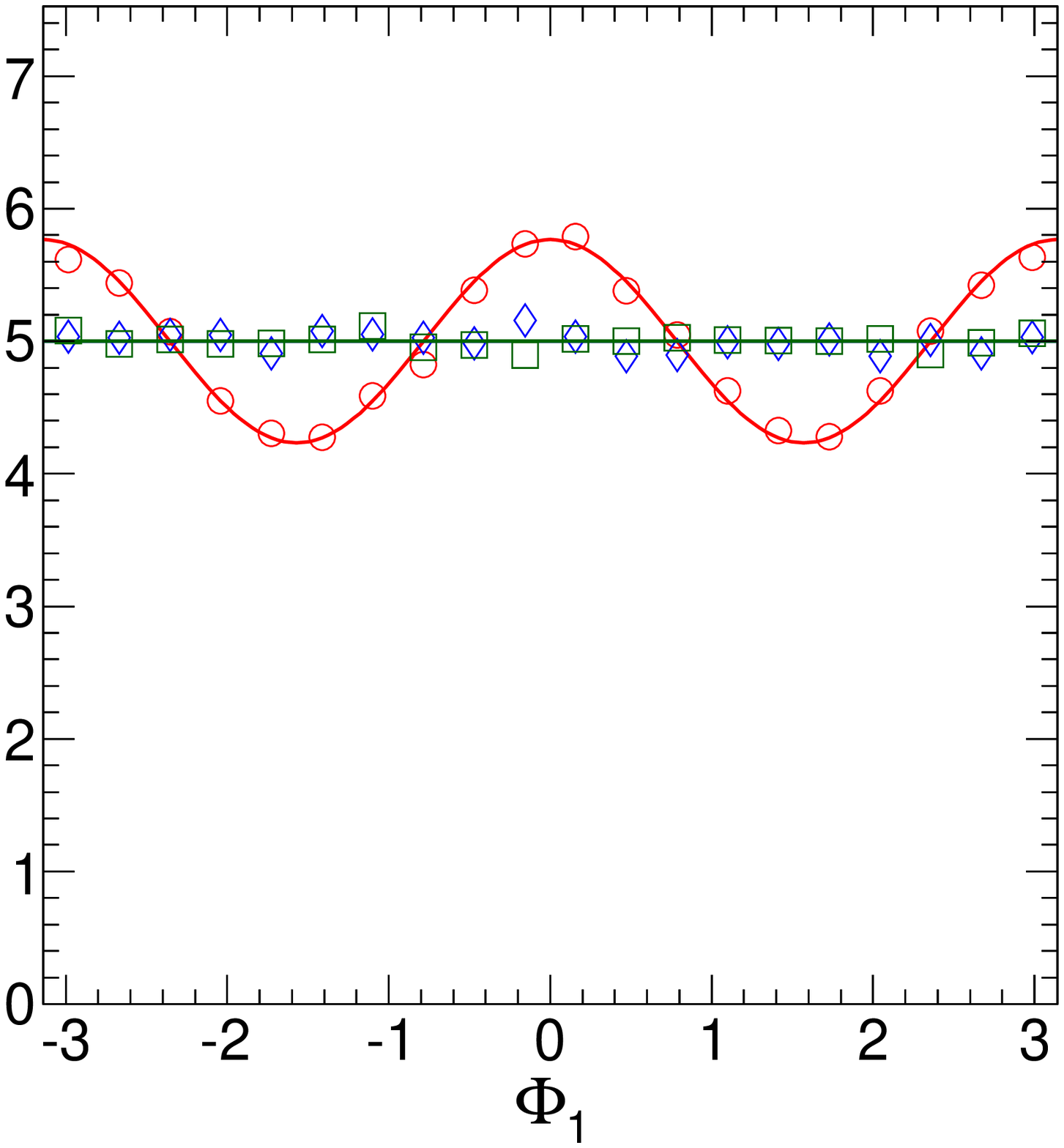,width=0.25\linewidth}
}
\centerline{
\epsfig{figure=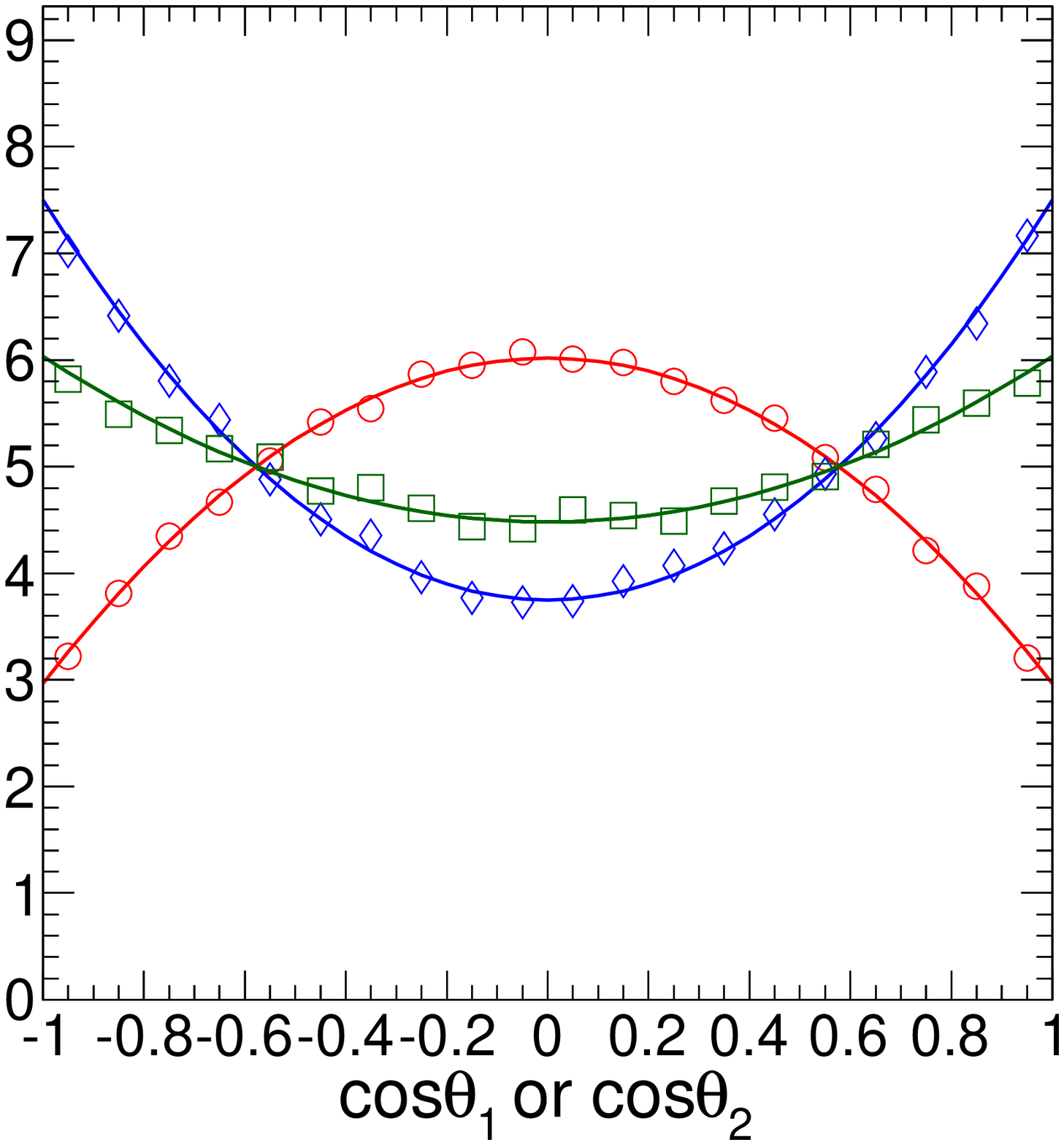,width=0.25\linewidth}
\epsfig{figure=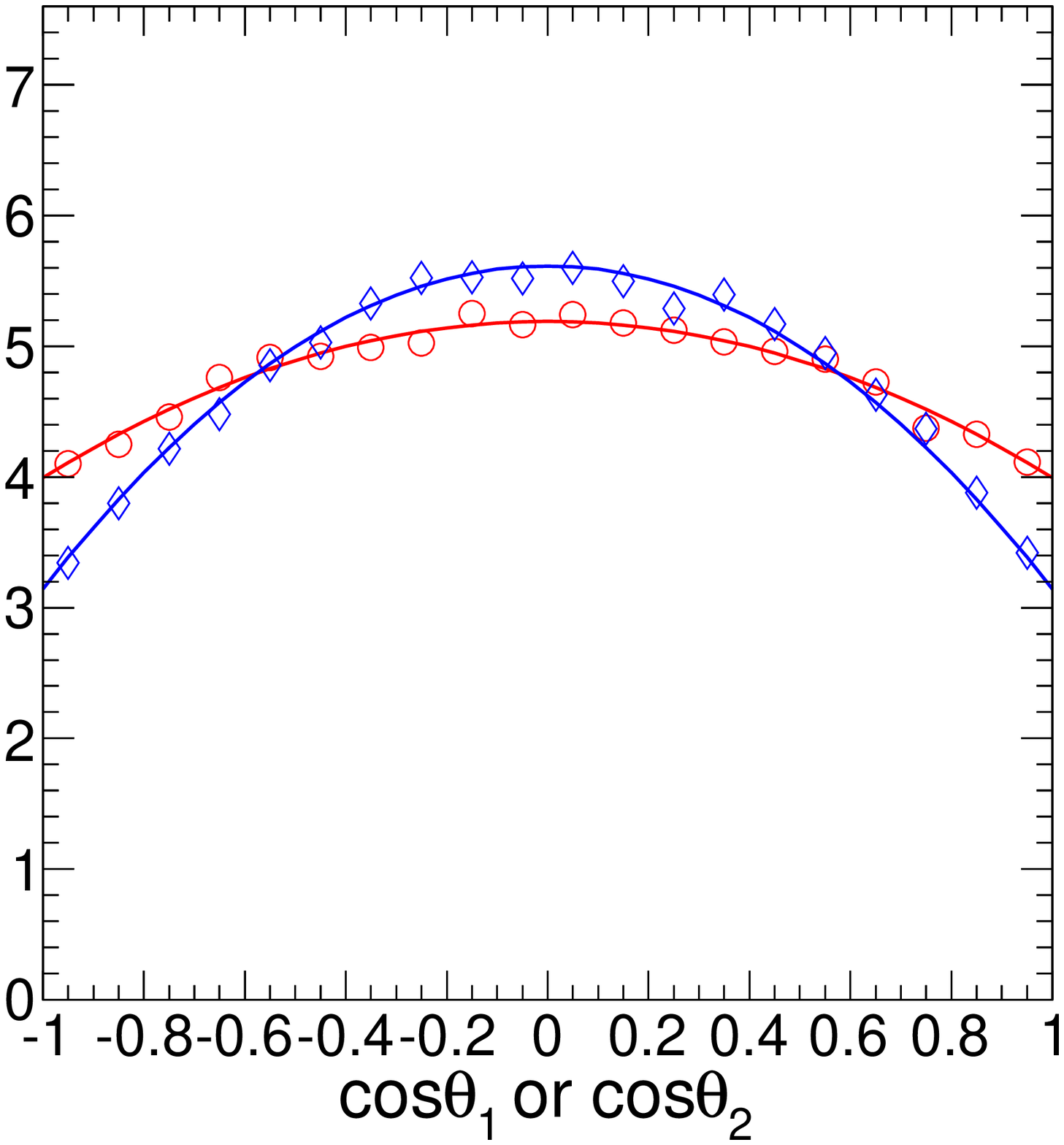,width=0.25\linewidth}
\epsfig{figure=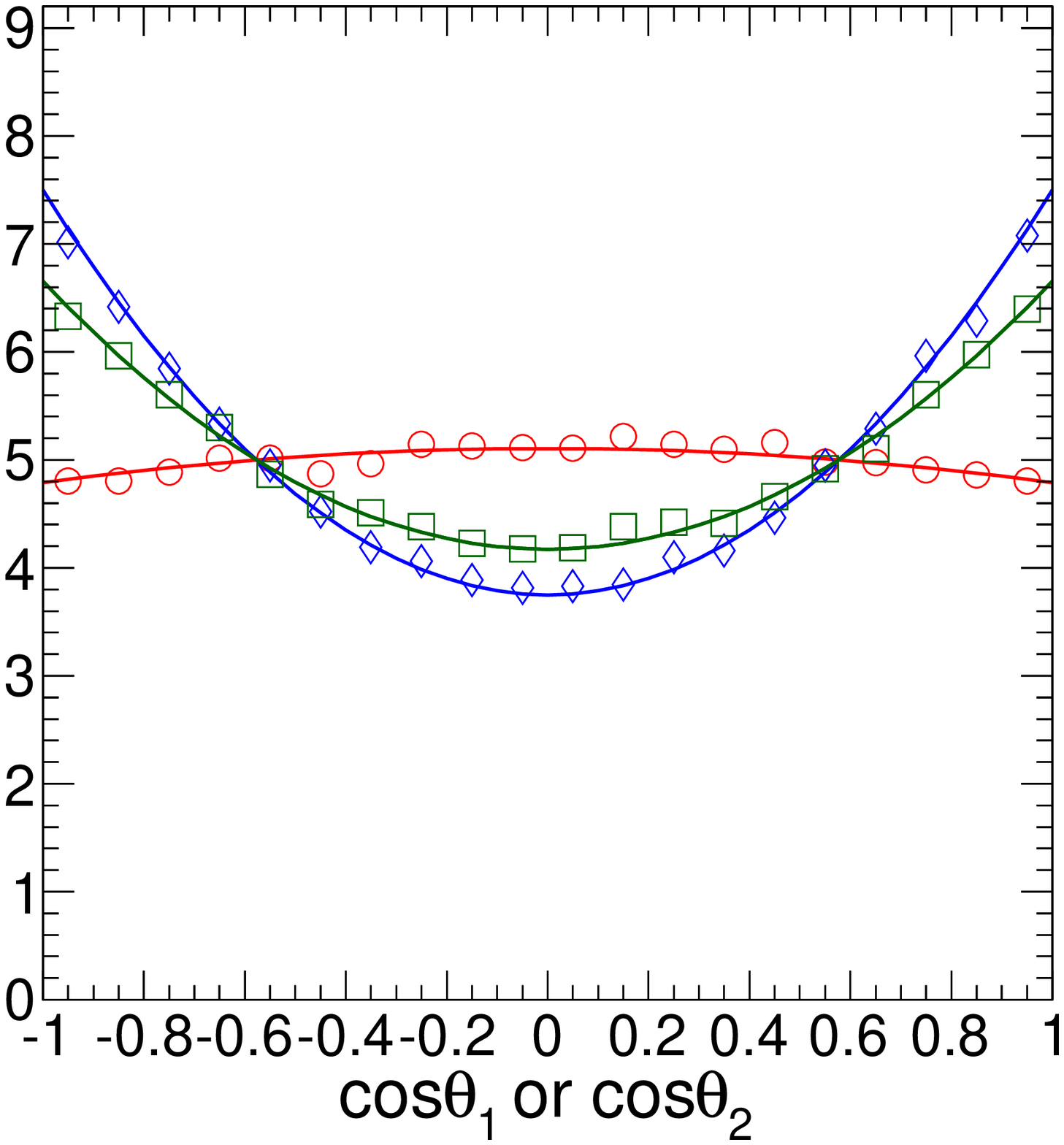,width=0.25\linewidth}
}
\centerline{
\epsfig{figure=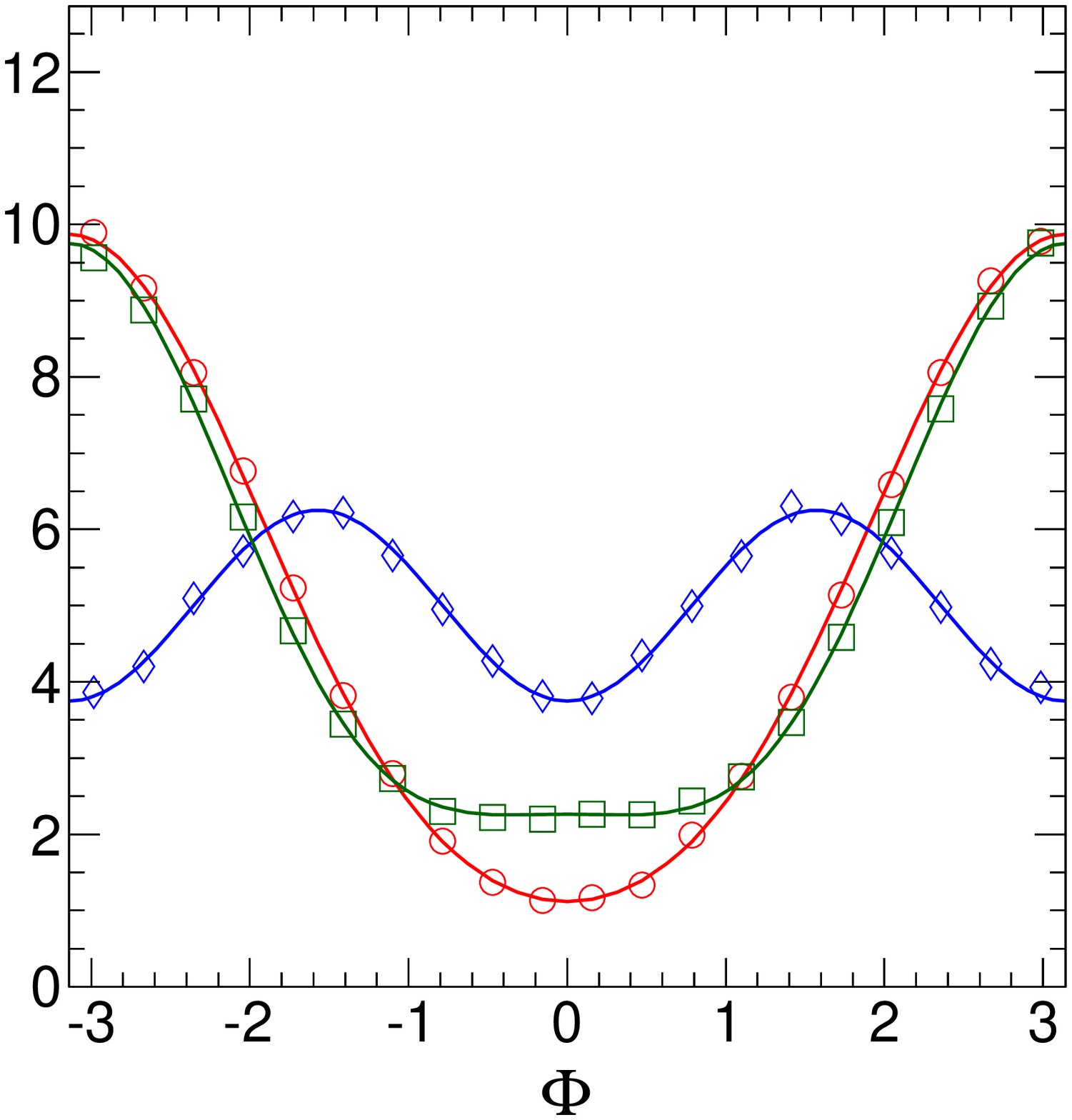,width=0.25\linewidth}
\epsfig{figure=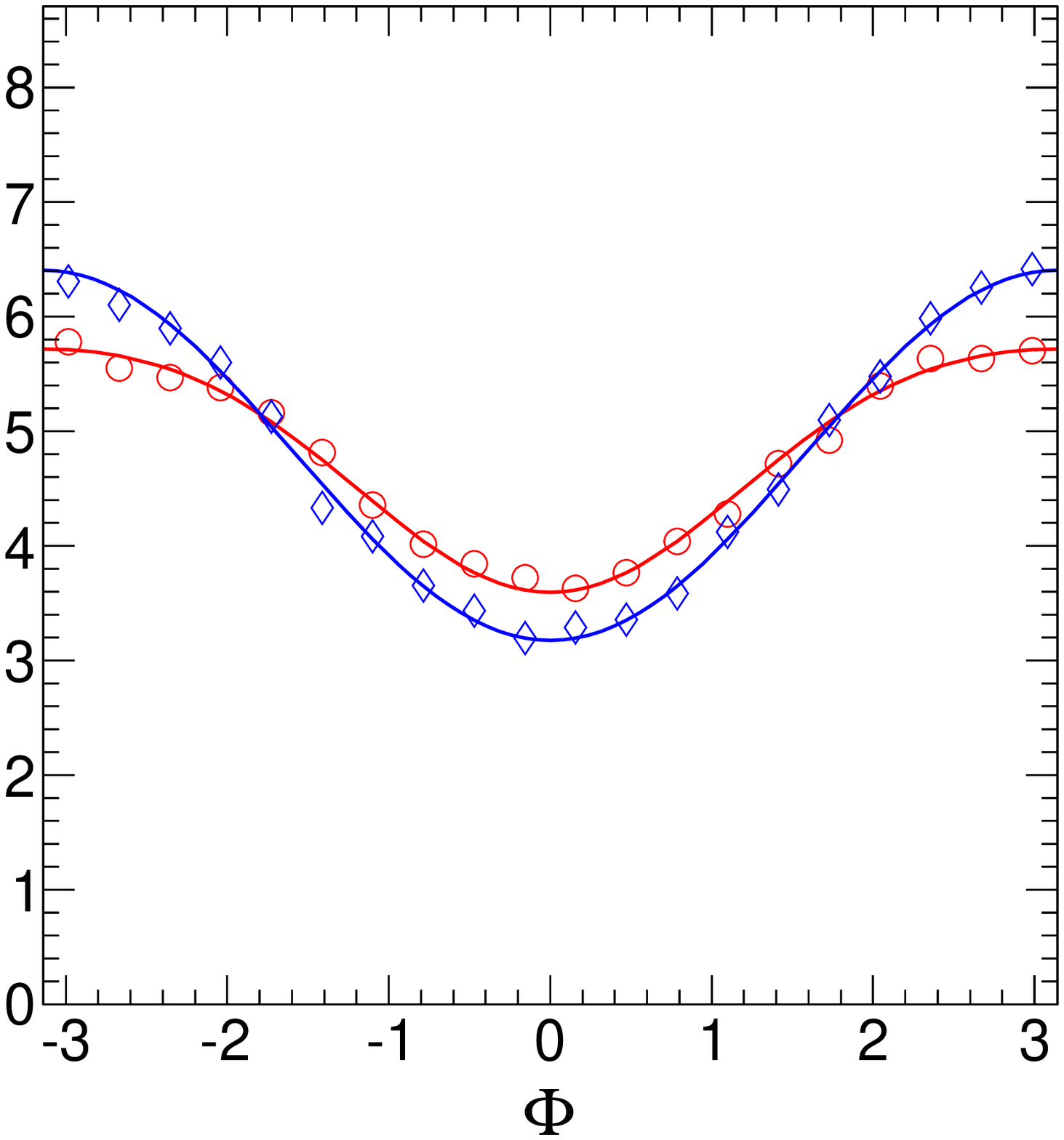,width=0.25\linewidth}
\epsfig{figure=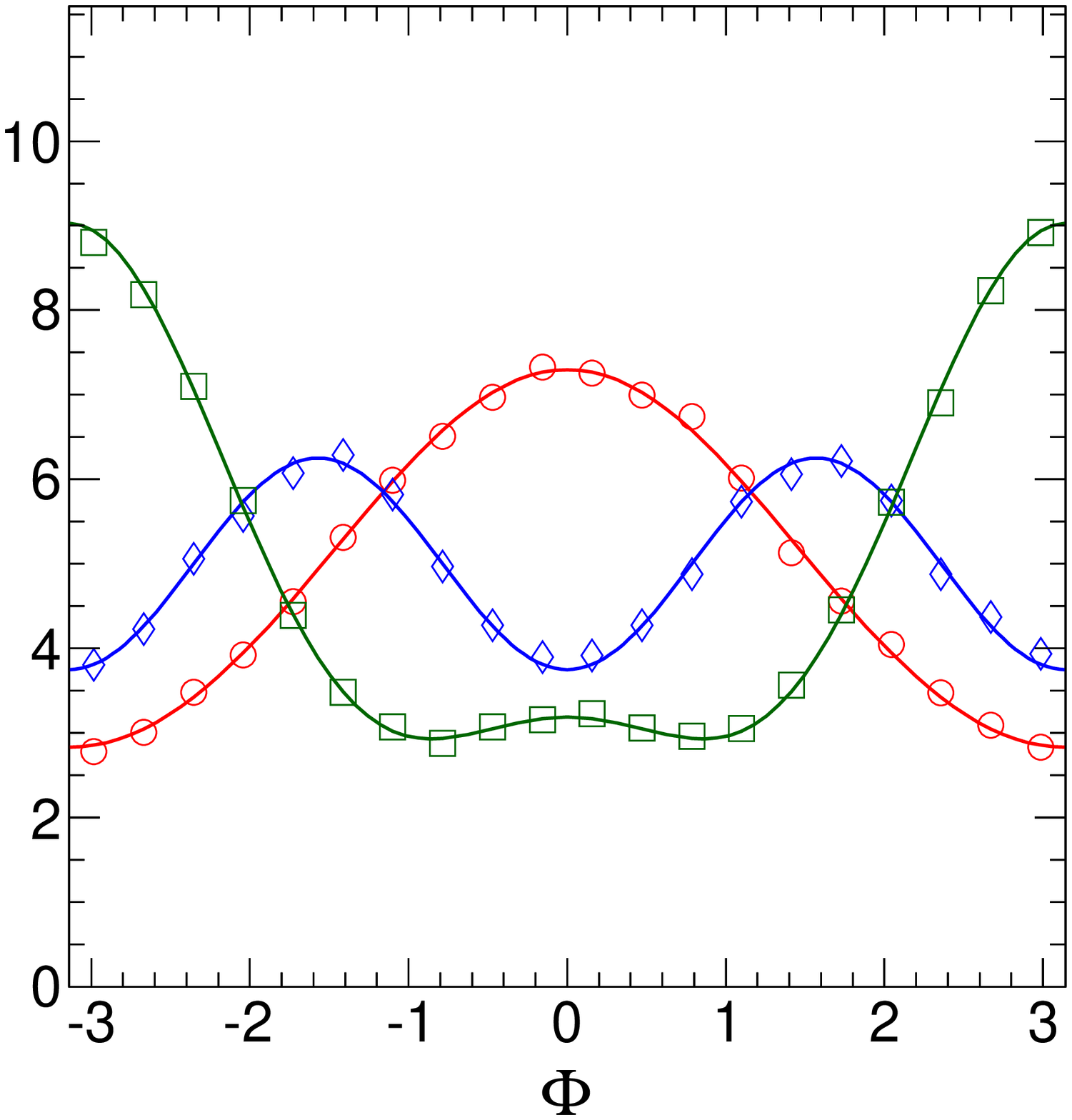,width=0.25\linewidth}
}
\caption{
Distributions of the observables in the $X\to WW$ analysis, from left to right:
spin-zero, spin-one, and spin-two signal.
The signal hypotheses shown are $J^+_m$ (red circles), $J^+_h$ (green squares), 
$J^-_h$ (blue diamonds), as defined in Table~\ref{table-scenarios}. 
The observables shown from top to bottom: 
$m_{1,2}$, $\cos\theta^*$, $\Phi_1$, $\cos\theta_{1,2}$, and $\Phi$.
Points show simulated events and lines show projections of analytical distributions.
}
\label{fig:simulated-ww}
\end{figure}





\begin{thebibliography}{99}

\clearpage


\bibitem{discovery-atlas} 
ATLAS collaboration, Phys. Lett. B  {\bf 716}, 1 (2012).

\bibitem{discovery-cms} 
CMS collaboration, Phys. Lett. B {\bf 716}, 30 (2012).

\bibitem{evidence}
CDF and D0 collaborations, Phys. Rev. Lett.  {\bf 109}, 071804 (2012).

\bibitem{landau}
L. D. Landau, Dokl. Akad. Nauk SSSR {\bf 60}, 207 (1948).

\bibitem{yang}
C. N. Yang, Phys. Rev. {\bf 77}, 242 (1950).

\bibitem{fits} 
D.~Carmi {\it et al.}, arXiv:1207.1718 [hep-ph]; 
P.~P.~Giardino {\it et al.},  arXiv:1207.1347 [hep-ph]; 
J.~Ellis and T.~You, arXiv:1207.1693 [hep-ph]; 
J.~R.~Espinosa {\it et al.}, arXiv:1207.1717 [hep-ph]; 
I.~Low, J.~Lykken, and G.~Shaughnessy, arXiv:1207.1093 [hep-ph];
T. Corbett {\it et al.}, arXiv:1207.1344 [hep-ph];
M. Montull and F. Riva, arXiv:1207.1716 [hep-ph];
S.~Banerjee, S. Mukhopadhyay, and B.~Mukhopadhyay, arXiv:1207.3588 [hep-ph];
F.~Bonnett {\it et al.}, arXiv:1207.4599 [hep-ph];  
T.~Plehn and M.~Rauch, arXiv:1207.6108 [hep-ph].

\bibitem{zeppenfeld}
B.~Coleppa, K.~Kumar and H.~E.~Logan, arXiv:1208.2692 [hep-ph]; 
see also plenary talk by D.~Zeppenfeld, 
SUSY 2012, Beijing, August 13--19 2012.

\bibitem{DellAquila1986}
J. R. Dell'Aquila and C. A. Nelson, Phys. Rev. D {\bf 33}, 80 (1986).

\bibitem{Nelson1988}
C. A. Nelson, Phys. Rev. D {\bf 37}, 1220 (1988).

\bibitem{Soni1993}
A. Soni and R. M. Xu, 
Phys. Rev. D {\bf 48}, 5259 (1993).

\bibitem{Barger1994}
V. Barger {\it et al.}, 
Phys. Rev. D {\bf 49}, 79 (1994).

\bibitem{Allanach2002}
B. C. Allanach {\it et al.}, 
JHEP {\bf 0212}, 039 (2002).

\bibitem{Choi2003}
S. Y. Choi, D. J. Miller, M. M. Muhlleitner, and P. M. Zerwas, Phys. Lett. B {\bf 553}, 61 (2003).

\bibitem{Buszello2004}
C. P. Buszello, I. Fleck, P. Marquard, and J. J. van der Bij, Eur. Phys. J. C {\bf 32}, 209 (2004).

\bibitem{Godbole2007}
R. M. Godbole, D. J. Miller, and M. M. Muhlleitner, J. High Energy Phys. {\bf 12}, 031 (2007).

\bibitem{Keung2008}
W. Y. Keung, I. Low, and J. Shu, Phys. Rev. Lett. {\bf 101}, 091802 (2008).

\bibitem{Antipin2008}
O. Antipin and A. Soni, J. High Energy Phys. {\bf 10}, 018 (2008).

\bibitem{Hagiwara2009}
K. Hagiwara, Q. Li, and K. Mawatari, J. High Energy Phys. {\bf 07}, 101(2009).

\bibitem{Cao2010}
Q.-H. Cao {\it et al.}, Phys. Rev. D {\bf 81}, 015010 (2010).

\bibitem{Gao:2010qx}
Y.Y. Gao {\it et al.}, Phys. Rev. D {\bf 81}, 075022 (2010).

\bibitem{DeRujula2010}
A. De Rujula {\it et al.}, Phys. Rev. D {\bf 82}, 013003 (2010).

\bibitem{Englert2010}
C. Englert, C. Hackstein, and M. Spannowsky, Phys. Rev. D {\bf 82}, 114024 (2010).

\bibitem{Gainer2011}
J. S. Gainer {\it et al.}, 
JHEP  {\bf 11}, 027 (2011).

\bibitem{Ellis2012}
J. Ellis and D. S. Hwang,
JHEP {\bf 09}, 071 (2012). 

\bibitem{cmshzz2l2q}
CMS collaboration,
JHEP {\bf 04}, 036 (2012).

\bibitem{dalitz}
R. H. Dalitz, Proc. Phys. Soc. London Sect. A {\bf 64}, 667 (1951).

\bibitem{Jacob1959}
M. Jacob and G. C. Wick,
Ann. Phys. {\bf 7}, 404 (1959).

\bibitem{Cabibbo1965}
N. Cabibbo and A. Maksymowicz,
Phys. Rev. {\bf 137}, 438 (1965).

\bibitem{Dunietz1991}
I. Dunietz  {\it et al.}, 
Phys. Rev. D {\bf 43}, 2193 (1991).

\bibitem{Kramer1992}
G. Kramer  and W. F. Palmer,
Phys. Rev. D {\bf 45}, 193 (1992).

\bibitem{pdg}
Particle Data Group, Phys. Rev. D {\bf 86}, 010001 (2012).

\bibitem{bpolarization}
A. V. Gritsan and J. G. Smith,
see review "Polarization in $B$ decays" in Ref.~\cite{pdg}.

\bibitem{Collins1977} 
J. C. Collins and D. E. Soper,
Phys. Rev. D  {\bf 16}, 2219 (1977).

\bibitem{Gunion:2012gc} 
  J.~F.~Gunion, Y.~Jiang and S.~Kraml,
  arXiv:1207.1545 [hep-ph].

\bibitem{support} 
The Monte-Carlo generator, the manual, and supporting material can be downloaded from 
http://www.pha.jhu.edu/spin/

\bibitem{pythia}
T. Sjostrand, S. Mrenna, and P. Skands,
JHEP {\bf 05}, 026 (2006).

\bibitem{cteq}
J.~Pumplin  {\it et al.},
JHEP {\bf 0207}, 012 (2002);
P.~M.~Nadolsky {\it et al.},
Phys.\ Rev.\  D {\bf 78}, 013004 (2008).

\bibitem{powheg}
P. Nason,
JHEP {\bf 11}, 040 (2004);
S. Frixione, P. Nason, and C. Oleari, 
JHEP {\bf 11}, 070 (2007);
S. Alioli {\it et al.},
JHEP {\bf 07},  060 (2008).

\bibitem{madgraph}
J. Alwall {\it et al.},
JHEP {\bf 0709}, 028 (2007).

\bibitem{incandela2012}
Signal hypothesis separation discussion and kinematic distributions in the $ZZ$ channel
are not included in Ref.~\cite{discovery-cms}, for more details see
J. Incandela, for CMS collaboration,
CERN seminar, July 4, 2012;
A. V. Gritsan, for CMS collaboration,
FNAL seminar, July 9, 2012.

\bibitem{drellyan-cms} 
CMS collaboration,
Phys. Rev. D  {\bf 84}, 112002 (2011).

\end{thebibliography}
\end{document}